\documentclass[a4paper,11pt]{article}
\usepackage{amssymb}
\usepackage{graphicx}
\usepackage{amsmath}

\DeclareFontFamily{U}{rsf}{}
\DeclareFontShape{U}{rsf}{m}{n}{
  <5> <6> rsfs5 <7> <8> <9> rsfs7 <10-> rsfs10}{}
\DeclareMathAlphabet\Scr{U}{rsf}{m}{n} \makeatletter
\@addtoreset{equation}{section} \makeatother

\def\be{\begin{equation}}
\def\ee{\end{equation}}
\def\ba{\begin{array}}
\def\ea{\end{array}}

\newcommand{\bea}{\begin{eqnarray}}
\newcommand{\eea}{\end{eqnarray}}

\begin{document}

\begin{titlepage}

\begin{flushright}
CERN-PH-TH/2012-150
\end{flushright}

\vskip 2.0 cm
\begin{center}  {\huge{\bf      Super-Ehlers in Any Dimension}}

\vskip 1.5 cm

{\Large{\bf Sergio Ferrara$^{1,2,3}$}, {\bf Alessio Marrani$^1$}, {\bf Mario Trigiante$^4$}}

\vskip 1.0 cm

$^1${\sl Physics Department, Theory Unit, CERN,\\CH 1211, Geneva 23, Switzerland\\
\texttt{sergio.ferrara@cern.ch}\\
\texttt{alessio.marrani@cern.ch}}\\

\vskip 0.5 cm

$^2${\sl INFN - Laboratori Nazionali di Frascati,\\Via Enrico Fermi
40, I-00044 Frascati, Italy}\\

\vskip 0.5 cm

$^3${\sl Department of Physics and Astronomy,\\University of California, Los Angeles, CA USA}

\vskip 0.5 cm

$^4${\sl Laboratory of Theoretical Physics,\\Dipartimento di Scienze Applicate e Tecnologia, Politecnico di Torino,\\ C.so Duca degli Abruzzi, I-10129 Torino, Italy\\
\texttt{mario.trigiante@polito.it}}

 \end{center}

 \vskip 3.5 cm

\begin{abstract}

We classify the enhanced helicity symmetry of the Ehlers group to extended supergravity theories in any dimension.The vanishing character of the pseudo-Riemannian cosets occurring
in this analysis is explained in terms of Poincar\'{e} duality .The latter resides in the nature of  regularly
embedded  quotient subgroups which are noncompact rank preserving.

 \end{abstract}
\vspace{24pt} \end{titlepage}


\newpage \tableofcontents \newpage

\section{\label{Intro}Introduction}

Three decades ago it was shown \cite{GS} that the $D$-dimensional \textit{%
Ehlers group} $SL(D-2,\mathbb{R})$ is a symmetry of $D$-dimensional Einstein
gravity, provided that the theory is formulated in the light-cone gauge. For
any $D\geqslant 4$-dimensional Lorentzian space-time, this results enables
to identify the graviton degrees of freedom with the Riemannian coset%
\begin{equation}
\mathcal{M}_{grav}=\frac{SL\left( D-2,\mathbb{R}\right) }{SO\left(
D-2\right) },
\end{equation}%
even if the action of the theory is not simply the sigma model action on
this coset (with the exception of a theory reduced to $D=3$). In $D=4$, this
statement reduces to the well known fact that the massless graviton
described by the Einstein-Hilbert action with two degrees of freedom of $\pm
2$ helicity has an enhanced symmetry $SO(2)\rightarrow SL(2,\mathbb{R})$.

In $\mathcal{N}$-extended supergravity in $D$ dimensions, $U$-duality%
\footnote{%
Here $U$-duality is referred to as the \textquotedblleft
continuous\textquotedblright\ symmetries of \cite{CJ-1}. Their discrete
versions are the $U$-duality non-perturbative string theory symmetries
introduced by Hull and Townsend \cite{HT-1}.} symmetries play an important
role to uncover, in terms of geometrical constructions, the non-linear
structure of the theories, whose most symmetric one is the theory with
maximal supersymmetry ($2N=32$ supersymmetries). Furthermore, $U$-duality
symmetries get unified with the Ehlers space-time symmetry if one descends
to $D=3$ \cite{CJLP,LP}. In the maximal case, the $D=3$ $U$-duality group is
$E_{8(8)}$, with \textit{maximal compact subgroup} (\textit{mcs}) $SO(16)$,
which is also the underlying Clifford algebra for massless supermultiplets
of maximal supersymmetry. As a consequence, the bosonic sector of the theory
is described by the sigma model $E_{8(8)}/SO(16)$ \cite%
{Cremmer-Lects,Julia-Lects,MS}.

Following these preliminaries, it comes as no surprise that it was further
discovered that in light-cone Hamiltonian formulation maximal supergravity
exhibits $E_{7(7)}$ symmetry in $D=4$ \cite{Br-2} and $E_{8(8)}$ symmetry in
$D=3$ \cite{Br-3} (for the $D=11$ theory, see \cite{Br-1}). Indeed, in any
space-time dimension $D$ and for any number of supersymmetries $\mathcal{N}%
=2N$, it is known that the $D=3$ $U$-duality group $G_{N}^{3}$ \cite%
{Tollsten} embeds (through a rank-preserving embedding; for some basic
definitions, see the start of App. A) the Ehlers group $SL(D-2,\mathbb{R})$
as a commutant of the $U$-duality group $G_{N}^{D}$ \cite{Keu-1,Keu-2}:%
\begin{equation}
G_{N}^{3}\supset G_{N}^{D}\times SL(D-2,\mathbb{R}).  \label{emb}
\end{equation}

It is then natural to conjecture that in a suitable light-cone formulation
of any $\mathcal{N}$-extended supergravity theories $G_{N}^{D}\times SL(D-2,%
\mathbb{R})$ (which we dub \textit{super-Ehlers group}) is a manifest
symmetry of the theory. Even if the super-Ehlers group is a bosonic
extension of the Ehlers group itself, the presence of the $U$-duality
commutant $G_{N}^{D}$ in (\ref{emb}) is closely related to supersymmetry. It
is intriguing to notice that the super-Ehlers symmetries, which we classify
below in any dimension, sometimes exhibit an \textit{\textquotedblleft
enhancement"} into some larger group\footnote{%
For enhancement to infinite symmetries, see \cite{MSS}.}; this occurs
whenever the embedding (\ref{emb}) is \textit{non}-maximal, and in $D=10$
type IIB supergravity. Furthermore, it sometimes occurs that the embedding (%
\ref{emb}) is maximal but \textit{non-symmetric}, as in $D=11$ supergravity.

At any rate, we will show that the common features of the embedding (\ref%
{emb}) are \textit{at least} two (\textit{cfr.} the start of App. A):

\begin{itemize}
\item It is \emph{regular} and preserves the rank of the group. Indeed, it
generally holds that%
\begin{equation}  \label{rank1}
\text{rank}\left( G_{N}^{3}\right) =\text{rank}\left( G_{N}^{D}\right) +%
\text{rank}\left( SL(D-2,\mathbb{R})\right) =\text{rank}\left(
G_{N}^{D}\right) +D-3.
\end{equation}%
The same relation holds for the \emph{non-compact rank} of these groups,
namely the rank of the corresponding symmetric Riemannian manifolds of which
the groups encode the isometries:
\begin{equation}  \label{rank2}
\text{rank}\left(\frac{ G_{N}^{3}}{H_N^3}\right) =\text{rank}\left(\frac{
G_{N}^{D}}{H_N^D}\right) +\text{rank}\left( \frac{SL(D-2,\mathbb{R})}{%
\mathrm{SO}(D-2)}\right) =\text{rank}\left(\frac{ G_{N}^{D}}{H_N^D}\right)
+D-3\,,
\end{equation}%
where $H_N^3$ and $H_N^D$ are the maximal compact subgroups of $G_{N}^{3}$
and $G_{N}^{D}$, respectively. As mentioned above, this does not imply the
embeddings to be in general maximal nor symmetric.

\medskip

\item The pseudo-Riemannian coset resulting from (\ref{emb}) has always zero
\textit{character} \cite{Helgason,Gilmore}, namely it has the same number of
compact and non-compact generators. We will show that this latter property
is related to the \textit{Poincar\'{e} duality} (\textit{alias}
electric-magnetic duality) of the spectrum of massless $p>0$ forms of the
theory, which can essentially be traced back to the existence of an \textit{%
Hodge involution} in the cohomology of the scalar manifold, singling out
only the physical forms and their duals in the cohomology of the $\left(
D-2\right) $-dimensional transverse space. This property also follows from
the regularity of the embedding of $G_{N}^{D}\times SL(D-2)$ inside $%
G_{N}^{3}$, the semisimplicity of the two groups and properties (\ref{rank1}%
),(\ref{rank2}), as it will be shown in Appendix \ref{GF}.
\end{itemize}

There is also another aspect of interest in the present analysis : the role
played by exceptional Lie groups and their relation to Jordan algebras and
Freudenthal triple systems \cite{GST,G-Lects}. In particular, a mathematical
construction, called \textit{Jordan pairs} (see e.g. \cite{Truini-1} for a
recent treatment, and a list of Refs.) corresponds to the maximal
non-symmetric embedding%
\begin{equation}
E_{8(8)}\supset E_{6(6)}\times SL(3,\mathbb{R}),  \label{UCCC}
\end{equation}%
which is nothing but (\ref{emb}) specified for maximal supersymmetry ($N=16$%
) and $D=5$. We point out that the \textit{Jordan pairs} relevant for
supergravity theories always pertain to suitable non-compact real forms of
Lie algebras, differently \textit{e.g.} from the treatment given in \cite%
{Truini-1}.

Moreover, it is worth observing that in $D=11$ supergravity $G_{16}^{11}$ is
empty, and thus (\ref{emb}) is the following maximal non-symmetric embedding
\cite{CJLP}:%
\begin{equation}
E_{8(8)}\supset SL(9,\mathbb{R}),  \label{UCCC-1}
\end{equation}%
which in fact was used long time ago \cite{Nahm} in order to construct the
gravity multiplet of this theory \cite{CJS}. For maximal supergravity ($%
N=N_{\max }=16$), (\ref{emb}) reads \footnote{%
This embedding was considered, but not proved, in \cite{Keu-1}. A proof is
presented in App. \ref{App-Embeddings} of the present paper.}%
\begin{equation}
E_{8(8)}\supset E_{11-D\left( 11-D\right) }\times SL(D-2,\mathbb{R}),
\label{emb-N=16}
\end{equation}%
where $G_{16}^{D}=E_{11-D\left( 11-D\right) }$ denotes the so-called \textit{%
Cremmer-Julia sequence} \cite{Cremmer-Lects,Julia-Lects}. The unique
exception is provided by type IIB chiral $D=10$ supergravity, in which (\ref%
{emb}) is given by a two-step chain of maximal embeddings\footnote{%
\textquotedblleft $s$" and \textquotedblleft $ns$" stand for
\textquotedblleft symmetric" and "non-symmetric" (embedding), respectively.}:%
\begin{equation}
E_{8(8)}\supset _{s}SL(2,\mathbb{R})\times E_{7(7)}\supset _{s}SL(2,\mathbb{R%
})\times SL(8,\mathbb{R}),  \label{UCCC-3}
\end{equation}%
which preserves the group rank.\bigskip

The plan of the paper is as follows.

In Sec. \ref{Sec-Maximal} we start by recalling some basic facts on $SO(N)$
Clifford algebras relevant for the classification of massless multiplets of $%
\mathcal{N}$-extended supersymmetry in any dimension. Here $\mathcal{N}=2N$
denotes the number of supersymmetries, regardless of the dimension $D$.
Thus, for instance maximal supergravity corresponds to $\mathcal{N}=32$ ($8$
spinor supercharges in $D=4$), whereas the minimal supergravity we consider
has $\mathcal{N}=8$ ($2$ spinor supercharges in $D=4$). We then proceed to
considering the embedding (\ref{emb-N=16}) pertaining to maximal
supergravity in any dimension $D\geqslant 4$ (in $D=10$ both IIA and IIB
theories are considered). The embedding (\ref{emb}), which can be regarded
as the \textit{\textquotedblleft non-compact enhancement"} of Nahm's
analysis \cite{Nahm}, in all cases consistently provides the massless
spectrum of the corresponding theory with the correct spin-statistics
content; illustrative analysis is worked out for $D=11$ and $D=10$ maximal
theories. Other theories which do not exhibit matter coupling are also
considered, namely $N=10,12$ in $D=4$ and $N=12$ in $D=5$.

In Sec. \ref{Sec-Half-Maximal} we consider half-maximal supergravity
theories ($N=8$), which can be matter coupled and exist in all $D\leqslant
10 $ dimensions; for $D=6$ we consider both inequivalent theories, namely
the chiral $(2,0)$ (type IIB) and the non-chiral $(1,1)$ (type IIA) ones.
Theories with $N=6$, living in $D=4$, are also considered.

Then, in Sec. \ref{Sec-Quarter-Maximal} we consider quarter-maximal theories
($N=4$), which live in $D=4,5,6$ and admit two different kinds of matter
multiplets. We confine ourselves to theories with symmetric scalar
manifolds, which (apart from the \textit{minimally coupled} models in $D=4$
and the \textit{non-Jordan} symmetric sequence in $D=5$) admit an
interpretation in terms of Euclidean Jordan algebras.

Pseudo-Riemannian cosets associated to the maximal-rank embeddings (\ref{emb}%
) are then analyzed in Sec. \ref{Sec-Chi=0-Cosets}. All such cosets enjoy
the property of having the same number of compact and non--compact
generators. This is also proven, using general group theoretical arguments,
in Appendix \ref{GF}. In\ Subsec. \ref{PD} this property is related to the
invariance of the spectrum of massless bosonic $p>0$ forms under \textit{%
Poincar\'{e}-duality}, or equivalently in Subsec. \ref{Hodge} in terms of an
\textit{Hodge involution} acting on the coset cohomology.

Final remarks and outlook are given in\ Sec. \ref{Sec-Conclusion}.

Three Appendices conclude the paper. In App. \ref{App-Embeddings}, some
embeddings of non-compact, real forms relevant for our analysis are
rigorously proved, while in App. \ref{App-Spinor-Polarizations} the issue of
inequivalent \textquotedblleft dual pairs\textquotedblright\ of subalgebras
of the $U$-duality algebra is discussed (see also \cite{BT-1}). The related
notions of $T$-dualities as $\mathfrak{so}(8,8)$ outer-automorphisms are
also dealt with. In App. \ref{pdld} the issue of Poincar\'{e} duality is
revisited with an explicit algebraic construction which makes use of
appropriate \emph{level decompositions}.

\section{\label{Sec-Maximal}Clifford Algebras and \textquotedblleft Pure"
Theories}

In the seminal paper by Nahm \cite{Nahm}, it was shown how massless
multiplets of supergravity are built in terms of irreps. of $SO(D-2)$, the
little group (\textit{spin}) of massless particles in $D$ dimensions. The
number of supersymmetries $2N$ is encoded in the Clifford algebra of $SO(N)$%
, and therefore the supermultiplets can be regarded as $SO\left( N\right) $
spinors decomposed into $SO\left( D-2\right) $ irreps (for theories with
particles with spin $s\leqslant 2$, which we consider throughout, $N_{\max
}=16$). Bosons and fermions thus correspond to the two semi-spinors (or
chiral spinors) of\footnote{%
Note that $N$ is always even, since for $D\geqslant 4$ spinor charges have
real dimensions multiples of $4$.
\par
Furthermore, it should be remarked that the cases $D=4$, $N=2$ and $D=10$, $%
N=8$ are somewhat particular, because $N=D-2$, so the two Clifford spinors
directly provide bosonic and fermionic supermultiplets' representations.} $%
SO(N)$.

In any dimension $D\geqslant 4$, $SO\left( N\right) $ exhibits a certain
commuting factor with the massless little group $SO\left( D-2\right) $. For
\textquotedblleft pure" supergravities, in which only the gravity multiplet
is present, such a commuting factor is the so-called $\mathcal{R}$-symmetry
of the theory. Then the question arises as to which is the non-compact group
commuting with the $SL(D-2,\mathbb{R})$ Ehlers group (which thus extends the
massless little group including the $\mathcal{R}$-symmetry), and furthermore
which is the non-compact group which extends the $SO(N)$ of the $N$%
-dimensional Clifford algebra pertaining to $2N$ local supersymmetries.

In describing massless multiplets of theories with $\mathcal{N}=2N$ local
supersymmetries, one consider the the rest-frame supersymmetry algebra
without central extension. Since the momentum squares to zero ($P^{\mu
}P_{\mu }=0$), only half of the supersymmetry charges survive, and the
creation operators of $N$ charges describe an $SO(N)$ Clifford algebra.
Moreover, due to the fact that in $D\geqslant 4$ spinors always have real
dimension multiple of $4$, $N$ is always even : $N=4,6,8,10,12,16$ (we do
not consider here $N=2$ at $D=4$, namely minimal $4$-dimensional
supergravity with $1$ spinor supercharge). It thus comes as no surprise that
$U$-duality groups $G_{N}^{3}$ in $D=3$ (in which there is only distinction
between bosons and fermions, but no spin is present for massless states)
contain in their \textit{mcs} the Clifford algebra symmetry $SO(N)$.

Supersymmetry dictates that massless bosons and fermions are simply the two
(chiral, semi-) spinor irreps. of $SO(N)$, while their spin $s$ content in $%
D $ space-time dimensions is obtained by suitably branching such irreps.
into $SO(D-2)$, which is the little group (spin) for massless particles in $%
D $ dimensions.

In the present Section we consider \textit{\textquotedblleft pure"} theories
in which the matter coupling is not allowed; they include \textit{maximally
supersymmetric} ($N=16$) theories in any dimension $D\leqslant 11$, as well
as \textit{non-maximal} theories with $N=10,12$ in $D=4$ and $N=12$ in $D=5$%
. For such theories, the Clifford algebra $SO(N)$ is nothing but the $mcs$
of the $U$-duality group $G_{N}^{3}$ in $D=3$; for \textit{non-maximal}
theories ($N<16$), this is true up to the presence of the so-called \textit{%
Clifford vacuum} factor group, which expresses further degeneracy of the
Clifford algebra symmetry. Moreover, the group $H_{N}^{D}=mcs\left(
G_{N}^{D}\right) $ which commutes with $SO(D-2)$ inside $SO(N)$ is the $%
\mathcal{R}$-symmetry, providing the degeneracy of the spin $s$
representations in the decomposition of the chiral spinors under the
embedding\footnote{%
Further commuting factor group occurs in the l.h.s. of (\ref{emb-mcs}) in
non-maximal ($N\leqslant 16$) theories; see analysis below.} \cite{Nahm}%
\begin{equation}
SO\left( N\right) \supset H_{N}^{D}\times SO\left( D-2\right) _{J}
\label{emb-mcs}
\end{equation}%
which is the (not necessarily maximal-rank, nor maximal nor symmetric)
counterpart of (\ref{emb}) at the level of \textit{mcs}. The subscript
\textquotedblleft $J$" denotes the spin group throughout.

\subsection{$N=16$ (\textit{Maximal})}

\label{maxim}

For \textit{maximal }($N=16$) supergravity theories with massless particles,
the $D=3$ $U$-duality group is $G_{3}^{16}=E_{8(8)}$, with \textit{mcs} $%
SO(16)$, which is the Clifford algebra for massless particles with $\mathcal{%
N}=32$ supersymmetries. (\ref{emb-N=16}) provides the rank-preserving
embedding of $D$-dimensional Ehlers group $SL\left( D-2,\mathbb{R}\right) $
into $E_{8(8)}$. The group commuting with $SL\left( D-2,\mathbb{R}\right) $
inside $E_{8(8)}$ is nothing but the $D$-dimensional $U$-duality group $%
G_{16}^{D}=E_{11-D\left( 11-D\right) }$, belonging to the so-called the
Cremmer-Julia sequence. All cases in $4\leqslant D\leqslant 11$ dimensions
are reported in Table 1 (non-compact level (\ref{emb})-(\ref{emb-N=16})) and
in Table 2 (\textit{mcs} level (\ref{emb-mcs})). In particular, in Table 2
also the decomposition of the vector irrep. $\mathbf{16}$ of the Clifford
algebra $SO(16)=mcs\left( E_{8(8)}\right) $ of maximal ($N=16\rightarrow
\mathcal{N}=32$) supersymmetry is reported for the embedding (\ref{emb-mcs})
pertaining to this case, namely \cite{Nahm} (see also \cite{Spinor-Algebras}%
):%
\begin{equation}
SO(16)\supset \mathcal{R}_{D}^{16}\times SO(D-2)_{J},  \label{emb-mcs-N=16}
\end{equation}%
where, as mentioned above, $\mathcal{R}_{D}^{16}\equiv mcs\left(
G_{D}^{16}\right) \equiv H_{D}^{16}$ is the $\mathcal{R}$-symmetry of the
maximal supergravity in $D$ (Lorentzian) space-time dimensions. Note that
the irrep. of $SO(D-2)$ occurring in the branching of the $\mathbf{16}$
along (\ref{emb-mcs-N=16}) are all spinors, and the $\mathcal{R}$-symmetry $%
\mathcal{R}_{D}^{16}$ is real, pseudo-real (quaternionic), complex,
depending on whether such spinor irrep. is real, pseudo-real or complex,
respectively.

\begin{table}[t!]
\begin{center}
\begin{tabular}{|c||c|c|}
\hline
$D$ & $%
\begin{array}{c}
\\
E_{8(8)}\supset E_{11-D\left( 11-D\right) }\times SL(D-2,\mathbb{R}) \\
~~%
\end{array}%
$ & $%
\begin{array}{c}
\\
\text{type} \\
~~%
\end{array}%
$ \\ \hline\hline
$%
\begin{array}{c}
\\
11 \\
~%
\end{array}%
$ & $E_{8(8)}\supset SL(9,\mathbb{R})~$ & $max$,~$ns$ \\ \hline
$%
\begin{array}{c}
\\
10,~IIA \\
~%
\end{array}%
$ & $E_{8(8)}\supset SO(1,1)\times SL(8,\mathbb{R})$ & $nm$,~$ns$ \\ \hline
$%
\begin{array}{c}
\\
10,~IIB \\
~%
\end{array}%
$ & $E_{8(8)}\supset SL(2,\mathbb{R})\times SL(8,\mathbb{R})$ & $nm$,~$ns$
\\ \hline
$%
\begin{array}{c}
\\
9 \\
~%
\end{array}%
$ & $E_{8(8)}\supset GL(2,\mathbb{R})\times SL(7,\mathbb{R})$ & $nm$,~$ns~$
\\ \hline
$%
\begin{array}{c}
\\
8 \\
~%
\end{array}%
$ & $E_{8(8)}\supset SL(2,\mathbb{R})\times SL(3,\mathbb{R})\times SL(6,%
\mathbb{R})$ & $nm$,~$ns$ \\ \hline
$%
\begin{array}{c}
\\
7 \\
~%
\end{array}%
$ & $E_{8(8)}\supset SL(5,\mathbb{R})\times SL(5,\mathbb{R})$ & $max$,~$ns$
\\ \hline
$%
\begin{array}{c}
\\
6 \\
~%
\end{array}%
$ & $E_{8(8)}\supset SO(5,5)\times SL(4,\mathbb{R})$ & $nm$,~$ns$ \\ \hline
$%
\begin{array}{c}
\\
5 \\
~%
\end{array}%
$ & $E_{8(8)}\supset E_{6(6)}\times SL(3,\mathbb{R})$ & $max$,~$ns$ \\ \hline
$%
\begin{array}{c}
\\
4 \\
~%
\end{array}%
$ & $E_{8(8)}\supset E_{7(7)}\times SL(2,\mathbb{R})$ & $max$,~$s$ \\ \hline
\end{tabular}%
\end{center}
\caption{Embedding $G_{N}^{3}\supset G_{N}^{D}\times SL(D-2,\mathbb{R})$ (%
\protect\ref{emb}) for \textit{maximal} supergravity theories ($N=16$) in $%
11\geqslant D\geqslant 4$ Lorentzian space-time dimensions \protect\cite%
{Keu-1,Keu-2}. $G_{N}^{D}$ is the $U$--duality group in $D$ dimensions for
the theory with $\mathcal{N}=2N$ supersymmetries. $SL(D-2,\mathbb{R})$ is
the Ehlers group in $D$ dimensions. For $N=16$, $G_{16}^{3}=E_{8(8)}$, and $%
G_{N}^{D}=E_{11-D(11-D)}$ belongs to the Cremmer-Julia sequence; thus, (%
\protect\ref{emb-N=16}) is obtained. The type ($max$(imal), $n$(ext-to-)$m$%
(aximal), $s$(ymmetric), $n$(on-)$s$(ymmetric)) of embedding is indicated.
Explicit proofs are given in App. \protect\ref{App-Embeddings}. }
\end{table}

Let us scan them briefly (as anticipated, for $D=11$ and $D=10$ the massless
spectrum analysis is also worked out, as an example of the consistence of
the embeddings with the massless spectrum of the corresponding theory). For
convenience of the reader, we anticipate that the embeddings (\ref{emb}) and
(\ref{emb-mcs}) are \textit{maximal} in $D=11,7,5$ (non-symmetric) and $4$
(symmetric), while they are \textit{next-to-maximal} in $D=10,9,8,6$; in
these latter cases, an \textit{\textquotedblleft enhancement"} of $%
E_{11-D\left( 11-D\right) }\times SL\left( D-2,\mathbb{R}\right) $ occurs
(see analysis below).

\begin{table}[t!]
\begin{center}
\begin{tabular}{|c||c|c|}
\hline
$D$ & $%
\begin{array}{c}
\\
SO\left( 16\right) \supset H_{16}^{D}\times SO(D-2)_{J} \\
~~%
\end{array}%
$ & $%
\begin{array}{c}
\\
\text{type} \\
~~%
\end{array}%
$ \\ \hline\hline
$%
\begin{array}{c}
\\
11 \\
~%
\end{array}%
$ & $%
\begin{array}{c}
\\
SO\left( 16\right) \supset SO(9) \\
\mathbf{16}=\mathbf{16}~~%
\end{array}%
$ & $max$,~$ns$ \\ \hline
$%
\begin{array}{c}
\\
10,~IIA \\
~%
\end{array}%
$ & $%
\begin{array}{c}
\\
SO\left( 16\right) \supset SO(8) \\
\mathbf{16}=\mathbf{8}_{s}+\mathbf{8}_{c}%
\end{array}%
$ & $nm$,~$ns$ \\ \hline
$%
\begin{array}{c}
\\
10,~IIB \\
~%
\end{array}%
$ & $%
\begin{array}{c}
\\
SO\left( 16\right) \supset SO(2)\times SO(8) \\
\mathbf{16}=\left( \mathbf{2},\mathbf{8}_{s}\right)%
\end{array}%
$ & $nm$,~$ns$ \\ \hline
$%
\begin{array}{c}
\\
9 \\
~%
\end{array}%
$ & $%
\begin{array}{c}
\\
SO\left( 16\right) \supset SO(2)\times SO(7) \\
\mathbf{16}=\left( \mathbf{2},\mathbf{8}\right)%
\end{array}%
$ & $nm$,~$ns~$ \\ \hline
$%
\begin{array}{c}
\\
8 \\
~%
\end{array}%
$ & $%
\begin{array}{c}
\\
SO(16)\supset U(1)\times SU(2)\times SU(4) \\
\mathbf{16}=\left( \mathbf{2},\mathbf{4}\right) +\left( \overline{\mathbf{2}}%
,\overline{\mathbf{4}}\right)%
\end{array}%
$ & $nm$,~$ns$ \\ \hline
$%
\begin{array}{c}
\\
7 \\
~%
\end{array}%
$ & $%
\begin{array}{c}
\\
SO(16)\supset USp(4)\times USp(4) \\
\mathbf{16}=\left( \mathbf{4},\mathbf{4}\right)%
\end{array}%
$ & $max$,~$ns$ \\ \hline
$%
\begin{array}{c}
\\
6 \\
~%
\end{array}%
$ & $%
\begin{array}{c}
\\
SO(16)\supset USp(4)_{L}\times USp(4)_{R}\times SU(2)_{L}\times SU(2)_{R} \\
\mathbf{16}=(\mathbf{4},\mathbf{1},\mathbf{2},\mathbf{1})+(\mathbf{1},%
\mathbf{4},\mathbf{1},\mathbf{2})%
\end{array}%
$ & $nm$,~$ns$ \\ \hline
$%
\begin{array}{c}
\\
5 \\
~%
\end{array}%
$ & $%
\begin{array}{c}
\\
SO(16)\supset USp(8)\times SU(2) \\
\mathbf{16}=(\mathbf{8},\mathbf{2})%
\end{array}%
$ & $max$,~$ns$ \\ \hline
$%
\begin{array}{c}
\\
4 \\
~%
\end{array}%
$ & $%
\begin{array}{c}
\\
SO(16)\supset SU(8)\times U(1) \\
\mathbf{16}=\mathbf{8}_{1}+\overline{\mathbf{8}}_{-1}%
\end{array}%
$ & $max$,~$s$ \\ \hline
\end{tabular}%
\end{center}
\caption{Embedding $H_{N}^{3}\supset H_{N}^{D}\times SO(D-2)$ (\protect\ref%
{emb-mcs}) \protect\cite{Nahm} for \textit{maximal} supergravity theories ($%
N=16$) in $11\geqslant D\geqslant 4$ Lorentzian space-time dimensions. In
this case, as for all \textit{\textquotedblleft pure"} theories, $H_{N}^{D}$
is the $\mathcal{R}$--symmetry for the theory with $\mathcal{N}=2N$
supersymmetries. $SO(D-2)$ is the little group (spin group) for massless
particles. In this case, $H_{16}^{3}=SO(16)$ is the Clifford algebra of
maximal supersymmetry. }
\end{table}

\begin{enumerate}
\item $D=11$ ($M$-theory). There is no continuous $U$-duality (and thus $%
\mathcal{R}$-symmetry) group, and (\ref{emb-N=16}) specifies to (\ref{UCCC-1}%
), namely the maximal non-symmetric embedding of the Ehlers group $SL(9,%
\mathbb{R})$ only:
\begin{equation}
\begin{array}{c}
E_{8(8)}\supset _{ns}SL(9,\mathbb{R}); \\
\mathbf{248}=\mathbf{80}+\mathbf{84}+\mathbf{84}^{\prime },%
\end{array}
\label{D=11}
\end{equation}%
where $\mathbf{84}$ and $\mathbf{84}^{\prime }$ are the $3$-fold
antisymmetric of $SL(9,\mathbb{R})$ and its dual; they correspond to gauge
fields coupling to $M2$ branes and $M5$ branes, respectively. The
corresponding \textit{mcs} level is given by the specification of (\ref%
{emb-mcs}) to the following non-symmetric embedding of the massless spin
group $SO\left( 9\right) $ only:%
\begin{equation}
SO\left( 16\right) \supset _{ns}SO\left( 9\right) .  \label{D=11-mcs}
\end{equation}%
For what concerns the massless spectrum, one considers the maximal symmetric
embedding\footnote{%
For further subtleties concerning exceptional Lie algebras, see \cite{BT-1}
and App. \ref{App-Spinor-Polarizations} further below.}%
\begin{equation}
E_{8(8)}\overset{mcs}{\supset }_{s}SO\left( 16\right) :\mathbf{248}=\mathbf{%
120}+\mathbf{128},
\end{equation}%
where $\mathbf{128}$ is one of the two chiral spinor irreps. of $SO\left(
16\right) $. Under (\ref{D=11-mcs}), such two chiral irreps. $\mathbf{128}$
and $\mathbf{128}^{\prime }$ further decompose as follows:%
\begin{equation}
SO\left( 16\right) \supset _{ns}SO\left( 9\right) :\left\{
\begin{array}{l}
\mathbf{128}=\mathbf{84}+\mathbf{44;} \\
\\
\mathbf{128}^{\prime }=\mathbf{128},%
\end{array}%
\right.  \label{D=11-decomp}
\end{equation}%
where, on the right-hand side, $\mathbf{44}$, $\mathbf{84}$ and $\mathbf{128}
$ are the rank-$2$ symmetric traceless, the rank-$3$ antisymmetric and the
gamma-traceless vector-spinor irreps. of the massless spin group $SO(9)$,
respectively. Thus, (\ref{D=11-decomp}) establishes the chiral spinor irrep.
$\mathbf{128}$ of the Clifford algebra $SO(16)$ to be irrep. pertaining to
the massless \textit{bosonic} spectrum (it branches into the graviton $%
\mathbf{44}$ and the $3$-form $\mathbf{84}$), whereas its conjugate
semi-spinor irrep. $\mathbf{128}^{\prime }$ pertains to the massless \textit{%
fermionic} spectrum of $M$-theory (it corresponds to the $D=11$ gravitino).

\item In $D=10$ type IIA theory the $U$-duality is $G_{16}^{10~IIA}=SO(1,1)$
(and thus no \texttt{continuous} $\mathcal{R}$-symmetry); since this theory
is obtained as the Kaluza-Klein $S^{1}$-reduction of $M$-theory, the
relevant chain of maximal embeddings reads
\begin{equation}
E_{8(8)}\supset _{ns}SL(9,\mathbb{R})\supset _{s}SO(1,1)\times SL(8,\mathbb{R%
});  \label{IIA-chain-1}
\end{equation}%
note the \textit{\textquotedblleft enhancement"} to $SL(9,\mathbb{R})$,
consistent with the $M$-theoretical origin of IIA theory. The corresponding
\textit{mcs} level is%
\begin{equation}
SO\left( 16\right) \supset _{ns}SO(9)\supset _{s}SO(8),
\label{IIA-chain-1-mcs}
\end{equation}%
where $SO(8)$ is the massless spin group. Throughout our analysis, we dub
\textit{\textquotedblleft next-to-maximal"} (\textit{nm}) those embeddings
given by a chain of two maximal embeddings; note that all \textit{nm}
embeddings considered in the present investigation are of \textit{maximal }%
rank, namely they preserve the rank of the original group. For what concerns
the IIA massless spectrum, one considers the branchings of $\mathbf{128}$
(bosons) and $\mathbf{128}^{\prime }$ (fermions) of the Clifford algebra $%
SO\left( 16\right) $ under the \textit{nm} embedding (\ref{IIA-chain-1-mcs}):%
\begin{eqnarray}
\mathbf{128} &=&\mathbf{84}+\mathbf{44}=\mathbf{56}_{v}+\mathbf{28}+\mathbf{%
35}_{v}+\mathbf{8}_{v}+\mathbf{1;}  \label{128-branching} \\
\mathbf{128}^{\prime } &=&\mathbf{128}=\mathbf{56}_{s}+\mathbf{56}_{c}+%
\mathbf{8}_{s}+\mathbf{8}_{c},  \label{128'-branching}
\end{eqnarray}%
where the subscripts \textquotedblleft $v$", \textquotedblleft $s$" and
\textquotedblleft $c$" respectively stand for \textit{vector}, \textit{spinor%
}, \textit{conjugate spinor}, and they pertain to the \textit{triality} of $%
SO(8)$, the little group (spin group) of massless particles in $D=10$. $%
\mathbf{56}_{i}$, $\mathbf{28}$, $\mathbf{35}_{i}$ and $\mathbf{8}_{i}$ ($%
i=v,s,c$) are the rank-$3$ antisymmetric, adjoint, rank-$2$ symmetric
traceless and vector/spinor irreps. of $SO(8)$, respectively. Thus, the
branching (\ref{128-branching}) consistently pertains to the IIA massless
\textit{bosonic} spectrum : $3$-form $C_{\mu \nu \rho }^{(3)}$ ($\mathbf{56}%
_{v}$), $B$-field $B_{\mu \nu }$ ($\mathbf{28}$), graviton $g_{\mu \nu }$ ($%
\mathbf{35}_{v}$), graviphoton $C_{\mu }^{(1)}$ ($\mathbf{8}_{v}$) and
dilaton scalar field $\phi _{10}$ ($\mathbf{1}$). On the other hand, the
branching (\ref{128'-branching}) pertains to the IIA massless \textit{%
fermionic} spectrum : gravitinos $\mathbf{56}_{s}$\textbf{\ }and $\mathbf{56}%
_{c}$ ($s=3/2$ Majorana-Weyl spinors of opposite chirality), and gauginos $%
\mathbf{8}_{s}$ and $\mathbf{8}_{c}$ ($s=1/2$ Majorana-Weyl spinors of
opposite chirality). This \textit{non-chiral} spectrum can also be deduced
by dimensional reduction of the maximal supersymmetric supermultiplet of $%
D=11$ supergravity ($M$-theory).

\item On the other hand, in $D=10$ type IIB theory the $U$-duality is $%
G_{16}^{10~IIB}=SL(2,\mathbb{R})$, and its \textit{mcs} is the $\mathcal{R}$%
-symmetry $U(1)$, and the relevant \textit{nm} embedding is given by (\ref%
{UCCC-3}), which we report here:%
\begin{equation}
E_{8(8)}\supset _{s}SL(2,\mathbb{R})\times E_{7(7)}\supset _{s}SL(2,\mathbb{R%
})\times SL(8,\mathbb{R});  \label{IIB-chain-1}
\end{equation}%
\begin{equation}
SO\left( 16\right) \supset _{s}U(1)\times SU(8)\supset _{s}U(1)\times SO(8);
\label{IIB-chain-1-mcs}
\end{equation}%
note the \textit{\textquotedblleft exceptional enhancement"} to $E_{7(7)}$
in (\ref{IIB-chain-1}). For what concerns the IIB massless spectrum, one
considers the branching of $\mathbf{128}$ (bosons) and $\mathbf{128}^{\prime
}$ (fermions) of the Clifford algebra $SO\left( 16\right) $ under the
\textit{nm} embedding (\ref{IIB-chain-1-mcs}). Under the decomposition%
\begin{equation}
\begin{array}{c}
SU(8)\supset _{s}SO(8) \\
\mathbf{8}=\mathbf{8}_{s},%
\end{array}
\label{br-1}
\end{equation}%
one obtains (disregarding $U(1)$ charges)%
\begin{eqnarray}
\mathbf{128} &=&\mathbf{70}_{0}+\mathbf{28+}\overline{\mathbf{28}}+\mathbf{1}%
+\mathbf{1}=\mathbf{35}_{v}+\mathbf{35}_{c}+\mathbf{28+28}+\mathbf{1}+%
\mathbf{1;}  \label{128-branching-IIB} \\
\mathbf{128}^{\prime } &=&\mathbf{56}+\overline{\mathbf{56}}+\mathbf{8}+%
\overline{\mathbf{8}}=\mathbf{56}_{s}+\mathbf{56}_{s}+\mathbf{8}_{s}+\mathbf{%
8}_{s}.  \label{128'-branching-IIB}
\end{eqnarray}%
Note that, upon (\ref{br-1}), the rank-$4$ antisymmetric self-real irrep. $%
\mathbf{70}$ of $SU(8)$ breaks into $\mathbf{35}_{v}+\mathbf{35}_{c}$ of $%
SO(8)$. Thus, the branching (\ref{128-branching-IIB}) consistently pertains
to the IIB massless \textit{bosonic} spectrum: graviton $g_{\mu \nu }$ ($%
\mathbf{35}_{v}$), $4$-form $C^{(4)}$ ($\mathbf{35}_{c}$), $B$-field $B_{\mu
\nu }$ ($\mathbf{28}$), 2-form $C_{\mu \nu }^{(2)}$ ($\mathbf{28}$), and two
scalar fields, namely the dilaton $\phi _{10}$ and the axion $C^{(0)}$ ($%
\mathbf{1}+\mathbf{1}$). On the other hand, the branching (\ref%
{128'-branching-IIB}) pertains to the IIB massless \textit{fermionic}
spectrum : gravitinos $\mathbf{56}_{s}$\textbf{\ }and $\mathbf{56}_{s}$ ($%
s=3/2$ Majorana-Weyl spinors of same chirality) and gauginos $\mathbf{8}_{s}$
and $\mathbf{8}_{s}$ ($s=1/2$ Majorana-Weyl spinors of same chirality). This
spectrum is \textit{chiral} and hence cannot be obtained by dimensional
reduction of the $D=11$ $M$-theory supermultiplet.

\item In $D=9$ the $U$-duality is $G_{16}^{9}=GL(2,\mathbb{R})\equiv
E_{2(2)} $, and its \textit{mcs} is the $\mathcal{R}$-symmetry $U(1)$. There
are two possible chains of maximal embeddings, which are equivalent up to
redefinitions of $SO(1,1)$ weights. The first chain, pertinent to a
dimensional reduction of $M$-theory, gives rise to a \textit{nm} embedding:%
\begin{equation}
E_{8(8)}\supset _{ns}SL(9,\mathbb{R})\supset _{s}GL(2,\mathbb{R})\times SL(7,%
\mathbb{R});  \label{D=9-1}
\end{equation}%
\begin{equation}
SO(16)\supset _{ns}SO(9)\supset _{s}U(1)\times SO(7),  \label{D=9-1-mcs}
\end{equation}%
whereas the second, pertaining to a Kaluza-Klein $S^{1}$-reduction of $D=10$
IIB theory, determines a \textit{"next-to-next-to-maximal"} (\textit{nnm})
embedding, because it is $3$-stepwise (it is given by a further branching of
IIB chain (\ref{IIB-chain-1})):%
\begin{equation}
E_{8(8)}\supset _{s}SL(2,\mathbb{R})\times E_{7(7)}\supset _{s}SL(2,\mathbb{R%
})\times SL(8,\mathbb{R})\supset _{s}GL(2,\mathbb{R})\times SL(7,\mathbb{R});
\label{D=9-2}
\end{equation}%
\begin{equation}
SO\left( 16\right) \supset _{s}U(1)\times SU(8)\supset _{s}U(1)\times
SO(8)\supset _{s}U(1)\times SO(7).  \label{D=9-2-mcs}
\end{equation}%
Besides being equivalent, (\ref{D=9-1})-(\ref{D=9-1-mcs}) and (\ref{D=9-2})-(%
\ref{D=9-2-mcs}) are consistent, because type IIA and IIB theories are
equivalent in $D\leqslant 9$ dimensions (except for half-maximal
supergravity in $D=6$; see further below).

\item In $D=8$ the $U$-duality is $G_{16}^{8}=SL(2,\mathbb{R})\times SL(3,%
\mathbb{R})\equiv E_{3(3)}$, and its \textit{mcs} is the $\mathcal{R}$%
-symmetry $U(1)\times SU(2)\sim U(2)$. The relevant \textit{nm} embedding
reads\footnote{%
(\ref{D=8-mcs}) is the $n=4$ case of the maximal non-symmetric embedding
pattern%
\begin{equation*}
SO\left( 4n\right) \supset _{ns}SU\left( 2\right) \times USp\left( 2n\right)
;
\end{equation*}%
\begin{equation*}
\mathbf{Adj}_{SO(4n)}=\mathbf{Adj}_{SU(2)}+\mathbf{Adj}_{USp(2n)}+(\mathbf{3}%
,\mathbf{A}_{2,0}),
\end{equation*}%
where $\mathbf{A}_{2,0}$ is the rank-2 antisymmetric skew-traceless irrep.
of $USp\left( 2n\right) $. For the first appearance of such an embedding in
supersymmetry, see \cite{FSaZ}.} ($SO(6)\sim SU(4)$)%
\begin{equation}
E_{8(8)}\supset _{ns}E_{6(6)}\times SL(3,\mathbb{R})\supset _{s}SL(2,\mathbb{%
R})\times SL(3,\mathbb{R})\times SL(6,\mathbb{R});  \label{D=8}
\end{equation}%
\begin{equation}
SO(16)\supset _{ns}USp(8)\times SU(2)\supset _{s}U(1)\times SU(2)\times
SU(4);  \label{D=8-mcs}
\end{equation}%
note the \textit{\textquotedblleft exceptional enhancement"} to $E_{6(6)}$
in (\ref{D=8}).

\item In $D=7$ the $U$-duality is $G_{16}^{7}=SL(5,\mathbb{R})\equiv
E_{4(4)} $, and its \textit{mcs} is the $\mathcal{R}$-symmetry $SO(5)\sim
USp(4)$. The relevant embedding is maximal non-symmetric:%
\begin{equation}
E_{8(8)}\supset _{ns}SL\left( 5,\mathbb{R}\right) \times SL(5,\mathbb{R});
\label{D=7}
\end{equation}%
\begin{equation}
SO(16)\supset _{ns}USp(4)\times USp(4).  \label{D=7-mcs}
\end{equation}%
Note that in this case there is perfect symmetry between the $\mathcal{R}$%
-symmetry and the massless spin sectors.

\item In $D=6$ (non-chiral $(2,2)$) maximal theory, the $U$-duality is $%
G_{16}^{6}=SO(5,5)\equiv E_{5(5)}$, and its \textit{mcs} is the $\mathcal{R}$%
-symmetry\footnote{%
Subscripts \textquotedblleft $L$" and \textquotedblleft $R$" denote left and
right chirality, respectively.} $SO(5)\times SO(5)\sim USp(4)_{L}\times
USp(4)_{R}$. The relevant \textit{nm} embedding reads ($SO(4)\sim
SU(2)\times SU(2)$)%
\begin{equation}
E_{8(8)}\supset _{s}SO\left( 8,8\right) \supset _{s}SO\left( 5,5\right)
\times SO\left( 3,3\right) \sim SO\left( 5,5\right) \times SL\left( 4,%
\mathbb{R}\right) ;  \label{D=6}
\end{equation}%
\begin{equation}
\begin{array}{c}
SO(16)\supset _{s}SO(8)\times SO(8)\supset _{s}SO\left( 5\right) _{L}\times
SO\left( 3\right) \times SO\left( 5\right) _{R}\times SO\left( 3\right) \\
\sim USp(4)_{L}\times USp(4)_{R}\times SU(2)_{L}\times SU(2)_{R};%
\end{array}
\label{D=6-mcs}
\end{equation}%
note the \textit{\textquotedblleft enhancement"} to $SO(8,8)$ in (\ref{D=6}%
).Note that in this case both the $\mathcal{R}$-symmetry and massless spin
groups factorize in the direct product of opposite chiralities identical
factors. The corresponding Jordan algebra interpretation of (\ref{D=6}) is
as follows:%
\begin{equation}
QConf\left( J_{3}^{\mathbb{O}_{s}}\right) \supset Str_{0}\left( J_{2}^{%
\mathbb{O}_{s}}\right) \times SL\left( 4,\mathbb{R}\right) ,
\end{equation}%
where $J_{3}^{\mathbb{O}_{s}}$ and $J_{2}^{\mathbb{O}_{s}}\sim \mathbf{%
\Gamma }_{5,5}$ are the rank-$2$ and rank-$3$ Euclidean Jordan algebras over
the split octonions $\mathbb{O}_{s}$, and $QConf$ and $Str_{0}$ respectively
denote the \textit{quasi-conformal} and \textit{reduced structure} groups%
\footnote{%
In theories related to Euclidean Jordan algebras $J_{3}$ of rank $3$, the
\textit{quasi-conformal} $QConf\left( J_{3}\right) $, \textit{conformal} $%
Conf\left( J_{3}\right) $ and \textit{reduced structure} $Str_{0}\left(
J_{3}\right) $ groups are the $U$-duality groups in $D=3$, $4$ and $5$
dimensions, respectively. In particular, $Conf\left( J_{3}\right) $ is
nothing but the automorphism group $Aut\left( \mathfrak{M}\left(
J_{3}\right) \right) $ of the corresponding Freudenthal triple system \cite%
{GST,G-Lects}.} (see \textit{e.g.} \cite{G-Lects} and Refs. therein).

\item In $D=5$ the $U$-duality undergoes an exceptional enhancement : $%
G_{16}^{5}=E_{6(6)}$, and its \textit{mcs} is the $\mathcal{R}$-symmetry $%
USp(8)$. The relevant embedding is maximal non-symmetric, and it is given by
(\ref{UCCC}), which we report here (note that it is the first step of
\textit{nm} embedding (\ref{D=8})-(\ref{D=8-mcs})):%
\begin{equation}
E_{8(8)}\supset _{ns}E_{6(6)}\times SL(3,\mathbb{R});  \label{D=5}
\end{equation}%
\begin{equation}
SO(16)\supset _{ns}USp(8)\times SU(2).  \label{D=5-mcs}
\end{equation}%
The corresponding Jordan algebra interpretation of (\ref{D=5}) is as follows:%
\begin{equation}
QConf\left( J_{3}^{\mathbb{O}_{s}}\right) \supset Str_{0}\left( J_{3}^{%
\mathbb{O}_{s}}\right) \times SL\left( 3,\mathbb{R}\right) ,
\end{equation}%
and it is a particular non-compact, real version of the \textit{Jordan-pair }%
embeddings of exceptional Lie algebras recently considered in \cite{Truini-1}%
. Note that the $SU(2)$ in (\ref{D=5-mcs}) is the \textit{principal} $SU(2)$
in $SL(3,\mathbb{R})$ in (\ref{D=5}).

\item In $D=4$ the $U$-duality is $G_{16}^{4}=E_{7(7)}$, and its \textit{mcs}
is the $\mathcal{R}$-symmetry $SU(8)$. The relevant embedding is maximal
symmetric (note that it is the first step of chains (\ref{IIB-chain-1})-(\ref%
{IIB-chain-1-mcs}) and (\ref{D=9-2})-(\ref{D=9-2-mcs})):%
\begin{equation}
E_{8(8)}\supset _{s}E_{7(7)}\times SL(2,\mathbb{R});  \label{D=4}
\end{equation}%
\begin{equation}
SO(16)\supset _{s}SU(8)\times U(1).  \label{D=4-mcs}
\end{equation}%
The corresponding Jordan algebra interpretation of (\ref{D=4}) is as follows:%
\begin{equation}
QConf\left( J_{3}^{\mathbb{O}_{s}}\right) \supset Conf\left( J_{3}^{\mathbb{O%
}_{s}}\right) \times SL\left( 2,\mathbb{R}\right) ,
\end{equation}%
where $Conf$ denotes the \textit{conformal} group of $J_{3}^{\mathbb{O}_{s}}$
(see \textit{e.g.} \cite{G-Lects} and Refs. therein). Similar
Jordan-algebraic interpretations can be given for other supergravities in
various dimensions.
\end{enumerate}

\subsection{$N=12$}

In the \textit{\textquotedblleft pure"} theory with $N=12$, the $D=3$ $U$%
-duality group is $G_{12}^{3}=E_{7(-5)}$, with \textit{mcs} $SO(12)\times
SU(2)_{CV}$, where $SO(12)$ is the Clifford algebra for massless particles
with $\mathcal{N}=24$ supersymmetries. The $SU(2)_{CV}$ factor pertains to
the so-called \textit{Clifford vacuum }(\textit{CV}), which is generally
present for \textit{non-maximal} theories ($N<16$), and it indicates further
degeneracy of the Clifford algebra symmetry. In this case, $SU(2)_{CV}$ can
be also explained by recalling that this theory shares the very same bosonic
sector of a \textit{matter-coupled} supergravity with $N=4$ \cite{GST}, in
which it is the $\mathcal{R}$-symmetry of the hypermultiplets' sector.

This theory can consistently be uplifted only to $D=4$ and $D=5$.

\begin{enumerate}
\item In $D=5$ the $U$-duality is $G_{12}^{5}=SU^{\ast }(6)$, and its
\textit{mcs} is the $\mathcal{R}$-symmetry $USp(6)$. The relevant embedding
is maximal non-symmetric:%
\begin{equation}
E_{7(-5)}\supset _{ns}SU^{\ast }(6)\times SL\left( 3,\mathbb{R}\right) ;
\label{D=5-N=6}
\end{equation}%
\begin{equation}
SO(12)\times SU(2)_{CV}\supset _{ns}USp(6)\times SU(2)_{J},
\label{D=5-mcs-N=6}
\end{equation}%
where we introduced the subscript \textquotedblleft $J$" in order to
discriminate between the Clifford vacuum $SU(2)_{CV}$ and the $SU(2)_{J}$
pertaining to the massless spin group in $D=5$. Note that the embedding (\ref%
{D=5-N=6}) is maximal non-symmetric, while the embedding (\ref{D=5-mcs-N=6})
is non-maximal non-symmetric.

\item In $D=4$ the $U$-duality is $G_{12}^{4}=SO^{\ast }(12)$, and its
\textit{mcs} is the $\mathcal{R}$-symmetry $U(6)$. The relevant embedding is
maximal symmetric:%
\begin{equation}
E_{7(-5)}\supset _{s}SO^{\ast }(12)\times SL\left( 2,\mathbb{R}\right) ;
\label{D=4-N=6}
\end{equation}%
\begin{equation}
SO(12)\times SU(2)_{CV}\supset _{s}SU(6)\times U(1)\times U(1)_{J},
\label{D=4-mcs-N=6}
\end{equation}%
and it pertains to the so-called $c^{\ast }$\textit{-map} (see \textit{e.g.}
\cite{Trig-1}, and Refs. therein).
\end{enumerate}

\subsection{$N=10$}

In the \textit{\textquotedblleft pure"} theory with $N=10$, the $D=3$ $U$%
-duality group is $G_{10}^{3}=E_{6(-14)}$, with \textit{mcs} $SO(10)\times
SO(2)_{CV}$, where $SO(10)$ is the Clifford algebra for massless particles
with $\mathcal{N}=20$ supersymmetries. In this case, $SO(2)_{CV}$ can be
also explained as \texttt{[add...]}

This theory can be uplifted only to $D=4$, in which the $U$-duality is $%
G_{10}^{4}=SU(5,1)$, and its \textit{mcs} is the $\mathcal{R}$-symmetry $%
U(5) $. The relevant embedding is maximal symmetric:%
\begin{equation}
E_{6(-14)}\supset _{s}SU(5,1)\times SL\left( 2,\mathbb{R}\right) ;
\label{D=4-N=5}
\end{equation}%
\begin{equation}
SO(10)\times SO(2)_{CV}\supset _{s}SU(5)\times U(1)\times U(1)_{J}.
\label{D=4-mcs-N=5}
\end{equation}

\section{\label{Sec-Half-Maximal}$N=8,6$ Matter Coupled Theories}

\subsection{$N=8$}

\textit{Half-maximal} theories with $N=8$ exist in $3\leqslant D\leqslant 10$%
; moreover, for $D=6$ two inequivalent theories exist, \textit{i.e.} the
non-chiral IIA $(1,1)$ and the chiral IIB $(2,0)$.

The $D=3$ $U$-duality group is $G_{8}^{3}=SO\left( 8,D-2+m\right) $, where $%
m $ is the number of matter multiplets in $D=3$ other than those coming from
the reduction of the gravity multiplet in $D$ dimensions. Furthermore,
\textit{mcs}$\left( G_{8}^{3}\right) =SO(8)\times SO\left( D-2+m\right)
_{CV} $, where $SO(8)$ is the Clifford algebra for massless particles with $%
\mathcal{N}=16$ supersymmetries, and $SO\left( D-2+m\right) _{CV}$ is the
\textit{Clifford vacuum} symmetry.

The relevant chain of maximal embeddings leading to the embedding of the $D$%
-dimensional Ehlers group $SL\left( D-2,\mathbb{R}\right) $ into $SO\left(
8,D-2+m\right) $ depends on the dimension and on the type of theory. We
anticipate that embeddings (\ref{emb}) and (\ref{emb-mcs}) are \textit{%
maximal} in $D=4$ (symmetric) and \textit{next-to-maximal} in $5\leqslant
D\leqslant 10$.

\begin{itemize}
\item For $D\geqslant 5$ (and $D=6$ type IIA $(1,1)$), it is given by the
following chain of two maximal symmetric steps:%
\begin{equation}
\begin{array}{r}
SO\left( 8,D-2+m\right) \supset _{s}SO\left( D-2,D-2\right) \times SO\left(
10-D,m\right) \\
\supset _{s}SL\left( D-2,\mathbb{R}\right) \times SO(1,1)\times SO\left(
10-D,m\right)%
\end{array}
\label{zero}
\end{equation}%
The group commuting with $SL\left( D-2,\mathbb{R}\right) $ inside $SO\left(
8,D-2+m\right) $ is nothing but the $D$-dimensional $U$-duality group $%
G_{8}^{D}=SO(1,1)\times SO\left( 10-D,m\right) $. Note the \textit{%
\textquotedblleft enhancement"} to $SO\left( D-2,D-2\right) \times SO\left(
10-D,m\right) $. Furthermore, it is worth remarking that also for $m=0$ the
\textit{Clifford vacuum} degeneracy is still present with an $SO\left(
D-2\right) _{CV}$ factor; this is an extra spin quantum number carried by
the $SO(8)$ Clifford algebra spinor. In fact, by considering the \textit{mcs}
level of the chain (\ref{zero}), one obtains%
\begin{equation}
\begin{array}{r}
SO\left( 8\right) _{\text{Clifford}}\times SO\left( D-2+m\right)
_{CV}\supset _{s}SO\left( D-2\right) \times SO\left( D-2\right) _{CV}\times
SO\left( 10-D\right) \times SO\left( m\right) _{CV} \\
\supset _{s}SO\left( D-2\right) _{J}\times SO\left( 10-D\right) _{\mathcal{R}%
}\times SO\left( m\right) _{CV},%
\end{array}
\label{zero-mcs}
\end{equation}%
where the $D$-dimensional massless spin group $SO\left( D-2\right)
_{J}=mcs\left( SL\left( D-2,\mathbb{R}\right) \right) $ is \textit{diagonally%
} embedded into $SO\left( D-2\right) \times SO\left( D-2\right) _{CV}$, and
the $\mathcal{R}$-symmetry is $SO(10-D)$. $SO\left( m\right) _{CV}$ is the
part of \textit{Clifford vacuum} symmetry due to matter coupling.

\item For $D=4$, the maximal symmetric embedding reads:%
\begin{equation}
SO\left( 8,2+m\right) \supset _{s}SO\left( 2,2\right) \times SO\left(
6,m\right) \sim SL(2,\mathbb{R})_{\text{Ehlers}}\times SL(2,\mathbb{R}%
)\times SO\left( 6,m\right) ,  \label{zero-2}
\end{equation}%
and it pertains to the so-called $c^{\ast }$\textit{-map} (see \textit{e.g.}
\cite{Trig-1}, and Refs. therein). The group commuting with $SL(2,\mathbb{R}%
)_{\text{Ehlers}}$ inside $SO\left( 8,2+m\right) $ is the $4$-dimensional $U$%
-duality group $G_{8}^{4}=SL(2,\mathbb{R})\times SO\left( 6,m\right) $. Also
in this case for $m=0$ the \textit{Clifford vacuum} degeneracy is still
present with an $SO\left( 2\right) _{CV}$ factor. In fact, by considering
the \textit{mcs} level of (\ref{zero-2}), one obtains the following maximal
symmetric embedding ($SO(6)\sim SU(4)$):%
\begin{equation}
\begin{array}{l}
SO\left( 8\right) _{\text{Clifford}}\times SO\left( 2+m\right) _{CV} \\
\supset _{s}SO\left( 2\right) _{J}\times SO\left( 2\right) _{CV}\times
SO\left( 6\right) \times SO(m)_{CV}\sim U\left( 1\right) _{J}\times U\left(
4\right) _{\mathcal{R}}\times SO(m)_{CV},%
\end{array}
\label{zero-2-mcs}
\end{equation}%
where $SO\left( 2\right) _{J}=mcs\left( SL(2,\mathbb{R})_{\text{Ehlers}%
}\right) $, and the $\mathcal{R}$-symmetry is $SO\left( 2\right) _{CV}\times
SO\left( 6\right) \sim U(4)_{\mathcal{R}}$. Moreover, $SO\left( m\right)
_{CV}$ is the part of \textit{Clifford vacuum} symmetry due to matter
coupling.

\item For $D=6$ type IIB $(2,0)$, it suffices to start with $SO\left(
8,3+m\right) $, and the maximal symmetric embedding reads as follows:%
\begin{equation}
SO\left( 8,3+m\right) \supset _{s}SO\left( 3,3\right) \times SO\left(
5,m\right) \sim SL(4,\mathbb{R})\times SO\left( 5,m\right) .  \label{zero-3}
\end{equation}%
The group commuting with $SL\left( 4,\mathbb{R}\right) $ inside $SO\left(
8,4+m\right) $ is the $6$-dimensional type IIB $U$-duality group $%
G_{8}^{6,IIB}=SO\left( 5,m\right) $. The corresponding \textit{mcs} level
reads%
\begin{equation}
\begin{array}{l}
SO\left( 8\right) _{\text{Clifford}}\times SO\left( 3+m\right) _{CV} \\
\supset _{s}\left( SO\left( 3\right) \times SO\left( 3\right) \right)
_{J}\times SO\left( 5\right) _{\mathcal{R}}\times SO(m)_{CV}\sim
SO(4)_{J}\times USp\left( 4\right) _{\mathcal{R}}\times SO(m)_{CV},%
\end{array}
\label{zero-3-mcs}
\end{equation}%
where $SO(4)_{J}=mcs\left( SL(4,\mathbb{R})_{\text{Ehlers}}\right) $, and
the $\mathcal{R}$-symmetry is $SO(5)\sim USp(4)$. Furthermore, $SO\left(
m\right) _{CV}$ is the part of \textit{Clifford vacuum} symmetry due to
matter coupling.
\end{itemize}

All cases in $4\leqslant D\leqslant 10$ dimensions are reported in Tables 3
and 4.

\begin{table}[t!]
\begin{center}
\begin{tabular}{|c||c|c|}
\hline
$D$ & $%
\begin{array}{c}
\\
SO\left( 8,D-2+m\right) \supset G_{D}^{8}\left( m\right) \times SL(D-2,%
\mathbb{R}) \\
~~%
\end{array}%
$ & $%
\begin{array}{c}
\\
\text{type} \\
~~%
\end{array}%
$ \\ \hline\hline
$%
\begin{array}{c}
\\
10 \\
~%
\end{array}%
$ & $SO\left( 8,8+m\right) \supset SO(1,1)\times SO\left( m\right) \times
SL\left( 8,\mathbb{R}\right) $ & $nm$,~$ns$ \\ \hline
$%
\begin{array}{c}
\\
9 \\
~%
\end{array}%
$ & $SO\left( 8,7+m\right) \supset SO(1,1)\times SO\left( 1,m\right) \times
SL\left( 7,\mathbb{R}\right) $ & $nm$,~$ns$ \\ \hline
$%
\begin{array}{c}
\\
8 \\
~%
\end{array}%
$ & $SO\left( 8,6+m\right) \supset SO(1,1)\times SO\left( 2,m\right) \times
SL\left( 6,\mathbb{R}\right) $ & $nm$,~$ns~$ \\ \hline
$%
\begin{array}{c}
\\
7 \\
~%
\end{array}%
$ & $SO\left( 8,5+m\right) \supset SO(1,1)\times SO\left( 3,m\right) \times
SL\left( 5,\mathbb{R}\right) $ & $nm$,~$ns$ \\ \hline
$%
\begin{array}{c}
\\
6,~IIA \\
~%
\end{array}%
$ & $SO\left( 8,4+m\right) \supset SO(1,1)\times SO\left( 4,m\right) \times
SL(4,\mathbb{R})$ & $nm$,~$ns$ \\ \hline
$%
\begin{array}{c}
\\
6,~IIB \\
~%
\end{array}%
$ & $SO\left( 8,3+m\right) \supset SO\left( 5,m\right) \times SL(4,\mathbb{R}%
)$ & $max$,~$s$ \\ \hline
$%
\begin{array}{c}
\\
5 \\
~%
\end{array}%
$ & $SO\left( 8,3+m\right) \supset SO(1,1)\times SO\left( 5,m\right) \times
SL(3,\mathbb{R})$ & $nm$,~$ns$ \\ \hline
$%
\begin{array}{c}
\\
4 \\
~%
\end{array}%
$ & $SO\left( 8,2+m\right) \supset \left( SL(2,\mathbb{R})\times SO\left(
6,m\right) \right) \times SL(2,\mathbb{R})$ & $max$,~$s$ \\ \hline
\end{tabular}%
\end{center}
\caption{Embedding $G_{8}^{3}\supset G_{8}^{D}\times SL(D-2,\mathbb{R})$ (%
\protect\ref{emb}) for \textit{half-maximal} supergravity theories ($N=8$)
in $10\geqslant D\geqslant 4$ Lorentzian space-time dimensions.}
\end{table}

\begin{table}[t!]
\begin{center}
\begin{tabular}{|c||c|c|}
\hline
$D$ & $%
\begin{array}{c}
\\
SO\left( 8\right) \times SO\left( D-2+m\right) \supset H_{D}^{8}\left(
m\right) \times SO(D-2) \\
~~%
\end{array}%
$ & $%
\begin{array}{c}
\\
\text{type} \\
~~%
\end{array}%
$ \\ \hline\hline
$%
\begin{array}{c}
\\
10 \\
~%
\end{array}%
$ & $SO\left( 8\right) \times SO\left( 8+m\right) \supset SO\left( m\right)
\times SO\left( 8\right) $ & $nm$,~$ns$ \\ \hline
$%
\begin{array}{c}
\\
9 \\
~%
\end{array}%
$ & $SO\left( 8\right) \times SO\left( 7+m\right) \supset SO\left( m\right)
\times SO\left( 7\right) $ & $nm$,~$ns$ \\ \hline
$%
\begin{array}{c}
\\
8 \\
~%
\end{array}%
$ & $SO\left( 8\right) \times SO\left( 6+m\right) \supset SO\left( 2\right)
\times SO\left( m\right) \times SO\left( 6\right) $ & $nm$,~$ns~$ \\ \hline
$%
\begin{array}{c}
\\
7 \\
~%
\end{array}%
$ & $SO\left( 8\right) \times SO\left( 5+m\right) \supset SO\left( 3\right)
\times SO\left( m\right) \times SO\left( 5\right) $ & $nm$,~$ns$ \\ \hline
$%
\begin{array}{c}
\\
6,~IIA \\
~%
\end{array}%
$ & $SO\left( 8\right) \times SO\left( 4+m\right) \supset SO\left( 4\right)
\times SO\left( m\right) \times SO(4)$ & $nm$,~$ns$ \\ \hline
$%
\begin{array}{c}
\\
6,~IIB \\
~%
\end{array}%
$ & $SO\left( 8\right) \times SO\left( 3+m\right) \supset SO\left( 5\right)
\times SO\left( m\right) \times SO(4)$ & $max$,~$s$ \\ \hline
$%
\begin{array}{c}
\\
5 \\
~%
\end{array}%
$ & $SO\left( 8\right) \times SO\left( 3+m\right) \supset SO\left( 5\right)
\times SO\left( m\right) \times SO(3)$ & $nm$,~$ns$ \\ \hline
$%
\begin{array}{c}
\\
4 \\
~%
\end{array}%
$ & $SO\left( 8\right) \times SO\left( 2+m\right) \supset \left( SO(2)\times
SO\left( 6\right) \times SO\left( m\right) \right) \times SO(2)$ & $max$,~$s$
\\ \hline
\end{tabular}%
\end{center}
\caption{Embedding $H_{8}^{3}\supset H_{8}^{D}\times SO(D-2)$ (\protect\ref%
{emb-mcs}) \protect\cite{Nahm} for \textit{half}-\textit{maximal}
supergravity theories ($N=8$) in $10\geqslant D\geqslant 4$ Lorentzian
space-time dimensions. Since \textit{matter coupling} is allowed, $H_{8}^{3}$
and in general $H_{8}^{D}$ entail both \textit{half-maximal} $\mathcal{R}$%
-symmetry and \textit{Clifford vacuum} symmetry.}
\end{table}

\subsection{$N=6$}

Theories with $N=6$ exist only in $D=3,4$.

The $D=3$ $U$-duality group is $G_{6}^{3}=SU(4,m+1)$, where $m$ is the
number of matter multiplets in $D=3$ other than those coming from the
reduction of the gravity multiplet in $4$ dimensions. Furthermore, \textit{%
mcs}$\left( G_{6}^{3}\right) =SU(4)\times U\left( m+1\right) _{CV}$, where $%
SU(4)\sim SO(6)$ is the Clifford algebra for massless particles with $%
\mathcal{N}=12$ supersymmetries, and $U\left( m+1\right) _{CV}$ is the
\textit{Clifford vacuum} symmetry.

The embedding of the $4$-dimensional Ehlers group $SL\left( 2,\mathbb{R}%
\right) $ into $SU(4,m+1)$ is maximal and symmetric:%
\begin{equation}
SU(4,m+1)\supset _{s}SL(2,\mathbb{R})\times U(3,m),
\end{equation}%
and at the \textit{mcs} level:%
\begin{equation}
SU(4)\times SU\left( m+1\right) \times U(1)\supset _{s}U(1)_{J}\times
U(3)\times U(m),
\end{equation}%
where $D=4$ $U$-duality group is $G_{6}^{4}=SU(3,m)$, and the $\mathcal{R}$%
-symmetry is $U(3)$. $U(m)$ is the $D=4$ \textit{Clifford vacuum} symmetry,
which is related to the number of matter multiplets.

\section{\label{Sec-Quarter-Maximal}$N=4$ Matter Coupled \textit{Symmetric}
Theories}

\textit{Quarter-maximal} theories with $N=4$ exist in $3\leqslant D\leqslant
6$; in particular, in $D=6$ they are chiral $(1,0)$ theories. The new
feature of $N=4$ theories is the possible existence of two different types
of matter multiplets, namely vector and hyper multiplets, transforming in
different ways under the $\mathcal{R}$-symmetry, which is $U(2)$ in $D=4$
and $USp(2)$ in $D=5,6$.

In the following treatment, we will only consider theories based on \textit{%
symmetric} Abelian-vector multiplets' scalar manifolds, which is a
restriction to $D=4$ (K\"{a}hler) and $D=5$ (real) special geometry; these
theories will be denoted as\footnote{%
We will not consider here the so-called \textit{non-Jordan symmetric sequence%
} (see \textit{e.g.} \cite{ICL-2-Phys} and Refs. therein) in $D=5$, based on
vector multiplets' real special symmetric scalar manifolds $\frac{SO(1,n)}{%
SO(n)}$, which gives rise to \textit{non-symmetric} coset manifolds in $D=4$
and in $D=3$.} \textit{symmetric} $N=4$ theories.

In $D=4,5$, \textit{symmetric} theories are classified by two infinite
sequences, as well as by isolated cases given by the so-called \textit{%
\textquotedblleft magical" }models.

We will also not consider the ($D$-independent) hypermultiplets'
quaternionic scalar manifolds.

For $N=4$, we recall that the Clifford algebra decomposes as%
\begin{equation}
SO\left( 4\right) \sim SU(2)_{v}\times SU(2)_{h},
\end{equation}%
where $SU(2)_{v}$ pertains to the $D=3$ reduction of $D=4$ vector
multiplets, while $SU(2)_{h}$ is related to the hypermultiplet sector, which
is insensitive to the number of space-time dimensions in which the \textit{%
quarter-maximal} theory is defined (namely, $3\leqslant D\leqslant 6$).
Since we disregard hypermultiplets, in the treatment below we only consider $%
SU(2)_{v}$ (and thus we remove the subscript \textquotedblleft $v$"), which
will be a commuting factor in the \textit{mcs} of the $D=3$ $U$-duality
group $G_{4}^{4}$.

\subsection{\textit{Minimal Coupling} Infinite Sequence and \textit{%
\textquotedblleft Pure"} $D=4$ Supergravity}

We start by considering the infinite sequence of $D=3$ \textit{quaternionic K%
\"{a}hler} symmetric spaces%
\begin{equation}
\frac{SU(2,1+n)}{SU(2)\times SU(1+n)\times U(1)},  \label{mc}
\end{equation}%
which can be uplifted only to $D=4$, giving rise to Maxwell-Einstein
supergravity models \textit{minimally coupled} to $n$ vector multiplets \cite%
{Luciani}. The $D=3$ $U$-duality group is $G_{4}^{3}=SU(2,1+n)$.

The embedding of the $4$-dimensional Ehlers group $SL\left( 2,\mathbb{R}%
\right) $ into $SU(2,n+1)$ is maximal and symmetric:%
\begin{equation}
SU(2,1+n)\supset _{s}SL(2,\mathbb{R})\times U(1,n),  \label{pre-c}
\end{equation}%
and at the \textit{mcs} level:%
\begin{equation}
SU(2)\times SU\left( 1+n\right) \times U(1)\supset _{s}U(1)_{J}\times
U(n)\times U(1)_{\mathcal{R}},  \label{c}
\end{equation}%
where $D=4$ $U$-duality group is $G_{4}^{4}=U(1,n)$. $U(1)_{\mathcal{R}}$ in
(\ref{c}) is the part of $D=4$ $\mathcal{R}$-symmetry $U(2)$ under which the
$D=4$ vector multiplets are charged, whereas the $U(n)$ factor correspond to
$D=4$ \textit{Clifford vacuum} symmetry (completely due to \textit{matter
coupling}).

By merging (\ref{pre-c}) and (\ref{c}), the following $c$-map is obtained
\cite{CFG}:%
\begin{equation}
\mathbb{CP}^{n}\equiv \frac{SU(1,n)}{U(n)}\overset{c}{\longrightarrow }\frac{%
SU(2,1+n)}{SU(2)\times SU\left( 1+n\right) \times U(1)},  \label{c-c}
\end{equation}%
where $\mathbb{CP}^{n}$ denotes the \textit{complex projective}
(non-compact) \textit{spaces}.

Note that for $n=0$ the quaternionic manifold (\ref{mc}) is not only K\"{a}%
hler, but also \textit{special K\"{a}hler}, and it is an example of Einstein
space with self-dual Weyl curvature (see \textit{e.g.} \cite{BW}, and Refs.
therein). It is usually called the \textit{universal hypermultiplet}, and it
corresponds to the $c$-map of \textit{\textquotedblleft pure"} $\mathcal{N}%
=2 $ supergravity in $D=4$, obtained as \textquotedblleft $n=0$ limit" of
the $\mathbb{CP}^{n}$ sequence; namely, by specifying $n=0$ in (\ref{c-c})
\cite{CFG}:%
\begin{equation}
\mathbb{\varnothing }\overset{c}{\longrightarrow }\frac{SU(2,1)}{SU(2)\times
U(1)}.  \label{univ-hyper}
\end{equation}%
Correspondingly, for $n=0$ (\ref{pre-c}) and (\ref{c}) respectively read%
\begin{equation}
SU(2,1)\supset _{s}SL(2,\mathbb{R})\times U(1)\sim U(1,1);
\end{equation}%
\begin{equation}
SU(2)\times U(1)\supset _{s}U(1)_{J}\times U(1)_{\mathcal{R}},
\end{equation}%
and thus the $4$ bosonic massless states of $\mathcal{N}=2$, $D=4$ \textit{%
\textquotedblleft pure"} supergravity are in the $\mathbf{2}_{\mathbb{C}}$
of $SU(2)\times U(1)\sim U(2)=mcs\left( SU(2,1)\right) $.

\subsection{ \textit{\textquotedblleft Pure"} $D=5,6$ Supergravity and $%
T^{3} $ and $ST^{2}$ Models in $D=4$}

\subsubsection{$D=5$}

Within the framework under consideration, \textit{\textquotedblleft pure"} $%
D=5$ supergravity can be obtained as $D=5$ uplift of the so-called $\mathcal{%
N}=2$, $D=4$ $T^{3}$ model, whose vector multiplet's scalar span the
symmetric special K\"{a}hler manifold $SL(2,\mathbb{R})/U(1)$ (with Ricci
scalar curvature $R=-2/3$ \cite{CVP}), and whose $D=3$ $U$-duality group is $%
G_{4,T^{3}}^{3}=G_{2(2)}$.

The embedding of the $5$-dimensional Ehlers group $SL\left( 3,\mathbb{R}%
\right) $ into $G_{4,T^{3}}^{3}$ is maximal and \textit{non-symmetric} (see
\textit{e.g.} \cite{Compere} and Refs. therein):%
\begin{equation}
G_{2(2)}\supset _{ns}SL(3,\mathbb{R}),  \label{pre-c-2}
\end{equation}%
and at the \textit{mcs} level:%
\begin{equation}
SU(2)\times SU(2)\supset _{s}SO(3)_{J}\sim SU(2)_{J},  \label{c-2}
\end{equation}%
where the $D=5$ massless spin group $SO(3)_{J}$ is \textit{diagonally}
embedded into $SU(2)\times SU(2)=mcs\left( G_{2(2)}\right) $. The $8$
bosonic massless states of $\mathcal{N}=2$, $D=5$ \textit{\textquotedblleft
pure"} supergravity are in the $\left( \mathbf{4},\mathbf{2}\right) $ of $%
mcs\left( G_{2(2)}\right) $ itself.

By merging (\ref{pre-c-2}) and (\ref{c-2}), the following $c$-map is obtained%
\footnote{%
Attention should be paid to distinguish $\left. \frac{SL(2,\mathbb{R})}{U(1)}%
\right\vert _{T^{3}}$ ($R=-2/3$) from the $n=1$ element of the $\mathbb{CP}%
^{n}$ infinite sequence treated above, namely the $\mathbb{CP}^{1}$ space
(axio-dilatonic $\mathcal{N}=2$, $D=4$ supergravity), which has $R=-2$. Note
that $R=-2$ and $R=-2/3$ are the unique two values for which the K\"{a}hler
manifold $\frac{SL(2,\mathbb{R})}{U(1)}$ is a \textit{special} K\"{a}hler
manifold \cite{CVP}.} \cite{CFG}:%
\begin{equation}
\left. \frac{SL(2,\mathbb{R})}{U(1)}\right\vert _{T^{3}}\overset{c}{%
\longrightarrow }\frac{G_{2(2)}}{SU(2)\times SU(2)}.  \label{c-c-2}
\end{equation}

The corresponding Jordan algebra interpretation of (\ref{pre-c-2}) reads%
\begin{equation}
QConf\left( \mathbb{R}\right) \supset _{s}SL(3,\mathbb{R}),
\end{equation}%
because the $T^{3}$ model is related to the (non-generic) simple rank-$3$
Euclidean Jordan algebra given by the reals $\mathbb{R}$ (see Tables 5-8).

\subsubsection{$D=6$}

Analogously, \textit{\textquotedblleft pure"} $D=6$ $(1,0)$ chiral
supergravity\footnote{%
We here disregard the various conditions to be fulfilled for \textit{%
anomaly-free} chiral supergravity theories in $D=6$ (see \textit{e.g. }\cite%
{An}).} can be obtained as $D=6$ uplift of the so-called $\mathcal{N}=2 $, $%
D=4$ $ST^{2}$ model, whose vector multiplets' scalars span the symmetric
special K\"{a}hler manifold $\left[ SL(2,\mathbb{R})/U(1)\right] ^{2}$, and
whose $D=3$ $U$-duality group is $G_{4,ST^{2}}^{3}=SO(4,3)$.

The embedding of the $6$-dimensional Ehlers group $SL\left( 4,\mathbb{R}%
\right) $ into $G_{4,ST^{2}}^{3}$ is maximal and \textit{symmetric}:%
\begin{equation}
SO(4,3)\supset _{s}SO\left( 3,3\right) \sim SL(4,\mathbb{R}),
\label{pre-c-3}
\end{equation}%
and at the \textit{mcs} level:%
\begin{equation}
SO(4)\times SO(3)\supset _{s}SO(3)\times SO(3)\sim SO(4)_{J},  \label{c-3}
\end{equation}%
where the $D=6$ massless spin group is $SO(4)_{J}$. The $12$ bosonic
massless states of \textit{\textquotedblleft pure"} $D=6$ $(1,0)$
supergravity are in the $\left( \mathbf{4},\mathbf{3}\right) $ of $%
SO(4)\times SO(3)=mcs\left( SO\left( 4,3\right) \right) $.

By merging (\ref{pre-c-3}) and (\ref{c-3}), the following $c$-map is
obtained \cite{CFG}:%
\begin{equation}
\left[ \frac{SL(2,\mathbb{R})}{U(1)}\right] ^{2}\overset{c}{\longrightarrow }%
\frac{SO\left( 4,3\right) }{SO(4)\times SO(3)}.  \label{c-c-3}
\end{equation}

The corresponding Jordan algebra interpretation of (\ref{pre-c-3}) reads%
\begin{equation}
QConf\left( \mathbb{R}\oplus \mathbf{\Gamma }_{1,0}\right) \supset _{s}SL(4,%
\mathbb{R}),
\end{equation}%
because the $ST^{2}$ model is related to the (non-generic) semi-simple rank-$%
3$ Euclidean Jordan algebra given by $\mathbb{R}\oplus \mathbf{\Gamma }%
_{1,0}\sim \mathbb{R}\oplus \mathbb{R}$.

\subsection{The Jordan Symmetric Infinite Sequence}

The aforementioned $ST^{2}$ model is actually the first element of the
so-called \textit{Jordan symmetric sequence} of \textit{quarter-maximal}
theories.

The $D=3$ $U$-duality group is $G_{4}^{3}=SO\left( 4,D-2+n\right) $, where $%
n $ is the number of matter multiplets in $D=3$ other than those coming from
the reduction of the gravity multiplet in $D$ dimensions. Furthermore,
\textit{mcs}$\left( G_{8}^{3}\right) =SO(4)\times SO\left( D-2+n\right)
_{CV} $; as mentioned, $SO(4)\sim SU(2)_{v}\times SU(2)_{h}$ is the Clifford
algebra for massless particles with $\mathcal{N}=8$ supersymmetries, and $%
SO\left( D-2+n\right) _{CV}$ is the \textit{Clifford vacuum} symmetry.

Let us consider the relevant chain of maximal embeddings leading to the
embedding of the $D$-dimensional Ehlers group\footnote{%
Note that, consistently, for $n=0$ (in $D=5$ and $D=6$) and $n=1$ (in $D=4$%
), one re-obtains the case of the $ST^{2}$ model treated above.} $SL\left(
D-2,\mathbb{R}\right) $ into $SO\left( 4,D-2+n\right) $.

\subsubsection{$D=6$}

In $D=6$, it suffices to start from $G_{4}^{3}=SO\left( 4,3+n\right) $, and
the corresponding maximal symmetric embedding reads%
\begin{equation}
SO\left( 4,3+n\right) \supset _{s}SO\left( 3,3\right) \times SO\left(
1,n\right) \sim SL\left( 4,\mathbb{R}\right) \times SO\left( 1,n\right) ,
\label{zero-zero}
\end{equation}%
and at the \textit{mcs} level:%
\begin{equation}
SO\left( 4\right) \times SO\left( 3+n\right) \supset _{s}SO\left( 3\right)
\times SO\left( 3\right) \times \times SO\left( n\right) \sim SO\left(
4\right) \times SO\left( n\right) ,  \label{zero-zero-2}
\end{equation}%
where $n$ is the number of matter (tensor) multiplets in $D=6$. The group
commuting with $SL\left( 4,\mathbb{R}\right) $ inside $SO\left( 4,3+n\right)
$ is nothing but the $6$-dimensional $U$-duality group of tensor multiplets $%
G_{4}^{6}=SO(1,n)$.

\subsubsection{$D=5$}

For $D=5$, one branches once more from (\ref{zero-zero}), getting:%
\begin{equation}
SO\left( 4,3+n\right) \supset _{s}SL\left( 4,\mathbb{R}\right) \times
SO\left( 1,n\right) \supset _{s}SL\left( 3,\mathbb{R}\right) \times
SO(1,1)\times SO\left( 1,n\right) ,  \label{pre-zero-zero}
\end{equation}%
and at the \textit{mcs} level:%
\begin{equation}
SO\left( 4\right) \times SO\left( 3+n\right) \supset _{s}SO\left( 4\right)
\times SO\left( n\right) \supset _{s}SO\left( 3\right) \times SO\left(
n\right) ,  \label{zero-zero-3}
\end{equation}%
where $n+1$ is the number of matter (vector) multiplets in $D=5$. The group
commuting with $SL\left( 3,\mathbb{R}\right) $ inside $SO\left( 4,3+n\right)
$ is nothing but the $5$-dimensional $U$-duality group $G_{4}^{5}=SO(1,1)%
\times SO(1,n)$. Note the \textit{\textquotedblleft enhancement"} to $%
SL\left( 4,\mathbb{R}\right) \times SO\left( 1,n\right) $ in (\ref%
{pre-zero-zero}).

\subsubsection{$D=4$}

For $D=4$, the embedding is maximal and symmetric :%
\begin{equation}
SO\left( 4,2+n\right) \supset _{s}SO(2,2)\times SO(2,n)\sim SL\left( 2,%
\mathbb{R}\right) _{\text{Ehlers}}\times SL\left( 2,\mathbb{R}\right) \times
SO(2,n),  \label{pre-zero-zero-2}
\end{equation}%
and at the \textit{mcs} level:%
\begin{equation}
SO\left( 4\right) \times SO\left( 2+n\right) \supset _{s}SO\left( 2\right)
_{J}\times SO\left( 2\right) \times SO\left( 2\right) \times SO\left(
n\right) ,  \label{zero-zero-4}
\end{equation}%
where $n$ is the number of matter (vector) multiplets in $D=4$. The group
commuting with $SL(2,\mathbb{R})_{\text{Ehlers}}$ inside $SO\left(
4,2+n\right) $ is nothing but the $4$-dimensional $U$-duality group $%
G_{4}^{4}=SL\left( 2,\mathbb{R}\right) \times SO(2,n)$. By merging (\ref%
{pre-zero-zero-2}) and (\ref{zero-zero-4}), one obtains the following $c$%
-map:%
\begin{equation}
\frac{SL\left( 2,\mathbb{R}\right) }{U(1)}\times \frac{SO(2,n)}{SO(2)\times
SO(n)}\overset{c}{\longrightarrow }\frac{SO(4,n+2)}{SO(4)\times SO(n+2)}.
\label{c-map-1}
\end{equation}

\subsection{\textit{Magical} Models}

Let us now consider the isolated cases of symmetric $N=8$ \textit{%
quarter-maximal} theories, the so-called \textit{magical} models \cite{GST}.
They are associated to rank-$2$ (in $D=6$) and rank-$3$ (in $D=5$) Euclidean
Jordan algebras over the four normed division algebras $\mathbb{O}$
(octonions), $\mathbb{H}$ (quaternions), $\mathbb{C}$ (complex numbers) and $%
\mathbb{R}$ (real numbers), and to the Freudenthal triple systems over such
algebras (in $D=4$). Consequently, they can be parametrized in terms of the
real dimension of the relevant division algebra, namely $q=8,4,2,1$ for $%
\mathbb{O}$, $\mathbb{H}$, $\mathbb{C}$ and $\mathbb{R}$, respectively. In
this respect, the $T^{3}$ model treated above corresponds to $q=-2/3$.

We will now analyze the relevant embeddings in $D=4$, $5$ and $6$.

\subsubsection{$D=4$}

In $D=4$, the magic models are related to the Freudenthal triple system $%
\mathfrak{M}\left( J_{3}^{\mathbb{A}}\right) $ over the rank-$3$ \textit{%
simple} Euclidean Jordan algebra $J_{3}^{\mathbb{A}}$ ($\mathbb{A}=\mathbb{O}%
,\mathbb{H},\mathbb{C},\mathbb{R}$). The $D=3$ and $D=4$ $U$-duality groups
are nothing but the \textit{quasi-conformal} and \textit{conformal} group of
$J_{3}^{\mathbb{A}}$, respectively, and they are related by the following
maximal \textit{symmetric} embedding:%
\begin{equation}
G_{4}^{3}\left( q\right) \supset _{s}SL(2,\mathbb{R})_{\text{Ehlers}}\times
G_{4}^{4}\left( q\right) ,  \label{pre-gen-1-3}
\end{equation}%
with \textit{mcs} level involving the $D=4$ massless spin group:%
\begin{equation}
mcs\left( G_{4}^{3}\left( q\right) \right) \supset _{s}SO(2)_{J}\times
mcs\left( G_{4}^{4}\left( q\right) \right) .  \label{gen-1-3}
\end{equation}%
(\ref{pre-gen-1-3})-(\ref{gen-1-3}) correspond to the following $c^{\ast }$%
-map symmetric embedding of the corresponding scalar manifolds in $D=3$
(para-quaternionic pseudo-Riemannian) and $D=4$ (special K\"{a}hler):%
\begin{equation}
\frac{G_{4}^{4}\left( q\right) }{mcs\left( G_{4}^{4}\left( q\right) \right) }%
\overset{c^{\ast }}{\longrightarrow }\frac{G_{4}^{3}\left( q\right) }{SL(2,%
\mathbb{R})\times G_{4}^{4}\left( q\right) }.
\end{equation}

\begin{table}[t!]
\begin{center}
\begin{tabular}{|c||c|c|}
\hline
$\mathfrak{M}\left( J_{3}^{\mathbb{A}}\right) $ & $%
\begin{array}{c}
\\
G_{4}^{3}\left( q\right) \supset _{s}G_{4}^{4}\left( q\right) \times
SL\left( 2,\mathbb{R}\right) \\
~~%
\end{array}%
$ & $%
\begin{array}{c}
\\
\text{type} \\
~~%
\end{array}%
$ \\ \hline\hline
$%
\begin{array}{c}
\\
\mathfrak{M}\left( J_{3}^{\mathbb{O}}\right) ~(q=8) \\
~%
\end{array}%
$ & $E_{8\left( -24\right) }\supset E_{7(-25)}\times SL\left( 2,\mathbb{R}%
\right) $ & max,~$s$ \\ \hline
$%
\begin{array}{c}
\\
\mathfrak{M}\left( J_{3}^{\mathbb{H}}\right) ~(q=4) \\
~%
\end{array}%
$ & $E_{7(-5)}\supset SO^{\ast }\left( 12\right) \times SL\left( 2,\mathbb{R}%
\right) $ & max,~$s$ \\ \hline
$%
\begin{array}{c}
\\
\mathfrak{M}\left( J_{3}^{\mathbb{C}}\right) ~(q=2) \\
~%
\end{array}%
$ & $E_{6(2)}\supset SU\left( 3,3\right) \times SL\left( 2,\mathbb{R}\right)
$ & max,~$s~$ \\ \hline
$%
\begin{array}{c}
\\
\mathfrak{M}\left( J_{3}^{\mathbb{R}}\right) ~(q=1) \\
~%
\end{array}%
$ & $F_{4(4)}\supset Sp\left( 6,\mathbb{R}\right) \times SL\left( 2,\mathbb{R%
}\right) $ & max,~$s$ \\ \hline
$%
\begin{array}{c}
\\
\mathfrak{M}\left( \mathbb{R}\right) ~(q=-2/3) \\
~%
\end{array}%
$ & $G_{2(2)}\supset SL\left( 2,\mathbb{R}\right) \times SL(2,\mathbb{R})$ &
max,~$s$ \\ \hline
\end{tabular}%
\end{center}
\caption{Embedding $G_{4}^{3}\left( q\right) \supset G_{4}^{4}\left(
q\right) \times _{s}SL(2,\mathbb{R})_{\text{Ehlers}}$ for \textit{magical}
Maxwell-Einstein supergravity theories ($N=8$) in $D=4$ Lorentzian
space-time dimensions. Also the case of $T^{3}$ model ($q=-2/3$) is
reported. }
\end{table}

\begin{table}[h!]
\begin{center}
\begin{tabular}{|c||c|c|}
\hline
$\mathfrak{M}\left( J_{3}^{\mathbb{A}}\right) $ & $%
\begin{array}{c}
\\
mcs\left( G_{4}^{3}\left( q\right) \right) \supset _{s}mcs\left(
G_{4}^{4}\left( q\right) \right) \times SO(2)_{J} \\
~~%
\end{array}%
$ & $%
\begin{array}{c}
\\
\text{type} \\
~~%
\end{array}%
$ \\ \hline\hline
$%
\begin{array}{c}
\\
\mathfrak{M}\left( J_{3}^{\mathbb{O}}\right) ~(q=8) \\
~%
\end{array}%
$ & $E_{7\left( -133\right) }\times SU(2)\supset E_{6(-78)}\times U(1)\times
SO\left( 2\right) _{J}$ & max,~$s$ \\ \hline
$%
\begin{array}{c}
\\
\mathfrak{M}\left( J_{3}^{\mathbb{H}}\right) ~(q=4) \\
~%
\end{array}%
$ & $SO(12)\times SU(2)\supset U(6)\times SO\left( 2\right) _{J}$ & max,~$s$
\\ \hline
$%
\begin{array}{c}
\\
\mathfrak{M}\left( J_{3}^{\mathbb{C}}\right) ~(q=2) \\
~%
\end{array}%
$ & $SU(6)\times SU(2)\supset S\left( U\left( 3\right) \times U(3)\right)
\times SO\left( 2\right) _{J}$ & max,~$s~$ \\ \hline
$%
\begin{array}{c}
\\
\mathfrak{M}\left( J_{3}^{\mathbb{R}}\right) ~(q=1) \\
~%
\end{array}%
$ & $USp(6)\times SU(2)\supset U(3)\times SO\left( 2\right) _{J}$ & max,~$s$
\\ \hline
$%
\begin{array}{c}
\\
\mathfrak{M}\left( \mathbb{R}\right) ~(q=-2/3) \\
~%
\end{array}%
$ & $SU(2)\times SU(2)\supset U(1)\times SO(2)_{J}$ & max,~$s$ \\ \hline
\end{tabular}%
\end{center}
\caption{Embedding $mcs\left( G_{4}^{3}\left( q\right) \right) \supset
_{s}mcs\left( G_{4}^{4}\left( q\right) \right) \times SO(2)_{J}$ for \textit{%
magical} Maxwell-Einstein supergravity theories ($N=8$) in $D=4$ Lorentzian
space-time dimensions. Also the case of $T^{3}$ model ($q=-2/3$) is
reported. }
\end{table}

The various cases are listed in Tables 5 and 6.

\subsubsection{$D=5$}

In $D=5$, the magic models are related to $J_{3}^{\mathbb{A}}$'s themselves.
The $D=5$ $U$-duality group is the \textit{reduced structure} group of $%
J_{3}^{\mathbb{A}}$, and the embedding of the $D=5$ Ehlers group $SL(3,%
\mathbb{R})$ into the $D=3$ $U$-duality group is maximal and \textit{%
non-symmetric}:%
\begin{equation}
G_{4}^{3}\left( q\right) \supset _{ns}SL(3,\mathbb{R})\times G_{4}^{5}\left(
q\right) ,  \label{ggen-1}
\end{equation}%
with \textit{mcs} level involving the $D=5$ massless spin group:%
\begin{equation}
mcs\left( G_{4}^{3}\left( q\right) \right) \supset _{s}SO(3)_{J}\times
mcs\left( G_{4}^{5}\left( q\right) \right) .  \label{ggen-3}
\end{equation}

The various cases are listed in Tables 7 and 8.

\begin{table}[t!]
\begin{center}
\begin{tabular}{|c||c|c|}
\hline
$J_{3}^{\mathbb{A}}$ & $%
\begin{array}{c}
\\
G_{4}^{3}\left( q\right) \supset _{s}G_{4}^{5}\left( q\right) \times
SL\left( 3,\mathbb{R}\right) \\
~~%
\end{array}%
$ & $%
\begin{array}{c}
\\
\text{type} \\
~~%
\end{array}%
$ \\ \hline\hline
$%
\begin{array}{c}
\\
J_{3}^{\mathbb{O}}~(q=8) \\
~%
\end{array}%
$ & $E_{8\left( -24\right) }\supset E_{6(-26)}\times SL\left( 3,\mathbb{R}%
\right) $ & max,~$ns$ \\ \hline
$%
\begin{array}{c}
\\
J_{3}^{\mathbb{H}}~(q=4) \\
~%
\end{array}%
$ & $E_{7(-5)}\supset SU^{\ast }\left( 6\right) \times SL\left( 3,\mathbb{R}%
\right) $ & max,~$ns$ \\ \hline
$%
\begin{array}{c}
\\
J_{3}^{\mathbb{C}}~(q=2) \\
~%
\end{array}%
$ & $E_{6(2)}\supset SL\left( 3,C\right) \times SL\left( 3,\mathbb{R}\right)
$ & max,~$ns~$ \\ \hline
$%
\begin{array}{c}
\\
J_{3}^{\mathbb{R}}~(q=1) \\
~%
\end{array}%
$ & $F_{4(4)}\supset SL\left( 3,R\right) \times SL\left( 3,\mathbb{R}\right)
$ & max,~$ns$ \\ \hline
$%
\begin{array}{c}
\\
\mathbb{R}~(q=-2/3) \\
~%
\end{array}%
$ & $G_{2(2)}\supset SL(3,\mathbb{R})$ & max,~$ns$ \\ \hline
\end{tabular}%
\end{center}
\caption{Embedding $G_{4}^{3}\left( q\right) \supset G_{4}^{5}\left(
q\right) \times _{s}SL(3,\mathbb{R})_{\text{Ehlers}}$ for \textit{magical}
Maxwell-Einstein supergravity theories ($N=8$) in $D=5$ Lorentzian
space-time dimensions. The $D=5$ uplift of $T^{3}$ model is \textit{%
\textquotedblleft pure"} minimal supergravity}
\end{table}

\begin{table}[h]
\begin{center}
\begin{tabular}{|c||c|c|}
\hline
$J_{3}^{\mathbb{A}}$ & $%
\begin{array}{c}
\\
mcs\left( G_{4}^{3}\left( q\right) \right) \supset _{s}mcs\left(
G_{4}^{5}\left( q\right) \right) \times SO\left( 3\right) _{J} \\
~~%
\end{array}%
$ & $%
\begin{array}{c}
\\
\text{type} \\
~~%
\end{array}%
$ \\ \hline\hline
$%
\begin{array}{c}
\\
J_{3}^{\mathbb{O}}~(q=8) \\
~%
\end{array}%
$ & $E_{7\left( -133\right) }\times SU(2)\supset F_{4(-52)}\times SO\left(
3\right) _{J}$ & max,~$ns$ \\ \hline
$%
\begin{array}{c}
\\
J_{3}^{\mathbb{H}}~(q=4) \\
~%
\end{array}%
$ & $SO(12)\times SU(2)\supset USp(6)\times SO\left( 3\right) _{J}$ & max,~$%
ns$ \\ \hline
$%
\begin{array}{c}
\\
J_{3}^{\mathbb{C}}~(q=2) \\
~%
\end{array}%
$ & $SU(6)\times SU(2)\supset SU\left( 3\right) \times SO\left( 3\right)
_{J} $ & max,~$ns~$ \\ \hline
$%
\begin{array}{c}
\\
J_{3}^{\mathbb{R}}~(q=1) \\
~%
\end{array}%
$ & $USp(6)\times SU(2)\supset SU(2)_{P}\times SO\left( 3\right) _{J}$ &
max,~$ns$ \\ \hline
$%
\begin{array}{c}
\\
\mathbb{R}~(q=-2/3) \\
~%
\end{array}%
$ & $SU(2)\times SU(2)\supset SO(3)_{J,D}$ & max,~$ns$ \\ \hline
\end{tabular}%
\end{center}
\caption{Embedding $mcs\left( G_{4}^{3}\left( q\right) \right) \supset
mcs\left( G_{4}^{5}\left( q\right) \right) \times _{s}SO(3)_{J}$ for \textit{%
magical} Maxwell-Einstein supergravity theories ($N=8$) in $D=5$ Lorentzian
space-time dimensions. $SU(2)_{P}$ denotes the \textit{principal} $SU(2)$,
whereas the subscript \textquotedblleft $D$" stands for \textit{diagonal
embedding}}
\end{table}

\subsubsection{$D=6$}

In $D=6$, the magic models are related to the rank-$2$ Jordan algebra $%
J_{2}^{\mathbb{A}}\sim \mathbf{\Gamma }_{1,q+1}$ (where \textquotedblleft $%
\sim $" here denotes a vector space isomorphism). Namely, the $D=6$ $U$%
-duality group is nothing but the \textit{reduced structure} group of $%
J_{2}^{\mathbb{A}}$ itself, with the exception of the cases corresponding to
$q=4$ and $q=2$, which have a further factor\footnote{%
We note that the non-triviality of the factor group $\mathcal{A}_{q}$ in the
$D=6$ $U$-duality group is related to the reality properties of the spinors
within the rank-$2$ Jordan algebras over the quaternions ($J_{2}^{\mathbb{H}%
}\sim \mathbf{\Gamma }_{1,5}$) and over the complex numbers ($J_{2}^{\mathbb{%
C}}\sim \mathbf{\Gamma }_{1,3}$), which are respectively pseudo-real
(quaternionic) and complex (see \textit{e.g.} Table 2 of \cite{G-gauge-D=6}).%
} $\mathcal{A}_{q=2}=SO(3)$ resp. $\mathcal{A}_{q=1}=SO(2)$ in the $U$%
-duality group. The embedding of the $D=6$ Ehlers group $SL(4,\mathbb{R})$
into the $D=3$ $U$-duality group is obtained by a two-steps chain of maximal
and \textit{symmetric} embeddings ($\mathcal{A}_{q}=Id$, $SO(3)$, $SO(2)$, $%
Id$ respectively for $q=8,4,2,1$):%
\begin{equation}
G_{4}^{3}\left( q\right) \supset _{s}SO\left( 4,q+4\right) \times \mathcal{A}%
_{q}\supset _{s}SL(4,\mathbb{R})\times SO(1,q+1)\times \mathcal{A}_{q},
\label{gggen-1}
\end{equation}%
with \textit{mcs} level involving the $D=6$ massless spin group:%
\begin{equation}
mcs\left( G_{4}^{3}\left( q\right) \right) \supset _{s}SO(4)_{J}\times
SO(q+1)\times mcs\left( \mathcal{A}_{q}\right) .  \label{gggen-3}
\end{equation}

Note the \textit{\textquotedblleft enhancement"} to $SO\left( 4,q+4\right)
\times \mathcal{A}_{q}$ in (\ref{gggen-1}). The various cases are listed in
Tables 9 and 10.

\begin{table}[t!]
\begin{center}
\begin{tabular}{|c||c|c|}
\hline
$J_{2}^{\mathbb{A}}$ & $%
\begin{array}{c}
\\
G_{4}^{3}\left( q\right) \supset SO(1,q+1)\times A_{q}\times SL(4,\mathbb{R})
\\
~~%
\end{array}%
$ & $%
\begin{array}{c}
\\
\text{type} \\
~~%
\end{array}%
$ \\ \hline\hline
$%
\begin{array}{c}
\\
J_{2}^{\mathbb{O}}~(q=8) \\
~%
\end{array}%
$ & $E_{8\left( -24\right) }\supset SO(1,9)\times SL\left( 4,\mathbb{R}%
\right) $ & $nm$,~$ns$ \\ \hline
$%
\begin{array}{c}
\\
J_{2}^{\mathbb{H}}~(q=4) \\
~%
\end{array}%
$ & $E_{7(-5)}\supset SO\left( 1,5\right) \times SO(3)\times SL\left( 4,%
\mathbb{R}\right) $ & $nm$,~$ns$ \\ \hline
$%
\begin{array}{c}
\\
J_{2}^{\mathbb{C}}~(q=2) \\
~%
\end{array}%
$ & $E_{6(2)}\supset SO\left( 1,3\right) \times SO(2)\times SL\left( 4,%
\mathbb{R}\right) $ & $nm$,~$ns~$ \\ \hline
$%
\begin{array}{c}
\\
J_{2}^{\mathbb{R}}~(q=1) \\
~%
\end{array}%
$ & $F_{4(4)}\supset SO\left( 1,2\right) \times SL\left( 4,\mathbb{R}\right)
$ & $nm$,~$ns$ \\ \hline
\end{tabular}%
\end{center}
\caption{Embedding $G_{4}^{3}\left( q\right) \supset _{ns}G_{4}^{6}\left(
q\right) \times SL(4,\mathbb{R})_{\text{Ehlers}}$ ($G_{4}^{6}\left( q\right)
=SO(1,q+1)\times \mathcal{A}_{q}$) for chiral \textit{magical}
Maxwell-Einstein supergravity theories ($N=8$) in $D=6$ Lorentzian
space-time dimensions. Recall $SO(1,5)\sim SU^{\ast }(4)$, $SO(1,3)\sim
SL\left( 3,\mathbb{C}\right) $, $SO(1,2)\sim SL(2,\mathbb{R})$.}
\end{table}

\begin{table}[h!]
\begin{center}
\begin{tabular}{|c||c|c|}
\hline
$J_{2}^{\mathbb{A}}$ & $%
\begin{array}{c}
\\
mcs\left( G_{4}^{3}\left( q\right) \right) \supset mcs\left( G_{4}^{6}\left(
q\right) \right) \times SO(4)_{J} \\
~~%
\end{array}%
$ & $%
\begin{array}{c}
\\
\text{type} \\
~~%
\end{array}%
$ \\ \hline\hline
$%
\begin{array}{c}
\\
J_{2}^{\mathbb{O}}~(q=8) \\
~%
\end{array}%
$ & $E_{7\left( -133\right) }\times SU(2)\supset SO(9)\times SO\left(
4\right) _{J}$ & $nm$,~$ns$ \\ \hline
$%
\begin{array}{c}
\\
J_{2}^{\mathbb{H}}~(q=4) \\
~%
\end{array}%
$ & $SO(12)\times SU(2)\supset SO\left( 5\right) \times SO(3)\times SO\left(
4\right) _{J}$ & $nm$,~$ns$ \\ \hline
$%
\begin{array}{c}
\\
J_{2}^{\mathbb{C}}~(q=2) \\
~%
\end{array}%
$ & $SU(6)\times SU(2)\supset SO\left( 3\right) \times SO(2)\times SO\left(
4\right) _{J}$ & $nm$,~$ns~$ \\ \hline
$%
\begin{array}{c}
\\
J_{2}^{\mathbb{R}}~(q=1) \\
~%
\end{array}%
$ & $USp(6)\times SU(2)\supset SO\left( 2\right) \times SO\left( 4\right)
_{J}$ & $nm$,~$ns$ \\ \hline
\end{tabular}%
\end{center}
\caption{Embedding $mcs\left( G_{4}^{3}\left( q\right) \right) \supset
_{ns}mcs\left( G_{4}^{5}\left( q\right) \right) \times SO(4)_{J}$ for chiral
\textit{magical} Maxwell-Einstein supergravity theories ($N=8$) in $D=6$
Lorentzian space-time dimensions. }
\end{table}

\section{\label{Sec-Chi=0-Cosets}Cosets with $\protect\chi =0$ and \textit{%
Poincar\'{e} Duality}}

From the previous treatment, a class of non-compact, pseudo-Riemannian
homogeneous spaces can be naturally constructed, with general structure:%
\begin{equation}
M_{N}^{D}\equiv \frac{G_{N}^{3}}{G_{N}^{D}\times SL(D-2,\mathbb{R})},
\label{Ggen}
\end{equation}%
determined by the embedding of the direct product of the $D$-dimensional
Ehlers group $SL(D-2,\mathbb{R})$ and of the $D$-dimensional $U$-duality
group $G_{N}^{D}$ of a supergravity with $\mathcal{N}=2N$ supersymmetries
into the $U$-duality group of the same theory reduced to $D=3$ (Lorentzian)
space-time dimensions. From previous Secs., such an embedding can be \textit{%
maximal} or non-maximal (namely, \textit{next-to-maximal}), and \textit{%
symmetric} or \textit{non-symmetric}, but, as mentioned, it always preserves
the rank of the group (\ref{rank1}), as well as the non-compact rank of the $%
D=3$ coset $G_{N}^{3}/H_{N}^{3}$ (\ref{rank2}).

Interestingly, the cosets $M_{N}^{D}$'s (\ref{Ggen}) all share the same
feature : they have an equal number of compact and non-compact generators,
thus implying the their \textit{coset character} $\chi $ \cite%
{Gilmore,Helgason} to be \textit{vanishing}:%
\begin{equation}
\chi \left( M_{N}^{D}\right) \equiv nc\left( M_{N}^{D}\right) -c\left(
M_{N}^{D}\right) =0.  \label{chichi}
\end{equation}

This property can also be related to the \textquotedblleft $mcs$
counterpart" of the class of cosets (\ref{Ggen}), given by the compact,
Riemannian homogeneous spaces with general structure%
\begin{equation}
\widehat{M}_{N}^{D}\equiv \frac{mcs\left( G_{N}^{3}\right) }{mcs\left(
G_{N}^{D}\right) \times SO(D-2)_{J}},  \label{Ggen-mcs}
\end{equation}%
determined by the embedding of the direct product of the $D$-dimensional
massless spin group $SO(D-2)=mcs\left( SL(D-2,\mathbb{R})\right) $ and of $%
H_{N}^{D}=mcs\left( G_{N}^{D}\right) $ into $H_{N}^{3}=mcs\left(
G_{N}^{3}\right) $. As the $M_{N}^{D}$'s (\ref{Ggen}), also the $\widehat{M}%
_{N}^{D}$'s (\ref{Ggen-mcs}) can be of various types, namely \textit{maximal}
or \textit{next-to-maximal}, \textit{symmetric} or \textit{non-symmetric}.

However, $\widehat{M}_{N}^{D}$'s (\ref{Ggen-mcs}) all share the same
property : the number of compact or non-compact generators of $M_{N}^{D}$'s (%
\ref{Ggen}) is always \textit{equal} to the (real) dimension of the
corresponding $\widehat{M}_{N}^{D}$'s themselves. This is a consequence of (%
\ref{chichi}) as well as the general formula on the signature of a
pseudo-Riemannian coset $G/H$ (see \textit{e.g.} \cite{Gilmore})%
\begin{equation}
\begin{array}{c}
c\left( G/H\right) =\text{dim}_{\mathbb{R}}\left( mcs\left( G\right) \right)
-\text{dim}_{\mathbb{R}}\left( mcs\left( H\right) \right) ; \\
\\
nc\left( G/H\right) =\text{dim}_{\mathbb{R}}\left( G\right) -\text{dim}_{%
\mathbb{R}}\left( H\right) -c\left( G/H\right) ,%
\end{array}%
\end{equation}%
from which thus follows that the compact generators of $M_{N}^{D}$ are the
very generators of the corresponding $\widehat{M}_{N}^{D}$:%
\begin{equation}
nc\left( M_{N}^{D}\right) =c\left( M_{N}^{D}\right) =dim_{\mathbb{R}}\left(
\widehat{M}_{N}^{D}\right)  \label{night-1}
\end{equation}

\texttt{\medskip }Along this line, further elaboration is possible. Indeed,
it generally holds that%
\begin{equation}
\text{dim}_{\mathbb{R}}\left[ \frac{G_{N}^{3}}{G_{N}^{D}\times SL(D-2,%
\mathbb{R})}\right] =2\text{dim}_{\mathbb{R}}\left[ \frac{H_{N}^{3}}{%
H_{N}^{D}\times SO(D-2)}\right] .
\end{equation}%
A possible interpretation of these results is as follows. In a supergravity
theory in $D$ space-time (Lorentzian) dimensions, the number of bosonic
massless degrees of freedom other than the scalar and graviton ones is given
by the difference between the dimension of the Clifford algebra and the sum
of the dimensions of the $D$-dimensional massless spin group and of the $D$%
-dimensional \textit{\textquotedblleft Clifford symmetry"} (\textit{i.e.}, $%
\mathcal{R}$-symmetry $+$ \textit{Clifford vacuum} degeneracy due to matter
coupling, if any).

Sec. \ref{Cosets} lists the cosets $M_{N}^{D}$'s (\ref{Ggen}) and their
\textquotedblleft $mcs$ counterparts" $\widehat{M}_{N}^{D}$'s (\ref{Ggen-mcs}%
) for all $N$'s and $D$'s treated in the present investigation. Then, in
Secs. \ref{PD} and \ref{Hodge} an interpretation of the vanishing character (%
\ref{chichi}) will be given in terms of \textit{Poincar\'{e} duality}, or
equivalently of \textit{Hodge involution} acting on the cohomology of $%
M_{N}^{D}$'s.

\subsection{\label{Cosets}The Cosets}

\subsubsection{$N=16$}

The specification of (\ref{Ggen}) and (\ref{Ggen-mcs}) to \textit{maximal
supergravity} ($N=16$) give rise the following spaces%
\begin{eqnarray}
M_{16}^{D} &\equiv &\frac{G_{16}^{3}}{G_{16}^{D}\times SL(D-2,\mathbb{R})}=%
\frac{E_{8(8)}}{G_{16}^{D}\times SL(D-2,\mathbb{R})};  \label{M-cosets} \\
\widehat{M}_{16}^{D} &\equiv &\frac{H_{16}^{3}}{H_{16}^{D}\times SO(D-2)}=%
\frac{SO(16)}{H_{16}^{D}\times SO(D-2)};  \label{M-hat-cosets}
\end{eqnarray}%
they are listed in Table 11, along with their number of compact and
non-compact generators. Among $M_{16}^{D}$'s, the unique maximal and \textit{%
symmetric} coset is the one pertaining to $D=4$ (\textit{cfr.} (\ref{D=4})):%
\begin{equation}
M_{16}^{4}\equiv \frac{G_{16}^{3}}{G_{16}^{4}\times SL(2,\mathbb{R})}=\frac{%
E_{8(8)}}{E_{7(7)}\times SL(2,\mathbb{R})},  \label{M-coset-D=4}
\end{equation}%
which is a rank-$4$ \textit{para-quaternionic} space, as resulting from the
classification of \cite{Ale}. Also the corresponding%
\begin{equation}
\widehat{M}_{16}^{4}=\frac{SO(16)}{SU(8)\times SO(2)}
\end{equation}%
is a maximal and \textit{symmetric} space among $\widehat{M}_{16}^{D}$'s.

\begin{table}[t!]
\begin{center}
\begin{tabular}{|c||c|c|l|}
\hline
$D$ & $%
\begin{array}{c}
\\
M_{16}^{D} \\
~~%
\end{array}%
$ & $%
\begin{array}{c}
\\
\widehat{M}_{16}^{D} \\
~~%
\end{array}%
$ & $%
\begin{array}{l}
c\left( M_{16}^{D}\right) = \\
nc\left( M_{16}^{D}\right) ~~%
\end{array}%
$ \\ \hline\hline
$%
\begin{array}{c}
\\
11 \\
~%
\end{array}%
$ & $\frac{E_{8(8)}}{SL(9,\mathbb{R})~}$ & $\frac{SO(16)}{SO(9)~}$ &
\multicolumn{1}{|c|}{$84$} \\ \hline
$%
\begin{array}{c}
\\
10,~IIA \\
~%
\end{array}%
$ & $\frac{E_{8(8)}}{SO(1,1)\times SL(8,\mathbb{R})}$ & $\frac{SO(16)}{SO(8)}
$ & \multicolumn{1}{|c|}{$92$} \\ \hline
$%
\begin{array}{c}
\\
10,~IIB \\
~%
\end{array}%
$ & $\frac{E_{8(8)}}{SL(2,R)\times SL(8,\mathbb{R})}$ & $\frac{SO(16)}{%
SO(2)\times SO(8)}$ & \multicolumn{1}{|c|}{$91$} \\ \hline
$%
\begin{array}{c}
\\
9 \\
~%
\end{array}%
$ & $\frac{E_{8(8)}}{GL(2,R)\times SL(7,\mathbb{R})}$ & $\frac{SO(16)}{%
SO(2)\times SO(7)}$ & \multicolumn{1}{|c|}{$98$} \\ \hline
$%
\begin{array}{c}
\\
8 \\
~%
\end{array}%
$ & $\frac{E_{8(8)}}{\left( SL(2,\mathbb{R})\times SL(3,\mathbb{R})\right)
\times SL(6,\mathbb{R})}$ & $\frac{SO(16)}{\left( SO(2)\times SO(3)\right)
\times SO(6)}$ & \multicolumn{1}{|c|}{$101$} \\ \hline
$%
\begin{array}{c}
\\
7 \\
~%
\end{array}%
$ & $\frac{E_{8(8)}}{SL(5,\mathbb{R})\times SL(5,\mathbb{R})}$ & $\frac{%
SO(16)}{SO(5)\times SO(5)}$ & \multicolumn{1}{|c|}{$100$} \\ \hline
$%
\begin{array}{c}
\\
6 \\
~%
\end{array}%
$ & $\frac{E_{8(8)}}{SO(5,5)\times SL(4,\mathbb{R})}$ & $\frac{SO(16)}{%
SO(5)\times SO(5)\times SO(4)}$ & \multicolumn{1}{|c|}{$94$} \\ \hline
$%
\begin{array}{c}
\\
5 \\
~%
\end{array}%
$ & $\frac{E_{8(8)}}{E_{6(6)}\times SL(3,\mathbb{R})}$ & $\frac{SO(16)}{%
USp(8)\times SO(3)}$ & \multicolumn{1}{|c|}{$81$} \\ \hline
$%
\begin{array}{c}
\\
4 \\
~%
\end{array}%
$ & $\frac{E_{8(8)}}{E_{7(7)}\times SL(2,\mathbb{R})}$ & $\frac{SO(16)}{%
SU(8)\times SO(2)}$ & \multicolumn{1}{|c|}{$56$} \\ \hline
\end{tabular}%
\end{center}
\caption{Pseudo-Riemannian non-compact $E_{8(8)}$-cosets $M_{16}^{D}$ (%
\protect\ref{M-cosets}) and Riemannian compact $SO(16)$-cosets $\protect%
\widehat{M}_{16}^{D}$ (\protect\ref{M-hat-cosets}) of \textit{maximal}
supergravity theories ($N=16$) in $11\geqslant D\geqslant 4$ Lorentzian
space-time dimensions. The number of compact generators $c$ (equal to the
number $nc$ of non-compact generators) of $M_{16}^{D}$ is also listed. All
cosets $M_{16}^{D}$ have \textit{vanishing character}.}
\end{table}

\subsubsection{$N=12$}

The specification of (\ref{Ggen}) and (\ref{Ggen-mcs}) to supergravity with $%
N=12$ in $D=5$ and in $D=4$ respectively reads
\begin{eqnarray}
M_{12}^{5} &\equiv &\frac{G_{12}^{3}}{G_{12}^{5}\times SL(3,\mathbb{R})}=%
\frac{E_{7(-5)}}{SU^{\ast }(6)\times SL(3,\mathbb{R})},~c=nc=45;  \label{uno}
\\
\widehat{M}_{12}^{5} &\equiv &\frac{H_{12}^{3}}{H_{12}^{5}\times SO(3)_{J}}=%
\frac{SO(12)\times SU(2)}{USp\left( 6\right) \times SO(3)_{J}}; \\
&&  \notag \\
M_{12}^{4} &\equiv &\frac{G_{12}^{3}}{G_{12}^{4}\times SL(2,\mathbb{R})}=%
\frac{E_{7(-5)}}{SO^{\ast }(12)\times SL(2,\mathbb{R})},~c=nc=32;
\label{uno-bis} \\
\widehat{M}_{12}^{4} &\equiv &\frac{H_{12}^{3}}{H_{12}^{4}\times SO(2)}=%
\frac{SO(12)\times SU(2)}{SU(6)\times U(1)\times SO(2)_{J}}.
\end{eqnarray}%
They all are maximal cosets, but $M_{12}^{5}$ and $\widehat{M}_{12}^{5}$ are
non-symmetric, whereas $M_{12}^{4}$ and $\widehat{M}_{12}^{4}$ are symmetric.

The values of $c=nc$ given in (\ref{uno}) and (\ref{uno-bis}) match the ones
of the magical quarter-maximal ($N=4$) theory for $q=4$ (see (\ref{one-1-1})
and (\ref{two-3-3}), respectively); indeed, these theories share the same
bosonic sector, and they are both related to $J_{3}^{\mathbb{H}}$.

\subsubsection{$N=10$}

The specification of (\ref{Ggen}) and (\ref{Ggen-mcs}) to supergravity with $%
N=10$ in $D=4$ gives rise to the following symmetric spaces%
\begin{eqnarray}
M_{10}^{5} &\equiv &\frac{G_{10}^{3}}{G_{10}^{4}\times SL\left( 2,\mathbb{R}%
\right) }=\frac{E_{6(-14)}}{SU(5,1)\times SL\left( 2,\mathbb{R}\right) }%
,~c=nc=20;  \label{uno-2} \\
\widehat{M}_{10}^{4} &\equiv &\frac{H_{10}^{3}}{H_{10}^{4}\times SO\left(
2\right) }=\frac{SO(10)\times U(1)}{SU(5)\times U(1)\times U(1)_{J}}.
\end{eqnarray}%
$M_{10}^{5}$ is a rank-$4$ \textit{para-quaternionic} coset.

\subsubsection{$N=8$}

The specification of (\ref{Ggen}) and (\ref{Ggen-mcs}) to \textit{%
half-maximal supergravity} ($N=8$) gives rise to the following spaces%
\begin{eqnarray}
M_{8}^{D} &\equiv &\frac{G_{8}^{3}}{G_{8}^{D}\times SL(D-2,\mathbb{R})}=%
\frac{SO\left( 8,D-2+m\right) }{G_{8}^{D}\times SL(D-2,\mathbb{R})};
\label{M-cosets-2} \\
\widehat{M}_{8}^{D} &\equiv &\frac{H_{8}^{3}}{H_{8}^{D}\times SO(D-2)}=\frac{%
SO\left( 8\right) \times SO\left( D-2+m\right) }{H_{8}^{D}\times SO(D-2)};
\label{M-hat-cosets-2}
\end{eqnarray}%
they are listed in Table 12, along with their number of compact and
non-compact generators.

Among $M_{8}^{D}$'s, the unique maximal and \textit{symmetric} cosets are
the ones pertaining to $D=6$ IIB and $D=4$(\textit{cfr.} (\ref{zero-2})):%
\begin{eqnarray}
M_{8}^{6,\text{~}IIB} &\equiv &\frac{G_{8}^{3}}{G_{8}^{6,~IIB}\times SL(2,%
\mathbb{R})}=\frac{SO\left( 8,3+m\right) }{SO\left( 5,m\right) \times SL(4,%
\mathbb{R})}; \\
M_{8}^{4} &\equiv &\frac{G_{8}^{3}}{G_{8}^{4}\times SL(2,\mathbb{R})}=\frac{%
SO\left( 8,2+m\right) }{SL(2,\mathbb{R})\times SO\left( 6,m\right) \times
SL(2,\mathbb{R})}.
\end{eqnarray}

\begin{table}[t!]
\begin{center}
\begin{tabular}{|c||c|c|l|}
\hline
$D$ & $%
\begin{array}{c}
\\
M_{8}^{D} \\
~~%
\end{array}%
$ & $%
\begin{array}{c}
\\
\widehat{M}_{8}^{D} \\
~~%
\end{array}%
$ & $%
\begin{array}{l}
c\left( M_{8}^{D}\right) = \\
nc\left( M_{8}^{D}\right) ~~%
\end{array}%
$ \\ \hline\hline
$%
\begin{array}{c}
\\
10 \\
~%
\end{array}%
$ & $\frac{SO\left( 8,8+m\right) }{\left( SO(1,1)\times SO\left( m\right)
\right) \times SL\left( 8,\mathbb{R}\right) }$ & $\frac{SO\left( 8\right)
\times SO\left( 8+m\right) }{SO\left( m\right) \times SO\left( 8\right) }$ &
\multicolumn{1}{|c|}{$8m+28$} \\ \hline
$%
\begin{array}{c}
\\
9 \\
~%
\end{array}%
$ & $\frac{SO\left( 8,7+m\right) }{\left( SO(1,1)\times SO\left( 1,m\right)
\right) \times SL\left( 7,\mathbb{R}\right) }$ & $\frac{SO\left( 8\right)
\times SO\left( 7+m\right) }{SO\left( m\right) \times SO\left( 7\right) }$ &
\multicolumn{1}{|c|}{$7m+28$} \\ \hline
$%
\begin{array}{c}
\\
8 \\
~%
\end{array}%
$ & $\frac{SO\left( 8,6+m\right) }{\left( SO(1,1)\times SO\left( 2,m\right)
\right) \times SL\left( 6,\mathbb{R}\right) }$ & $\frac{SO\left( 8\right)
\times SO\left( 6+m\right) }{\left( SO\left( 2\right) \times SO\left(
m\right) \right) \times SO\left( 6\right) }$ & \multicolumn{1}{|c|}{$6m+27$}
\\ \hline
$%
\begin{array}{c}
\\
7 \\
~%
\end{array}%
$ & $\frac{SO\left( 8,5+m\right) }{\left( SO(1,1)\times SO\left( 3,m\right)
\right) \times SL\left( 5,\mathbb{R}\right) }$ & $\frac{SO\left( 8\right)
\times SO\left( 5+m\right) }{\left( SO\left( 3\right) \times SO\left(
m\right) \right) \times SO\left( 5\right) }$ & \multicolumn{1}{|c|}{$5m+25$}
\\ \hline
$%
\begin{array}{c}
\\
6,~IIA \\
~%
\end{array}%
$ & $\frac{SO\left( 8,4+m\right) }{\left( SO(1,1)\times SO\left( 4,m\right)
\right) \times SL(4,\mathbb{R})}$ & $\frac{SO\left( 8\right) \times SO\left(
4+m\right) }{\left( SO\left( 4\right) \times SO\left( m\right) \right)
\times SO(4)}$ & \multicolumn{1}{|c|}{$4m+22$} \\ \hline
$%
\begin{array}{c}
\\
6,~IIB \\
~%
\end{array}%
$ & $\frac{SO\left( 8,3+m\right) }{SO\left( 5,m\right) \times SL(4,\mathbb{R}%
)}$ & $\frac{SO\left( 8\right) \times SO\left( 3+m\right) }{\left( SO\left(
5\right) \times SO\left( m\right) \right) \times SO(4)}$ &
\multicolumn{1}{|c|}{$3m+15$} \\ \hline
$%
\begin{array}{c}
\\
5 \\
~%
\end{array}%
$ & $\frac{SO\left( 8,3+m\right) }{\left( SO(1,1)\times SO\left( 5,m\right)
\right) \times SL(3,\mathbb{R})}$ & $\frac{SO\left( 8\right) \times SO\left(
3+m\right) }{\left( SO\left( 5\right) \times SO\left( m\right) \right)
\times SO(3)}$ & \multicolumn{1}{|c|}{$3m+18$} \\ \hline
$%
\begin{array}{c}
\\
4 \\
~%
\end{array}%
$ & $\frac{SO\left( 8,2+m\right) }{\left( SL(2,\mathbb{R})\times SO\left(
6,m\right) \right) \times SL(2,\mathbb{R})}$ & $\frac{SO\left( 8\right)
\times SO\left( 2+m\right) }{\left( SO(2)\times SO\left( 6\right) \times
SO\left( m\right) \right) \times SO(2)}$ & \multicolumn{1}{|c|}{$2m+12$} \\
\hline
\end{tabular}%
\end{center}
\caption{Pseudo-Riemannian non-compact $M_{8}^{D}$ (\protect\ref{M-cosets-2}%
) and Riemannian compact cosets $\protect\widehat{M}_{8}^{D}$ (\protect\ref%
{M-hat-cosets-2}) of \textit{half-maximal} supergravity theories ($N=8$) in $%
10\geqslant D\geqslant 4$ Lorentzian space-time dimensions. The number of
compact generators $c$ (equal to the number $nc$ of non-compact generators)
of $M_{8}^{D}$ is also listed. All cosets $M_{8}^{D}$ have \textit{vanishing
character}.}
\end{table}

\subsubsection{$N=6$}

The specification of (\ref{Ggen}) and (\ref{Ggen-mcs}) to supergravity with $%
N=6$ in $D=4$ gives rise to the following symmetric spaces%
\begin{eqnarray}
M_{6}^{4} &\equiv &\frac{G_{6}^{3}}{G_{6}^{4}\times SL(2,\mathbb{R})}=\frac{%
SU(4,m+1)}{SU(3,m)\times SL(2,\mathbb{R})},~c=nc=2m+7; \\
\widehat{M}_{6}^{4} &\equiv &\frac{H_{6}^{3}}{H_{6}^{4}\times SO(2)}=\frac{%
SU(4)\times SU\left( m+1\right) \times U(1)}{U(3)\times U(m)\times U(1)_{J}}.
\end{eqnarray}

\subsubsection{$N=4$ \textit{Symmetric}}

\paragraph{\textit{Minimal Coupling}}

The specification of (\ref{Ggen}) and (\ref{Ggen-mcs}) to \textit{minimally
coupled} Maxwell-Einstein supergravity with $N=4$ in $D=4$ gives rise to the
following symmetric spaces%
\begin{eqnarray}
M_{4}^{4} &\equiv &\frac{G_{4}^{3}}{G_{4}^{4}\times SL(2,\mathbb{R})}=\frac{%
SU(2,1+n)}{U(1,n)\times SL(2,\mathbb{R})},~c=nc=2n+2;  \label{M-cosets-4} \\
\widehat{M}_{4}^{4} &\equiv &\frac{H_{4}^{3}}{H_{4}^{4}\times SO(2)}=\frac{%
SU(2)\times SU(1+n)\times U(1)}{U(n)\times U(1)\times U(1)}.
\end{eqnarray}%
$M_{4}^{4}$ has rank $1$ for $n=0$, and rank $2$ for $n\geqslant 1$, and it
is \textit{para-quaternionic}. It is nothing but a suitable
pseudo-Riemannian form of the manifold (\ref{mc}) itself, namely the $%
c^{\ast }$-map of the rank-$1$ symmetric \textit{special K\"{a}hler} maximal
coset in $D=4$:%
\begin{equation}
\mathbb{CP}^{n}\equiv \frac{SU(1,n)}{U(n)}\overset{c^{\ast }}{%
\longrightarrow }\frac{SU(2,1+n)}{U(1,n)\times SL(2,\mathbb{R})}.
\end{equation}

\paragraph{$T^{3}$ Model}

The specification of (\ref{Ggen}) and (\ref{Ggen-mcs}) to the so-called $%
T^{3}$ model in $D=4$ gives rise to the following symmetric spaces%
\begin{eqnarray}
M_{4,T^{3}}^{4} &\equiv &\frac{G_{4,T^{3}}^{3}}{G_{4,T^{3}}^{4}\times SL(2,%
\mathbb{R})}=\frac{G_{2(2)}}{SL(2,\mathbb{R})\times SL(2,\mathbb{R})_{\text{%
Ehlers}}},~c=nc=4;  \label{M-cosets-5} \\
\widehat{M}_{4,T^{3}}^{4} &\equiv &\frac{mcs\left( G_{4,T^{3}}^{3}\right) }{%
H_{4,T^{3}}^{4}\times SO(2)}=\frac{SU(2)\times SU(2)}{U(1)\times U(1)}.
\end{eqnarray}%
$M_{4,T^{3}}^{4}$ is rank-$2$ \textit{para-quaternionic}. It is nothing but
a suitable pseudo-Riemannian form of the manifold in the r.h.s. of (\ref%
{c-c-2}), namely the $c^{\ast }$-map of the rank-$1$ symmetric \textit{%
special K\"{a}hler} maximal coset in $D=4$:%
\begin{equation}
\left. \frac{SL(2,\mathbb{R})}{U(1)}\right\vert _{T^{3}}\overset{c^{\ast }}{%
\longrightarrow }\frac{G_{2(2)}}{SL(2,\mathbb{R})\times SL(2,\mathbb{R})}.
\end{equation}

\paragraph{$ST^{2}$ Model}

The specification of (\ref{Ggen}) and (\ref{Ggen-mcs}) to the so-called $%
ST^{2}$ model in $D=4$ gives rise to the following symmetric spaces%
\begin{eqnarray}
M_{4,ST^{2}}^{4} &\equiv &\frac{G_{4,ST^{2}}^{3}}{G_{4,ST^{2}}^{4}\times
SL(2,\mathbb{R})_{\text{Ehlers}}}=\frac{SO(4,3)}{SL(2,\mathbb{R})\times SL(2,%
\mathbb{R})\times SL(2,\mathbb{R})_{\text{Ehlers}}},~c=nc=6;
\label{M-cosets-6} \\
\widehat{M}_{4,ST^{2}}^{4} &\equiv &\frac{mcs\left( G_{4,ST^{2}}^{3}\right)
}{H_{4,ST^{2}}^{4}\times SO(2)}=\frac{SO(4)\times SO(3)}{U(1)\times
U(1)\times U(1)}.
\end{eqnarray}

\paragraph{Jordan Symmetric Sequence}

As mentioned above, the $ST^{2}$ model can be regarded as the first element
of the so-called \textit{Jordan symmetric sequence} of \textit{%
quarter-maximal} theories. The specification of (\ref{Ggen}) and (\ref%
{Ggen-mcs}) to such a sequence in $D=6$, $D=5$ and $D=4$ respectively gives
rise to the following spaces:

\begin{itemize}
\item[$D=6:$]
\begin{eqnarray}
M_{4}^{6} &\equiv &\frac{G_{4}^{3}}{G_{4}^{6}\times SL(4,\mathbb{R})}=\frac{%
SO\left( 4,3+n\right) }{SO(1,n)\times SL(4,\mathbb{R})},~c=nc=3n+3;
\label{M-cosets-7} \\
\widehat{M}_{4}^{6} &\equiv &\frac{H_{4}^{3}}{H_{4}^{6}\times SO(4)}=\frac{%
SO(4)\times SO(3+n)}{SO(3)\times SO(n)\times SO(3)};
\end{eqnarray}%
$M_{4}^{6}$ and $\widehat{M}_{4}^{6}$ are maximal and \textit{symmetric}
spaces.

\item[$D=5:$]
\begin{eqnarray}
M_{4}^{5} &\equiv &\frac{G_{4}^{3}}{G_{4}^{5}\times SL(3,\mathbb{R})}=\frac{%
SO\left( 4,3+n\right) }{SO(1,1)\times SO\left( 1,n\right) \times SL(3,%
\mathbb{R})},~c=nc=3n+6; \\
\widehat{M}_{4}^{5} &\equiv &\frac{H_{4}^{3}}{H_{4}^{5}\times SO(3)}=\frac{%
SO(4)\times SO(3+n)}{SO(n)\times SO(3)};
\end{eqnarray}%
$M_{4}^{5}$ and $\widehat{M}_{4}^{5}$ are non-maximal and \textit{%
non-symmetric} spaces.

\item[$D=4:$]
\begin{eqnarray}
M_{4}^{4} &\equiv &\frac{G_{4}^{3}}{G_{4}^{4}\times SL(2,\mathbb{R})_{\text{%
Ehlers}}}=\frac{SO\left( 4,2+n\right) }{SL(2,\mathbb{R})\times SO(2,n)\times
SL(2,\mathbb{R})_{\text{Ehlers}}},~c=nc=2n+4; \\
\widehat{M}_{4}^{4} &\equiv &\frac{mcs\left( G_{4}^{3}\right) }{%
H_{4}^{4}\times SO(2)}=\frac{SO(4)\times SO(2+n)}{U(1)\times U(1)\times
SO(n)\times U(1)}.
\end{eqnarray}%
$M_{4}^{4}$ and $\widehat{M}_{4}^{4}$ are maximal and \textit{symmetric }%
spaces. $M_{4}^{4}$ is \textit{para-quaternionic} and it has rank $2$ in the
case $n=0$ and rank $3$ for $n\geqslant 1$; it is nothing but a suitable
pseudo-Riemannian form of the manifold in the r.h.s. of (\ref{c-map-1}),
namely the $c^{\ast }$-map of the symmetric \textit{special K\"{a}hler}
maximal coset in $D=4$:%
\begin{equation}
\frac{SL\left( 2,\mathbb{R}\right) }{U(1)}\times \frac{SO(2,n)}{SO(2)\times
SO(n)}\overset{c^{\ast }}{\longrightarrow }\frac{SO\left( 4,2+n\right) }{%
SL(2,\mathbb{R})\times SO(2,n)\times SL(2,\mathbb{R})_{\text{Ehlers}}}.
\end{equation}
\end{itemize}

\paragraph{\textit{Magical} Models}

\begin{itemize}
\item[$D=4:$] The specification of (\ref{Ggen}) and (\ref{Ggen-mcs}) to
\textit{magical} models in $D=4$ gives rise to maximal symmetric spaces.
Their general structure reads
\begin{eqnarray}
M_{4}^{4}\left( q\right) &\equiv &\frac{G_{4}^{3}\left( q\right) }{%
G_{4}^{4}\left( q\right) \times SL(2,\mathbb{R})_{\text{Ehlers}}};
\label{one-1} \\
\widehat{M}_{4}^{4}\left( q\right) &\equiv &\frac{H_{4}^{3}\left( q\right) }{%
H_{4}^{4}\left( q\right) \times SO(2)},  \label{two-2}
\end{eqnarray}%
listed in Table 13. The number of compact and non-compact generators of $%
M_{4}^{4}\left( q\right) $ can be $q$-parametrized as follows:%
\begin{equation}
c\left( M_{4}^{4}\left( q\right) \right) =nc\left( M_{4}^{4}\left( q\right)
\right) =6q+8=\text{dim}_{\mathbb{R}}\left( \mathbf{R}\left( G_{4}^{4}\left(
q\right) \right) \right) ,  \label{one-1-1}
\end{equation}%
where $\mathbf{R}$ is the symplectic irrep. of the $D=4$ $U$-duality group $%
G_{4}^{4}\left( q\right) $ in which the Abelian two-form field strengths
sit; see Subsec. \ref{PD} for further analysis. Thus, the split of the
generators of $M_{4}^{4}\left( q\right) $ into a signature $\left(
nc,c=nc\right) $ is consistent with the \textit{Ehlers-doublet} irrep. $%
\left( \mathbf{R},\mathbf{2}\right) $ of $G_{4}^{4}\left( q\right) \times
SL(2,\mathbb{R})_{\text{Ehlers}}$. Moreover, $M_{4}^{4}\left( q\right) $ is
a rank-$4$ \textit{pseudo-quaternionic} space, given by the $c^{\ast }$-map
of the corresponding symmetric \textit{special K\"{a}hler} maximal coset in $%
D=4$:%
\begin{equation}
\frac{G_{4}^{4}\left( q\right) }{mcs\left( G_{4}^{4}\left( q\right) \right) }%
\overset{c^{\ast }}{\longrightarrow }\frac{G_{4}^{3}\left( q\right) }{%
G_{4}^{4}\left( q\right) \times SL(2,\mathbb{R})}.
\end{equation}

\item[$D=5:$] The specification of (\ref{Ggen}) and (\ref{Ggen-mcs}) to
\textit{magical} models in $D=5$ gives rise to the maximal, \textit{%
non-symmetric} spaces listed in Table 14. Their general structure reads
\begin{eqnarray}
M_{4}^{5}\left( q\right) &\equiv &\frac{G_{4}^{3}\left( q\right) }{%
G_{4}^{5}\left( q\right) \times SL(3,\mathbb{R})};  \label{two-3} \\
\widehat{M}_{4}^{5}\left( q\right) &\equiv &\frac{H_{4}^{3}\left( q\right) }{%
H_{4}^{5}\left( q\right) \times SO(3)}.  \label{three-3}
\end{eqnarray}%
The number of compact and non-compact generators of $M_{4}^{5}\left(
q\right) $ can be $q$-parametrized as follows:%
\begin{equation}
c\left( M_{4}^{5}\left( q\right) \right) =nc\left( M_{4}^{5}\left( q\right)
\right) =9\left( q+1\right) =\text{dim}_{\mathbb{R}}\left( \mathcal{R},%
\mathbf{3}\right) ,  \label{two-3-3}
\end{equation}%
where $\left( \mathcal{R},\mathbf{3}\right) $ is the irrep. of $%
G_{4}^{5}\left( q\right) \times SL(3,\mathbb{R})_{\text{Ehlers}}$. Thus, the
split of the generators of $M_{4}^{5}\left( q\right) $ into a signature $%
\left( nc,c=nc\right) $ is consistent with a pair of \textit{Jordan-triplet}
irreps. $\left( \mathcal{R},\mathbf{3}\right) $ (see Subsec. \ref{PD} for
further analysis).

\item[$D=6:$] The specification of (\ref{Ggen}) and (\ref{Ggen-mcs}) to
\textit{magical }models in $D=6$ respectively gives rise to the non-maximal,
\textit{non-symmetric} spaces listed in Table 15\footnote{%
Note that the results on $c=nc$ for $q=8$ (magical\textit{\ exceptional}
supergravity) in $D=4,5,6$ match the results holding for maximal
supergravity in the same dimensions. This is not surprising, because maximal
($N=16$) and exceptional ($N=4$) theories are respectively related to $%
J_{3}^{\mathbb{O}_{s}}$ and $J_{3}^{\mathbb{O}}$, the unique difference
given by the split \textit{vs.} division form of the octonionic algebra $%
\mathbb{O}$.}. Their general structure reads
\begin{eqnarray}
M_{4}^{6}\left( q\right) &\equiv &\frac{G_{4}^{3}\left( q\right) }{%
G_{4}^{6}\left( q\right) \times SL(4,\mathbb{R})};  \label{four-4} \\
\widehat{M}_{4}^{6}\left( q\right) &\equiv &\frac{H_{4}^{3}\left( q\right) }{%
H_{4}^{6}\left( q\right) \times SO(4)},  \label{five-5}
\end{eqnarray}%
where the $U$-duality group $G_{4}^{6}\left( q\right) $ in $D=6$ reads $%
SO(1,q+1)\times \mathcal{A}_{q}$. The number of compact and non-compact
generators of $M_{4}^{6}\left( q\right) $ can be $q$-parametrized as follows:%
\begin{equation}
c\left( M_{4}^{6}\left( q\right) \right) =nc\left( M_{4}^{6}\left( q\right)
\right) =11q+6.  \label{four-4-4}
\end{equation}%
The meaning of $11q+6$ and the covariant split in terms of irreps. of $%
SO(1,q+1)\times mcs\left( \mathcal{A}_{q}\right) \times SO(4)$ will be
discussed in Subsec. \ref{PD}.
\end{itemize}

\begin{table}[t!]
\begin{center}
\begin{tabular}{|c||c|c|l|}
\hline
$\mathfrak{M}\left( J_{3}^{\mathbb{A}}\right) $ & $%
\begin{array}{c}
\\
M_{4}^{4}\left( q\right) \\
~~%
\end{array}%
$ & $%
\begin{array}{c}
\\
\widehat{M}_{4}^{4}\left( q\right) \\
~~%
\end{array}%
$ & $%
\begin{array}{l}
c\left( M_{4}^{4}\left( q\right) \right) = \\
nc\left( M_{4}^{4}\left( q\right) \right) ~~%
\end{array}%
$ \\ \hline\hline
$%
\begin{array}{c}
\\
\mathfrak{M}\left( J_{3}^{\mathbb{O}}\right) ~(q=8) \\
~%
\end{array}%
$ & $\frac{E_{8\left( -24\right) }}{E_{7(-25)}\times SL\left( 2,\mathbb{R}%
\right) }$ & $\frac{E_{7\left( -133\right) }\times SU(2)}{E_{6(-78)}\times
U(1)\times SO\left( 2\right) }$ & \multicolumn{1}{|c|}{$56$} \\ \hline
$%
\begin{array}{c}
\\
\mathfrak{M}\left( J_{3}^{\mathbb{H}}\right) ~(q=4) \\
~%
\end{array}%
$ & $\frac{E_{7(-5)}}{SO^{\ast }\left( 12\right) \times SL\left( 2,\mathbb{R}%
\right) }$ & $\frac{SO(12)\times SU(2)}{SU(6)\times U(1)\times SO(2)}$ &
\multicolumn{1}{|c|}{$32$} \\ \hline
$%
\begin{array}{c}
\\
\mathfrak{M}\left( J_{3}^{\mathbb{C}}\right) ~(q=2) \\
~%
\end{array}%
$ & $\frac{E_{6(2)}}{SU\left( 3,3\right) \times SL\left( 2,\mathbb{R}\right)
}$ & $\frac{SU(6)\times SU(2)}{S\left( U\left( 3\right) \times U(3)\right)
\times SO(2)}$ & \multicolumn{1}{|c|}{$20$} \\ \hline
$%
\begin{array}{c}
\\
\mathfrak{M}\left( J_{3}^{\mathbb{R}}\right) ~(q=1) \\
~%
\end{array}%
$ & $\frac{F_{4(4)}}{Sp\left( 6,\mathbb{R}\right) \times SL\left( 2,\mathbb{R%
}\right) }$ & $\frac{USp(6)\times SU(2)}{SU(3)\times U(1)\times SO(2)}$ &
\multicolumn{1}{|c|}{$14$} \\ \hline
$%
\begin{array}{c}
\\
\mathfrak{M}\left( \mathbb{R}\right) ~(q=-2/3) \\
~%
\end{array}%
$ & $\frac{G_{2(2)}}{SL\left( 2,\mathbb{R}\right) \times SL(2,\mathbb{R})}$
& $\frac{SU(2)\times SU(2)}{U(1)\times SO(2)_{J}}$ & \multicolumn{1}{|c|}{$4$%
} \\ \hline
\end{tabular}%
\end{center}
\caption{Pseudo-Riemannian, non-compact, maximal, \textit{para-quaternionic}
\textit{symmetric} cosets $M_{4}^{4}\left( q\right) $ (\protect\ref{one-1})
and Riemannian, compact, maximal cosets $\protect\widehat{M}_{4}^{4}\left(
q\right) $ (\protect\ref{two-2}) of \textit{magic quarter-maximal}
supergravity theories ($N=4$) in $D=4$ Lorentzian space-time dimensions.
Also the $T^{3}$ model ($q=-2/3$) is reported. The number of compact
generators $c$ (equal to the number $nc$ of non-compact generators) of $%
M_{4}^{4}\left( q\right) $ is also listed. All cosets $M_{4}^{4}\left(
q\right) $ have \textit{vanishing character}.}
\end{table}

\begin{table}[t!]
\begin{center}
\begin{tabular}{|c||c|c|l|}
\hline
$J_{3}^{\mathbb{A}}$ & $%
\begin{array}{c}
\\
M_{4}^{5}\left( q\right) \\
~~%
\end{array}%
$ & $%
\begin{array}{c}
\\
\widehat{M}_{4}^{5}\left( q\right) \\
~~%
\end{array}%
$ & $%
\begin{array}{l}
c\left( M_{4}^{5}\left( q\right) \right) = \\
nc\left( M_{4}^{5}\left( q\right) \right) ~~%
\end{array}%
$ \\ \hline\hline
$%
\begin{array}{c}
\\
J_{3}^{\mathbb{O}}~(q=8) \\
~%
\end{array}%
$ & $\frac{E_{8\left( -24\right) }}{E_{6(-26)}\times SL\left( 3,\mathbb{R}%
\right) }$ & $\frac{E_{7\left( -133\right) }\times SU(2)}{F_{4(-52)}\times
SU(2)\times SO\left( 3\right) _{J}}$ & \multicolumn{1}{|c|}{$81$} \\ \hline
$%
\begin{array}{c}
\\
J_{3}^{\mathbb{H}}~(q=4) \\
~%
\end{array}%
$ & $\frac{E_{7(-5)}}{SU^{\ast }\left( 6\right) \times SL\left( 3,\mathbb{R}%
\right) }$ & $\frac{SO(12)\times SU(2)}{USp(6)\times SO\left( 3\right) }$ &
\multicolumn{1}{|c|}{$45$} \\ \hline
$%
\begin{array}{c}
\\
J_{3}^{\mathbb{C}}~(q=2) \\
~%
\end{array}%
$ & $\frac{E_{6(2)}}{SL\left( 3,\mathbb{C}\right) \times SL\left( 3,\mathbb{R%
}\right) }$ & $\frac{SU(6)\times SU(2)}{SU\left( 3\right) \times SO\left(
3\right) }$ & \multicolumn{1}{|c|}{$27$} \\ \hline
$%
\begin{array}{c}
\\
J_{3}^{\mathbb{R}}~(q=1) \\
~%
\end{array}%
$ & $\frac{F_{4(4)}}{SL\left( 3,\mathbb{R}\right) \times SL\left( 3,\mathbb{R%
}\right) }$ & $\frac{USp(6)\times SU(2)}{SU(2)_{P}\times SO\left( 3\right)
_{J}}$ & \multicolumn{1}{|c|}{$18$} \\ \hline
$%
\begin{array}{c}
\\
\mathbb{R}~(q=-2/3) \\
~%
\end{array}%
$ & $G_{2(2)}\supset SL(3,\mathbb{R})$ & $\frac{SU(2)\times SU(2)}{%
SO(3)_{J,D}}$ & \multicolumn{1}{|c|}{$3$} \\ \hline
\end{tabular}%
\end{center}
\caption{Pseudo-Riemannian, non-compact, maximal, \textit{non-symmetric}
cosets $M_{4}^{5}\left( q\right) $ (\protect\ref{two-3}) and Riemannian,
compact, maximal cosets $\protect\widehat{M}_{4}^{5}\left( q\right) $ (%
\protect\ref{three-3}) of \textit{magic quarter-maximal} supergravity
theories ($N=4$) in $D=5$ Lorentzian space-time dimensions. Also the $D=5$
uplift of $T^{3}$ model ($q=-2/3$), namely \textit{minimal \textquotedblleft
pure"} supergravity, is reported. The number of compact generators $c$
(equal to the number $nc$ of non-compact generators) of $M_{4}^{5}\left(
q\right) $ is also listed. All cosets $M_{4}^{5}\left( q\right) $ have
\textit{vanishing character}.}
\end{table}

\begin{table}[t!]
\begin{center}
\begin{tabular}{|c||c|c|l|}
\hline
$J_{2}^{\mathbb{A}}$ & $%
\begin{array}{c}
\\
M_{4}^{6}\left( q\right) \\
~~%
\end{array}%
$ & $%
\begin{array}{c}
\\
\widehat{M}_{4}^{6}\left( q\right) \\
~~%
\end{array}%
$ & $%
\begin{array}{l}
c\left( M_{4}^{6}\left( q\right) \right) = \\
nc\left( M_{4}^{6}\left( q\right) \right) ~~%
\end{array}%
$ \\ \hline\hline
$%
\begin{array}{c}
\\
J_{2}^{\mathbb{O}}~(q=8) \\
~%
\end{array}%
$ & $\frac{E_{8\left( -24\right) }}{SO(1,9)\times SL\left( 4,\mathbb{R}%
\right) }$ & $\frac{E_{7\left( -133\right) }\times SU(2)}{SO(9)\times
SO\left( 4\right) _{J}}$ & \multicolumn{1}{|c|}{$94$} \\ \hline
$%
\begin{array}{c}
\\
J_{2}^{\mathbb{H}}~(q=4) \\
~%
\end{array}%
$ & $\frac{E_{7(-5)}}{SO\left( 1,5\right) \times SO(3)\times SL\left( 4,%
\mathbb{R}\right) }$ & $\frac{SO(12)\times SU(2)}{SO\left( 5\right) \times
SO(3)\times SO\left( 4\right) _{J}}$ & \multicolumn{1}{|c|}{$50$} \\ \hline
$%
\begin{array}{c}
\\
J_{2}^{\mathbb{C}}~(q=2) \\
~%
\end{array}%
$ & $\frac{E_{6(2)}}{SO\left( 1,3\right) \times SO(2)\times SL\left( 4,%
\mathbb{R}\right) }$ & $\frac{SU(6)\times SU(2)}{SO\left( 3\right) \times
SO(2)\times SO\left( 4\right) _{J}}$ & \multicolumn{1}{|c|}{$28$} \\ \hline
$%
\begin{array}{c}
\\
J_{2}^{\mathbb{R}}~(q=1) \\
~%
\end{array}%
$ & $\frac{F_{4(4)}}{SO\left( 1,2\right) \times SL\left( 4,\mathbb{R}\right)
}$ & $\frac{USp(6)\times SU(2)}{SO\left( 2\right) \times SO\left( 4\right)
_{J}}$ & \multicolumn{1}{|c|}{$17$} \\ \hline
\end{tabular}%
\end{center}
\caption{Pseudo-Riemannian, non-compact, non-maximal, \textit{non-symmetric}
cosets $M_{4}^{6}\left( q\right) $ (\protect\ref{four-4}) and Riemannian,
compact, non-maximal, \textit{non-symmetric} cosets $\protect\widehat{M}%
_{4}^{6}\left( q\right) $ (\protect\ref{five-5}) of \textit{magic }$\left(
1,0\right) $ chiral supergravity theories ($N=4$) in $D=6$ Lorentzian
space-time dimensions. The number of compact generators $c$ (equal to the
number $nc$ of non-compact generators) of $M_{4}^{6}\left( q\right) $ is
also listed. All cosets $M_{4}^{6}\left( q\right) $ have \textit{vanishing
character}.}
\end{table}

\subsection{\label{PD}\textit{Poincar\'{e} Duality}}

We are now going to analyze the signature split of the manifolds $M_{N}^{D}$
(\ref{Ggen}), focussing on the maximal ($N=32$) and magical quarter-maximal
cases ($N=8$).

Nicely, the split signature of $M_{N}^{D}$ covariantly decomposes under $%
mcs\left( G_{N}^{D}\right) \times SO(D-2)_{J}$ into a pair of sets of
irreps., which are related by \textit{Poincar\'{e} duality} (\textit{alias}
eletric-magnetic duality). In other words, the signature of the
pseudo-Riemannian manifolds $M_{N}^{D}$'s naturally arrange the spectrum of $%
p>0$ massless forms of the corresponding supergravity theory into a pair of
sets of irreps. of $mcs\left( G_{N}^{D}\right) \times SO(D-2)_{J}$, which
are interchanged under \textit{Poincar\'{e} duality}.

As a consequence, the $\chi =0$ feature of the manifolds $M_{N}^{D}$ (\ref%
{Ggen}) is actually \textit{Poincar\'{e}-duality-invariant} (or,
equivalently, \textit{electric-magnetic duality-invariant}).

\subsubsection{$N=16$}

\begin{enumerate}
\item $D=11$ ($M$-theory) : the relevant manifold is maximal non-symmetric:
\begin{equation}
M_{16}^{11}=\frac{E_{8(8)}}{SL\left( 9,\mathbb{R}\right) }:\left[
\begin{array}{lll}
& c & nc \\
E_{8(8)}: & 120 & 128 \\
SL(9,\mathbb{R}): & 36 & 44 \\
M_{16}^{11}: & 84 & 84%
\end{array}%
\right] .
\end{equation}%
Such a signature splitting is covariant with respect to $SO\left( 9\right)
=mcs\left( SL\left( 9,\mathbb{R}\right) \right) $:%
\begin{eqnarray}
&&%
\begin{array}{l}
E_{8(8)}\supset _{ns}SL(9,\mathbb{R}); \\
\mathbf{248}=\mathbf{80}+\mathbf{84+84}^{\prime };%
\end{array}
\\
&&%
\begin{array}{l}
Sl(9,\mathbb{R})\overset{mcs}{\supset }SO(9); \\
\mathbf{84}^{(\prime )}=\mathbf{84}.%
\end{array}%
\end{eqnarray}%
Therefore, the split $\left( c,nc\right) =\left( 84,84\right) $ can be
interpreted as the split into two Poincar\'{e}-dual $\mathbf{84}$'s of $%
SO(9) $; namely, the $3$-form potential (coupled to $M2$-brane) and its
\textit{Poincar\'{e} dual} $6$-form potential (coupled to $M5$-brane):%
\begin{equation}
(c,nc)=\left( 84,84\right) =\underset{M2}{\mathbf{84}}+\underset{M5}{\mathbf{%
84}}\text{~of~}SO(9)\text{.}
\end{equation}

\item $D=10$ IIA : the relevant manifold is non-maximal and non-symmetric:
\begin{equation}
M_{16}^{10~IIA}=\frac{E_{8(8)}}{SO(1,1)\times SL\left( 8,\mathbb{R}\right) }:%
\left[
\begin{array}{lll}
& c & nc \\
E_{8(8)}: & 120 & 128 \\
SO(1,1)\times SL(8,\mathbb{R}): & 28 & 36 \\
M_{16}^{10~IIA}: & 92 & 92%
\end{array}%
\right] .
\end{equation}%
Such a signature splitting is covariant with respect to $SO\left( 8\right)
=mcs\left( SO(1,1)\times SL\left( 8,\mathbb{R}\right) \right) $. Indeed,
disregarding $SO(1,1)$ weights, it holds that:%
\begin{eqnarray}
&&%
\begin{array}{l}
E_{8(8)}\supset _{nm}SO(1,1)\times SL(8,\mathbb{R}); \\
\mathbf{248}=\mathbf{63}+\mathbf{1+8}+\mathbf{8}^{\prime }+\mathbf{28}+%
\mathbf{28}^{\prime }+\mathbf{56}+\mathbf{56}^{\prime };%
\end{array}
\\
&&%
\begin{array}{l}
Sl(8,\mathbb{R})\overset{mcs}{\supset }SO(8); \\
\mathbf{8}^{(\prime )},\mathbf{28}^{(\prime )},\mathbf{56}^{(\prime )}=%
\mathbf{8}_{v}\mathbf{,28,56}_{v}.%
\end{array}%
\end{eqnarray}%
Therefore, the split $\left( c,nc\right) =\left( 92,92\right) $ can be
interpreted as the split into two sets of Poincar\'{e}-dual irreps. of $%
SO(8) $; namely, the graviphoton $C_{\mu }^{(1)}$ $\mathbf{8}_{v}$, the $2$%
-form $B_{\mu \nu }$ $\mathbf{28}$, the $3$-form $C_{\mu \nu \rho }^{(3)}$ $%
\mathbf{56}_{v}$ potentials, and their \textit{Poincar\'{e} duals}, namely
the $7$-form $\widetilde{C}_{\mu _{1}...\mu _{7}}$, $6$-form $\widetilde{B}%
_{\mu _{1}...\mu _{6}}$ and $5$-form $\widetilde{C}_{\mu _{1}...\mu _{5}}$
potentials:%
\begin{equation}
(c,nc)=\left( 92,92\right) =\left( \underset{C^{(1)}}{\mathbf{8}_{v}}+%
\underset{B^{(2)}}{\mathbf{28}}+\underset{C^{(3)}}{\mathbf{56}_{v}}\right)
+\left( \underset{\widetilde{C}^{(7)}}{\mathbf{8}_{v}}+\underset{\widetilde{B%
}^{(6)}}{\mathbf{28}}+\underset{\widetilde{C}^{(5)}}{\mathbf{56}_{v}}\right)
\text{~of~}SO(8)\text{.}
\end{equation}

\item $D=10$ IIB : the relevant manifold is non-maximal and non-symmetric:
\begin{equation}
M_{16}^{10~IIB}=\frac{E_{8(8)}}{SL(2,\mathbb{R})\times SL\left( 8,\mathbb{R}%
\right) }:\left[
\begin{array}{lll}
& c & nc \\
E_{8(8)}: & 120 & 128 \\
{\small SL(2,}\mathbb{R}{\small )\times SL(8,}\mathbb{R}{\small )}: & 29 & 37
\\
M_{16}^{10~IIB}: & 91 & 91%
\end{array}%
\right] .
\end{equation}%
Such a signature splitting is covariant with respect to $SO\left( 8\right)
\times SO(2)=mcs\left( SL\left( 8,\mathbb{R}\right) \times SL(2,\mathbb{R}%
)\right) $. Indeed, it holds that:%
\begin{eqnarray}
&&%
\begin{array}{l}
E_{8(8)}\supset _{nm}SL(8,\mathbb{R})\times SL(2,\mathbb{R}); \\
\mathbf{248}=\left( \mathbf{63,1}\right) +\left( \mathbf{1,3}\right) +\left(
\mathbf{70,1}\right) +\left( \mathbf{28,2}\right) +\left( \mathbf{28}%
^{\prime },\mathbf{2}\right) ;%
\end{array}
\\
&&%
\begin{array}{l}
SL(8,\mathbb{R})\times SL(2,\mathbb{R})\overset{mcs}{\supset }SO(8)\times
SO(2); \\
\left( \mathbf{8},\mathbf{1}\right) =\left( \mathbf{8}_{v},\mathbf{1}\right)
\\
\left( \mathbf{28}^{(\prime )},\mathbf{2}\right) =\left( \mathbf{28,2}%
\right) ; \\
\left( \mathbf{70,1}\right) =\left( \mathbf{35}_{s},\mathbf{1}\right)
+\left( \mathbf{35}_{c},\mathbf{1}\right) .%
\end{array}%
\end{eqnarray}%
Therefore, the split $\left( c,nc\right) =\left( 91,91\right) $ can be
interpreted as the split into two sets of Poincar\`{e}-dual irreps. of $%
SO(8)\times SO(2)$; namely, the $2$-form $C_{\mu \nu }^{(2)}$ $\left(
\mathbf{28,2}\right) $ and $4$-form $C_{\mu _{1}...\mu _{4}}^{(4)}$ $\left(
\mathbf{35}_{s},\mathbf{1}\right) $ potentials, and their \textit{Poincar%
\`{e} duals}, namely the $6$-form $\widetilde{C}_{\mu _{1}...\mu _{6}}$ $%
\left( \mathbf{28,2}\right) $ and the $4$-form $C_{\mu _{1}...\mu
_{4}}^{(4)} $ $\left( \mathbf{35}_{c},\mathbf{1}\right) $ potentials:%
\begin{equation}
(c,nc)=\left( 91,91\right) =\left( \underset{C^{(2)}}{\left( \mathbf{28,2}%
\right) }+\underset{C^{(4)}}{\left( \mathbf{35}_{s},\mathbf{1}\right) }%
\right) +\left( \underset{\widetilde{C}^{(6)}}{\left( \mathbf{28,2}\right) }+%
\underset{C^{(4)}}{\left( \mathbf{35}_{c},\mathbf{1}\right) }\right) \text{%
~of~}SO(8)\times SO(2)\text{.}
\end{equation}

\item $D=9$ : the relevant manifold is non-maximal and non-symmetric:
\begin{equation}
M_{16}^{9}=\frac{E_{8(8)}}{GL(2,\mathbb{R})\times SL\left( 7,\mathbb{R}%
\right) }:\left[
\begin{array}{lll}
& c & nc \\
E_{8(8)}: & 120 & 128 \\
{\small GL(2,}\mathbb{R}{\small )\times SL(7,}\mathbb{R}{\small )}: & 22 & 30
\\
M_{16}^{9}: & 98 & 98%
\end{array}%
\right] .
\end{equation}%
Such a signature splitting is covariant with respect to $SO\left( 7\right)
\times SO(2)=mcs\left( SL\left( 7,\mathbb{R}\right) \times GL(2,\mathbb{R}%
)\right) $. Indeed, disregarding $SO(1,1)$ weights, it holds that:%
\begin{eqnarray}
&&%
\begin{array}{l}
E_{8(8)}\supset _{nm}SL(7,\mathbb{R})\times GL(2,\mathbb{R}); \\
\mathbf{248}=\left( \mathbf{48,1}\right) +\left( \mathbf{1,1}\right) +\left(
\mathbf{1,3}\right) \\
+\left( \mathbf{7,1}\right) +\left( \mathbf{7}^{\prime },\mathbf{1}\right)
+\left( \mathbf{7,2}\right) +\left( \mathbf{7}^{\prime },\mathbf{2}\right)
+\left( \mathbf{21},\mathbf{2}\right) +\left( \mathbf{21}^{\prime },\mathbf{2%
}\right) +\left( \mathbf{35},\mathbf{1}\right) +\left( \mathbf{35}^{\prime },%
\mathbf{1}\right) ;%
\end{array}
\\
&&%
\begin{array}{l}
SL(7,\mathbb{R})\overset{mcs}{\supset }SO(7); \\
\left( \mathbf{7}^{(\prime )},\mathbf{21}^{(\prime )},\mathbf{35}\right)
=\left( \mathbf{7,21,35}\right) .%
\end{array}%
\end{eqnarray}%
Therefore, the split $\left( c,nc\right) =\left( 98,98\right) $ can be
interpreted as the split into two sets of Poincar\`{e}-dual irreps. of $%
SO(7)\times SO(2)$; namely, the graviphotons $\left( \mathbf{7,1}\right) $
and $\left( \mathbf{7,2}\right) $, the $2$-form $\left( \mathbf{21},\mathbf{2%
}\right) $ and the $3$-form $\left( \mathbf{35},\mathbf{1}\right) $
potentials, and their \textit{Poincar\`{e} duals}, namely the $6$-forms $%
\left( \mathbf{7,1}\right) $ and $\left( \mathbf{7,2}\right) $ duals of
graviphotons, the $5$-form $\left( \mathbf{21},\mathbf{2}\right) $ and the $%
4 $-form $\left( \mathbf{35},\mathbf{1}\right) $ potentials:%
\begin{equation}
(c,nc)=\left( 98,98\right) =\left\{
\begin{array}{c}
\left( \mathbf{7,1}\right) +\left( \mathbf{7,2}\right) +\left( \mathbf{21},%
\mathbf{2}\right) +\left( \mathbf{35},\mathbf{1}\right) \\
+ \\
\left( \mathbf{7,1}\right) +\left( \mathbf{7,2}\right) +\left( \mathbf{21},%
\mathbf{2}\right) +\left( \mathbf{35},\mathbf{1}\right)%
\end{array}%
\right. \text{~of~}SO(7)\times SO(2)\text{.}
\end{equation}

\item $D=8$ : the relevant manifold is non-maximal and non-symmetric:
\begin{equation}
M_{16}^{8}=\frac{E_{8(8)}}{SL(3,\mathbb{R})\times SL(2,\mathbb{R})\times
SL\left( 6,\mathbb{R}\right) }:\left[
\begin{array}{lll}
& c & nc \\
E_{8(8)}: & 120 & 128 \\
{\small SL(3,}\mathbb{R}{\small )\times SL(2,}\mathbb{R}{\small )\times SL(6,%
}\mathbb{R}{\small )}: & 19 & 27 \\
M_{16}^{8}: & 101 & 101%
\end{array}%
\right] .
\end{equation}%
Such a signature splitting is covariant with respect to%
\begin{equation}
SO\left( 6\right) \times SO(2)\times SO(3)=mcs\left( SL\left( 6,\mathbb{R}%
\right) \times SL(2,\mathbb{R})\times SL(3,\mathbb{R})\right) .
\end{equation}%
Indeed, it holds that:%
\begin{eqnarray}
&&%
\begin{array}{l}
E_{8(8)}\supset _{nm}SL(6,\mathbb{R})\times SL(2,\mathbb{R})\times SL(3,%
\mathbb{R}); \\
\mathbf{248}=\left( \mathbf{35,1,1}\right) +\left( \mathbf{1,3,1}\right)
+\left( \mathbf{1,1,8}\right) \\
+\left( \mathbf{20,2,1}\right) +\left( \mathbf{6}^{\prime },\mathbf{2,3}%
\right) +\left( \mathbf{6},\mathbf{2,3}^{\prime }\right) +\left( \mathbf{15},%
\mathbf{1,3}\right) +\left( \mathbf{15}^{\prime },\mathbf{1,3}^{\prime
}\right) ;%
\end{array}
\\
&&%
\begin{array}{l}
SL(6,\mathbb{R})\times SL(2,\mathbb{R})\times SL(3,\mathbb{R})\overset{mcs}{%
\supset }SO(6)\times SO(2)\times SO(3); \\
\left( \mathbf{6}^{(\prime )},\mathbf{2,3}^{(\prime )}\right) =\left(
\mathbf{6},\mathbf{2,3}\right) ; \\
\left( \mathbf{15}^{(\prime )},\mathbf{1,3}^{(\prime )}\right) =\left(
\mathbf{15},\mathbf{1,3}\right) ; \\
\left( \mathbf{20,2,1}\right) =\left( \mathbf{10,2,1}\right) +\left(
\overline{\mathbf{10}}\mathbf{,2,1}\right) .%
\end{array}%
\end{eqnarray}%
Therefore, the split $\left( c,nc\right) =\left( 101,101\right) $ can be
interpreted as the split into two sets of Poincar\`{e}-dual irreps. of $%
SO(6)\times SO(2)\times SO(3)$; namely, the $1$-form $\left( \mathbf{6},%
\mathbf{2,3}\right) $, the $2$-form $\left( \mathbf{15},\mathbf{1,3}\right) $%
, the $3$-form $\left( \mathbf{10,2,1}\right) $ potentials, and their
\textit{Poincar\`{e} duals}, namely the $5$-form $\left( \mathbf{6},\mathbf{%
2,3}\right) $, the $4$-form $\left( \mathbf{15},\mathbf{1,3}\right) $ and
the $3$-form $\left( \mathbf{10,2,1}\right) $:%
\begin{equation}
(c,nc)=\left( 101,101\right) =\left\{
\begin{array}{c}
\left( \mathbf{6},\mathbf{2,3}\right) +\left( \mathbf{15},\mathbf{1,3}%
\right) +\left( \mathbf{21},\mathbf{2}\right) +\left( \mathbf{10,2,1}\right)
\\
+ \\
\left( \mathbf{6},\mathbf{2,3}\right) +\left( \mathbf{15},\mathbf{1,3}%
\right) +\left( \mathbf{21},\mathbf{2}\right) +\left( \mathbf{10,2,1}\right)%
\end{array}%
\right. \text{~of~}SO(6)\times SO(3)\times SO(2)\text{.}
\end{equation}

\item $D=7$ : the relevant manifold is maximal and non-symmetric:
\begin{equation}
M_{16}^{7}=\frac{E_{8(8)}}{SL(5,\mathbb{R})\times SL(5,\mathbb{R})}:\left[
\begin{array}{lll}
& c & nc \\
E_{8(8)}: & 120 & 128 \\
{\small SL(5,}\mathbb{R}{\small )\times SL(5,}\mathbb{R}{\small )}: & 20 & 28
\\
M_{16}^{7}: & 100 & 100%
\end{array}%
\right] .
\end{equation}%
Such a signature splitting is covariant with respect to $SO\left( 5\right)
\times SO(5)=mcs\left( SL\left( 5,\mathbb{R}\right) \times SL(5,\mathbb{R}%
)\right) $. Indeed, it holds that:%
\begin{eqnarray}
&&%
\begin{array}{l}
E_{8(8)}\supset _{ns}SL(5,\mathbb{R})\times SL(5,\mathbb{R}); \\
\mathbf{248}=\left( \mathbf{24,1}\right) +\left( \mathbf{1,24}\right)
+\left( \mathbf{10,5}\right) +\left( \mathbf{10}^{\prime }\mathbf{,5}%
^{\prime }\right) +\left( \mathbf{5},\mathbf{10}^{\prime }\right) +\left(
\mathbf{5}^{\prime },\mathbf{10}\right) ;%
\end{array}
\\
&&%
\begin{array}{l}
SL(5,\mathbb{R})\times SL(5,\mathbb{R})\overset{mcs}{\supset }SO(5)\times
SO(5); \\
\left( \mathbf{10}^{(\prime )}\mathbf{,5}^{(\prime )}\right) =\left( \mathbf{%
10,5}\right) .%
\end{array}%
\end{eqnarray}%
Therefore, the split $\left( c,nc\right) =\left( 100,100\right) $ can be
interpreted as the split into two sets of Poincar\`{e}-dual irreps. of $%
SO(5)\times SO(5)\sim USp(4)\times USp(4)$; namely, the $1$-form $\left(
\mathbf{10,5}\right) $ and the $2$-form $\left( \mathbf{5},\mathbf{10}%
\right) $ potentials, and their \textit{Poincar\`{e} duals}, namely the $4$%
-form $\left( \mathbf{10,5}\right) $ and the $3$-form $\left( \mathbf{5},%
\mathbf{10}\right) $ potentials:%
\begin{equation}
(c,nc)=\left( 100,100\right) =\left( \left( \mathbf{10,5}\right) +\left(
\mathbf{5},\mathbf{10}\right) \right) +\left( \left( \mathbf{10,5}\right)
+\left( \mathbf{5},\mathbf{10}\right) \right) \text{~of~}SO(5)\times SO(5)%
\text{.}
\end{equation}

\item $D=6$ (non chiral $\left( 2,2\right) $): the relevant manifold is
non-maximal and non-symmetric:
\begin{equation}
M_{16}^{6}=\frac{E_{8(8)}}{SO(5,5)\times SL(4,\mathbb{R})}:\left[
\begin{array}{lll}
& c & nc \\
E_{8(8)}: & 120 & 128 \\
SO(5,5)\times SL(4,\mathbb{R}): & 26 & 34 \\
M_{16}^{6}: & 94 & 94%
\end{array}%
\right] .
\end{equation}%
Such a signature splitting is covariant with respect to%
\begin{equation}
mcs\left( SL\left( 5,\mathbb{R}\right) \times SL(5,\mathbb{R})\right)
=SO\left( 5\right) \times SO(5)\times SO(3)\times SO(3)\sim USp(4)_{L}\times
USp(4)_{R}\times SU(2)\times SU(2).
\end{equation}%
Indeed, it holds that:%
\begin{eqnarray}
&&%
\begin{array}{l}
E_{8(8)}\supset _{ns}SO(5,5)\times SL(4,\mathbb{R}); \\
\mathbf{248}=\left( \mathbf{45,1}\right) +\left( \mathbf{1,15}\right)
+\left( \mathbf{10,6}\right) +\left( \mathbf{16,4}\right) +\left( \mathbf{16}%
^{\prime },\mathbf{4}^{\prime }\right) ;%
\end{array}
\\
&&%
\begin{array}{l}
SO(5,5)\times SL(4,\mathbb{R})\overset{mcs}{\supset }USp(4)_{L}\times
USp(4)_{R}\times SU(2)\times SU(2); \\
\left( \mathbf{10,6}\right) =\left( \mathbf{1,5,1,3}\right) +\left( \mathbf{%
1,5,3,1}\right) +\left( \mathbf{5,1,1,3}\right) +\left( \mathbf{5,1,3,1}%
\right) ; \\
\left( \mathbf{16}^{(\prime )}\mathbf{,4}^{(\prime )}\right) =\left( \mathbf{%
4},\mathbf{4},\mathbf{2},\mathbf{2}\right) .%
\end{array}%
\end{eqnarray}%
Therefore, the split $\left( c,nc\right) =\left( 94,94\right) $ can be
interpreted as the split into two sets of Poincar\`{e}-dual irreps. of $%
USp(4)_{L}\times USp(4)_{R}\times SU(2)\times SU(2)$; namely, the $5$
self-dual $2$-forms $B_{\mu \nu \mid R}^{+}$ $\left( \mathbf{1,5,1,3}\right)
$, the $5$ anti-self-dual $2$-forms $B_{\mu \nu \mid L}^{-}$ $\left( \mathbf{%
5,1,3,1}\right) $ and the $16$ $1$-forms $A_{\mu }^{\alpha \dot{\alpha}}$ $%
\left( \mathbf{4,4},\mathbf{2,2}\right) $ potentials, and their \textit{%
Poincar\`{e} duals}, namely the $5$ anti-self-dual $2$-forms $B_{\mu \nu
\mid R}^{-}$ $\left( \mathbf{1,5,3,1}\right) $, the $5$ self-dual $2$-form $%
B_{\mu \nu \mid L}^{+}$ $\left( \mathbf{5,1,1,3}\right) $ and the $16$ $3$%
-form $\widetilde{A}_{\mu _{1}...\mu _{4}}^{\alpha \dot{\alpha}}$ $\left(
\mathbf{4,4},\mathbf{2,2}\right) $ potentials:%
\begin{equation}
(c,nc)=\left( 94,94\right) =\left\{
\begin{array}{c}
\left( \mathbf{1,5,1,3}\right) +\left( \mathbf{5,1,3,1}\right) +\left(
\mathbf{4,4},\mathbf{2,2}\right) \\
+ \\
\left( \mathbf{1,5,3,1}\right) +\left( \mathbf{5,1,1,3}\right) +\left(
\mathbf{4,4},\mathbf{2,2}\right)%
\end{array}%
\right. \text{~of~}USp(4)_{L}\times USp(4)_{R}\times SU(2)\times SU(2)\text{.%
}
\end{equation}

\item $D=5$ : the relevant manifold is maximal and non-symmetric:
\begin{equation}
M_{16}^{5}=\frac{E_{8(8)}}{E_{6(6)}\times SL(3,\mathbb{R})}:\left[
\begin{array}{lll}
& c & nc \\
E_{8(8)}: & 120 & 128 \\
E_{6(6)}\times SL(3,\mathbb{R}): & 39 & 47 \\
M_{16}^{5}: & 81 & 81%
\end{array}%
\right] .
\end{equation}%
Such a signature splitting is covariant with respect to $USp(8)\times
SO(3)=mcs\left( E_{6(6)}\times SL(3,\mathbb{R})\right) $. Indeed, it holds
that:%
\begin{eqnarray}
&&%
\begin{array}{l}
E_{8(8)}\supset _{ns}SL(3,\mathbb{R})\times E_{6(6)}; \\
\mathbf{248}=\left( \mathbf{8,1}\right) +\left( \mathbf{1,78}\right) +\left(
\mathbf{3,27}\right) +\left( \mathbf{3}^{\prime }\mathbf{,27}^{\prime
}\right) ;%
\end{array}
\\
&&%
\begin{array}{l}
SL(3,\mathbb{R})\times E_{6(6)}\overset{mcs}{\supset }SO(3)\times USp(8); \\
\left( \mathbf{3}^{(\prime )}\mathbf{,27}^{(\prime )}\right) =\left( \mathbf{%
3,27}\right) .%
\end{array}%
\end{eqnarray}%
Therefore, the split $\left( c,nc\right) =\left( 81,81\right) $, which is
related to the so-called \textit{Jordan pairs} (see \textit{e.g.} \cite%
{Truini-1}), can be interpreted as the split into two sets of Poincar\`{e}%
-dual irreps. of $SO(3)\times USp(8)$; namely, the $27$ graviphotons $A_{\mu
}$ $\left( \mathbf{3,27}\right) $, and their \textit{Poincar\`{e} duals},
namely the $27$ $2$-forms $\widetilde{A}_{\mu \nu }$ $\left( \mathbf{3,27}%
\right) $:%
\begin{equation}
(c,nc)=\left( 81,81\right) =\left( \mathbf{3,27}\right) +\left( \mathbf{3,27}%
\right) \text{~of~}SO(3)\times USp(8)\text{.}
\end{equation}%
Note that the $\mathbf{3}$ of the massless spin group $SO(3)\equiv SO(3)_{J}$
corresponds to the three physical polarizations of the graviphotons in $D=5$.

\item $D=4$ : the relevant manifold is para-quaternionic, maximal and
symmetric:
\begin{equation}
M_{16}^{4}=\frac{E_{8(8)}}{E_{7(7)}\times SL(2,\mathbb{R})}:\left[
\begin{array}{lll}
& c & nc \\
E_{8(8)}: & 120 & 128 \\
E_{7(7)}\times SL(2,\mathbb{R}): & 64 & 72 \\
M_{16}^{4}: & 56 & 56%
\end{array}%
\right] .
\end{equation}%
Such a signature splitting is covariant with respect to $SU(8)\times
SO(2)_{J}=mcs\left( E_{7(7)}\times SL(2,\mathbb{R})\right) $. Indeed, it
holds that:%
\begin{eqnarray}
&&%
\begin{array}{l}
E_{8(8)}\supset _{ns}SL(2,\mathbb{R})\times E_{7(7)}; \\
\mathbf{248}=\left( \mathbf{3,1}\right) +\left( \mathbf{1,133}\right)
+\left( \mathbf{2,56}\right) ;%
\end{array}
\\
&&%
\begin{array}{l}
SL(2,\mathbb{R})\times E_{7(7)}\overset{mcs}{\supset }SO(2)_{J}\times SU(8);
\\
\left( \mathbf{2,56}\right) =\left( \mathbf{2,28}\right) +\left( \mathbf{2,}%
\overline{\mathbf{28}}\right) .%
\end{array}%
\end{eqnarray}%
Therefore, the split $\left( c,nc\right) =\left( 56,56\right) $, which
corresponds to a pair of Freudenthal systems $\mathfrak{M}\left( J_{3}^{%
\mathbb{O}_{s}}\right) $, can be interpreted as the split into two sets of
Poincar\`{e}-dual irreps. of $SO(2)_{J}\times SU(8)$; namely, the $28$
graviphotons $A_{\mu }$ $\left( \mathbf{2,28}\right) $, and their \textit{%
Poincar\`{e}-Hodge duals}, namely the $28$ graviphotons $\widetilde{A}_{\mu
} $ $\left( \mathbf{2,}\overline{\mathbf{28}}\right) $:%
\begin{equation}
(c,nc)=\left( 56,56\right) =\left( \mathbf{2,28}\right) +\left( \mathbf{2,}%
\overline{\mathbf{28}}\right) \text{~of~}SO(2)_{J}\times SU(8)\text{.}
\end{equation}%
Note that the $\mathbf{2}$ of the massless spin group $SO(2)_{J}$
corresponds to the two physical polarizations of the graviphotons in $D=4$.
\end{enumerate}

\subsubsection{$N=4$ \textit{Magical} Models}

\begin{itemize}
\item[$D=4:$] the relevant manifold is para-quaternionic, maximal and
symmetric (recall (\ref{one-1}) and (\ref{one-1-1})):
\begin{equation}
M_{4}^{4}\left( q\right) \equiv \frac{G_{4}^{3}\left( q\right) }{%
G_{4}^{4}\left( q\right) \times SL(2,\mathbb{R})_{\text{Ehlers}}}:\left(
c,nc\right) =\left( 6q+8,6q+8\right) .
\end{equation}%
Such a signature splitting is covariant with respect to $mcs\left(
G_{4}^{4}\left( q\right) \right) \times SO(2)_{J}$. Indeed, it holds that:%
\begin{eqnarray}
&&%
\begin{array}{l}
G_{4}^{3}\left( q\right) \supset _{s}SL(2,\mathbb{R})\times G_{4}^{4}\left(
q\right) ; \\
\mathbf{Adj}_{G_{4}^{3}}=\left( \mathbf{3,1}\right) +\left( \mathbf{1,Adj}%
_{G_{4}^{4}}\right) +\left( \mathbf{2,R}\right) ;%
\end{array}
\\
&&%
\begin{array}{l}
SL(2,\mathbb{R})\times G_{4}^{4}\overset{mcs}{\supset }SO(2)_{J}\times
mcs\left( G_{4}^{4}\right) ; \\
\left( \mathbf{2,R}\right) =\left( \mathbf{2,1}\right) +\left( \mathbf{2,}%
\mathcal{R}\right) +\left( \mathbf{2,1}\right) +\left( \mathbf{2,}\overline{%
\mathcal{R}}\right) ,%
\end{array}%
\end{eqnarray}%
where the bar here denotes the conjugate irrep. $\mathbf{R}$ (dim$=6q+8$)
denotes the relevant symplectic irrep. of $G_{4}^{4}$ into which the vectors
sit, and $\mathcal{R}$ ($\overline{\mathcal{R}}$) is its electric (magnetic)
$D=5$ counterpart, of dimension $3q+3$. The irrep. $\mathbf{R}$ is given by%
\footnote{%
Actually, in the case $q=4$, $\mathbf{32}^{\prime }$ is the \textit{conjugate%
} of the irreps. $\mathbf{32}$ in which the vectors sit; see App. \ref%
{App-Spinor-Polarizations} for further detail.}:%
\begin{equation}
\begin{array}{cccccc}
q: & 8 & 4 & 2 & 1 & -2/3 \\
G_{4}^{4}: & E_{7(-25)} & SO^{\ast }(12) & SU(3,3) & Sp(6,\mathbb{R}) & SL(2,%
\mathbb{R}) \\
\mathbf{R}: & \mathbf{56} & \mathbf{32}^{\prime } & \mathbf{20} & \mathbf{14}%
^{\prime } & \mathbf{4}%
\end{array}%
\end{equation}%
On the other hand, the irrep. $\mathcal{R}$ is given by:%
\begin{equation}
\begin{array}{cccccc}
q: & 8 & 4 & 2 & 1 & -2/3 \\
G_{4}^{5}: & E_{6(-26)} & SU^{\ast }(6) & SL(3,\mathbb{C}) & SL(3,\mathbb{R})
& SL(2,\mathbb{R}) \\
\mathcal{R}: & \mathbf{27} & \mathbf{15} & \mathbf{9=}\left( \mathbf{3},%
\overline{\mathbf{3}}\right) & \mathbf{6}^{\prime } & \mathbf{1}%
\end{array}%
\end{equation}%
Therefore, the split of signature of $M_{4}^{4}\left( q\right) $, which
corresponds to a pair of Freudenthal systems $\mathfrak{M}\left( J_{3}^{%
\mathbb{A}}\right) $, can be interpreted as the split into two sets of
Poincar\`{e}-dual irreps. of $SO(2)_{J}\times mcs\left( G_{4}^{4}\right) $;
namely, the $D=4$ graviphoton $A_{\mu }$ $\left( \mathbf{2,1}\right) $ and
the $3q+3$ matter vectors $\left( \mathbf{2,}\mathcal{R}\right) $, and their
\textit{Poincar\`{e} duals}, namely the graviphoton $A_{\mu }$ $\left(
\mathbf{2,1}\right) $ and the $3q+3$ matter vectors $\left( \mathbf{2,}%
\overline{\mathcal{R}}\right) $:%
\begin{equation}
\left( c,nc\right) =\left( 6q+8,6q+8\right) =\left( \left( \mathbf{2,1}%
\right) +\left( \mathbf{2,}\mathcal{R}\right) \right) +\left( \left( \mathbf{%
2,1}\right) +\left( \mathbf{2,}\overline{\mathcal{R}}\right) \right) \text{%
~of~}SO(2)_{J}\times mcs\left( G_{4}^{4}\right) \text{.}
\end{equation}%
Note that the $\mathbf{2}$ of the massless spin group $SO(2)_{J}$
corresponds to the two physical polarizations of the graviphotons.

\item[$D=5:$] the relevant manifold is para-quaternionic, maximal and
non-symmetric (recall (\ref{two-3}) and (\ref{two-3-3})):
\begin{equation}
M_{4}^{5}\left( q\right) \equiv \frac{G_{4}^{3}\left( q\right) }{%
G_{4}^{5}\left( q\right) \times SL(3,\mathbb{R})}:\left( c,nc\right) =\left(
9\left( q+1\right) ,9\left( q+1\right) \right) .
\end{equation}%
Such a signature splitting is covariant with respect to $mcs\left(
G_{4}^{5}\right) \times SO(3)$. Indeed, it holds that:%
\begin{eqnarray}
&&%
\begin{array}{l}
G_{4}^{3}\left( q\right) \supset _{s}SL(3,\mathbb{R})\times G_{4}^{5}\left(
q\right) ; \\
\mathbf{Adj}_{G_{4}^{3}}=\left( \mathbf{8,1}\right) +\left( \mathbf{1,Adj}%
_{G_{4}^{5}}\right) +\left( \mathbf{3,}\mathcal{R}\right) +\left( \mathbf{3}%
^{\prime }\mathbf{,}\mathcal{R}^{\prime }\right) ;%
\end{array}
\\
&&%
\begin{array}{l}
SL(3,\mathbb{R})\times G_{4}^{5}\left( q\right) \overset{mcs}{\supset }%
SO(3)\times mcs\left( G_{4}^{5}\right) ; \\
\left( \mathbf{3}^{(\prime )}\mathbf{,}\mathcal{R}^{(\prime )}\right)
=\left( \mathbf{3,1}\right) +\left( \mathbf{3,}\mathfrak{R}\right) ,%
\end{array}%
\end{eqnarray}%
where the prime here denotes the non-compact analogue of conjugation. $%
\mathfrak{R}$ (dim$=3q+2$) denotes the relevant irrep. of $mcs\left(
G_{4}^{5}\right) $ into which the $D=5$ matter vectors sit. Therefore, the
split of signature of $M_{4}^{5}\left( q\right) $, which corresponds to a
\textit{Jordan pair} (see \textit{e.g.} \cite{Truini-1}), can be interpreted
as the split into two sets of Poincar\`{e}-dual irreps. of $SO(3)_{J}\times
mcs\left( G_{4}^{5}\right) $; namely, the $D=5$ graviphoton $A_{\mu }$ $%
\left( \mathbf{3,1}\right) $ and the $3q+2$ matter vectors $\left( \mathbf{3,%
}\mathfrak{R}\right) $, and their \textit{Poincar\`{e} duals}, namely the
graviphoton $A_{\mu }$ $\left( \mathbf{3,1}\right) $ and the $3q+2$ matter
vectors $\left( \mathbf{2,}\mathfrak{R}\right) $:%
\begin{equation}
\left( c,nc\right) =\left( 9\left( q+1\right) ,9\left( q+1\right) \right)
=\left( \left( \mathbf{3,1}\right) +\left( \mathbf{3,}\mathfrak{R}\right)
\right) +\left( \left( \mathbf{3,1}\right) +\left( \mathbf{3,}\mathfrak{R}%
\right) \right) \text{~of~}SO(3)\times mcs\left( G_{4}^{5}\right) \text{.}
\end{equation}%
Note that the $\mathbf{3}$ of the massless spin group $SO(3)\equiv SO(3)_{J}$
corresponds to the three physical polarizations of the vectors in $D=5$.

\item[$D=6:$] the relevant manifold is maximal and non-symmetric (recall (%
\ref{four-4}) and (\ref{four-4-4})):
\begin{equation}
M_{4}^{6}\left( q\right) \equiv \frac{G_{4}^{3}\left( q\right) }{SO\left(
1,q+1\right) \times \mathcal{A}_{q}\times SL(4,\mathbb{R})}:\left(
c,nc\right) =\left( 11q+6,11q+6\right) .
\end{equation}%
Such a signature splitting is covariant with respect to $SO\left( q+1\right)
\times mcs\left( \mathcal{A}_{q}\right) \times SO(4)$. Indeed, it holds that:%
\begin{eqnarray}
&&%
\begin{array}{l}
G_{4}^{3}\left( q\right) \supset _{s}SL(4,\mathbb{R})\times SO\left(
1,q+1\right) \times \mathcal{A}_{q}; \\
\mathbf{Adj}_{G_{4}^{3}}=\left( \mathbf{15,1,1}\right) +\left( \mathbf{1,Adj}%
_{SO\left( 1,q+1\right) },\mathbf{1}\right) +\left( \mathbf{1},\mathbf{1},%
\mathbf{Adj}_{\mathcal{A}_{q}}\right) \\
+\left( \mathbf{4,Spin,2}\right) +\left( \mathbf{4}^{\prime }\mathbf{,Spin}%
^{\prime }\mathbf{,2}\right) +\left( \mathbf{6,q+2,1}\right) ;%
\end{array}
\label{UCLA-1} \\
&&%
\begin{array}{l}
SL(4,\mathbb{R})\times SO\left( 1,q+1\right) \times \mathcal{A}_{q}\overset{%
mcs}{\supset }SU(2)\times SU(2)\times SO\left( q+1\right) \times mcs\left(
\mathcal{A}_{q}\right) ; \\
\left( \mathbf{4}^{(\prime )}\mathbf{,Spin}^{(\prime )}\mathbf{,2}\right)
=\left( \mathbf{2,2,Spin,2}\right) , \\
\left( \mathbf{6,q+2,1}\right) =\left( \mathbf{3},\mathbf{1},\mathbf{q+1},%
\mathbf{1}\right) +(\mathbf{1},\mathbf{3},\mathbf{q+1},\mathbf{1})+(\mathbf{3%
},\mathbf{1},\mathbf{1},\mathbf{1})+(\mathbf{1},\mathbf{3},\mathbf{1},%
\mathbf{1}).%
\end{array}
\label{UCLA-2}
\end{eqnarray}%
In (\ref{UCLA-1}), $\mathbf{Spin}$, $\mathbf{Spin}^{\prime }$ and $\mathbf{%
q+2}$ respectively denote the two conjugate chiral (semi)spinors and the
vector irreps. of $SO\left( 1,q+1\right) \sim SL(2,\mathbb{A})$ (with $%
\mathbb{A}=\mathbb{R},\mathbb{C},\mathbb{H},\mathbb{O}$ for $q=1,2,4,8$,
respectively), whereas in the right-hand side of (\ref{UCLA-2}) $\mathbf{Spin%
}$ and $\mathbf{q+1}$ respectively denote the spinor and vector irreps. of $%
SO\left( q+1\right) $. The irrep. $\mathbf{Spin}$ of $SO\left( q+1\right) $
is given by:%
\begin{equation}
\begin{array}{ccccc}
q: & 8 & 4 & 2 & 1 \\
SO\left( q+1\right) : & SO(9) & SO(5) & SO(3) & SO(2) \\
\mathbf{Spin}: & \mathbf{16} & \mathbf{4} & \mathbf{2} & \mathbf{2}%
\end{array}%
\end{equation}%
Thus, through these branchings, the resulting pair of Poincar\`{e}-dual
irreps. of $SU(2)\times SU(2)\times SO\left( q+1\right) \times mcs\left(
\mathcal{A}_{q}\right) $ irreps. is composed by: $i)$ the physical
polarizations $\left( \mathbf{2,2,Spin,2}\right) $ of massless $1$-forms and
the physical polarizations of $2$-forms, which (under the assumption of
gravity sector to be anti-self-dual) split into $(\mathbf{1},\mathbf{3},%
\mathbf{1},\mathbf{1})$ (anti-self-dual gravity sector) and $\left( \mathbf{3%
},\mathbf{1},\mathbf{q+1},\mathbf{1}\right) $ (self-dual matter sector); $%
ii) $ the physical polarizations $\left( \mathbf{2,2,Spin,2}\right) $ of
massless $3$-forms and the physical polarizations of Poincar\'{e}-dual $2$%
-forms, which split into $(\mathbf{3},\mathbf{1},\mathbf{1},\mathbf{1})$
(self-dual Poincar\'{e}-dual gravity sector) and $\left( \mathbf{1},\mathbf{3%
},\mathbf{q+1},\mathbf{1}\right) $ (anti-self-dual Poincar\'{e}-dual matter
sector). The real dimension of each set of such irreps. can be computed as
(here square brackets denote the integer part)%
\begin{equation}
2^{\left[ q/2\right] +2+\left( 1-\delta _{q,8}\right) }+3\left( q+2\right)
=11q+6,
\end{equation}%
and thus corresponds to the signature split of $M_{4}^{6}\left( q\right) $
in terms of irreps. of $SU(2)\times SU(2)\times SO\left( q+1\right) \times
mcs\left( \mathcal{A}_{q}\right) $:
\begin{equation}
\left( c,nc\right) =\left( 11q+6,11q+6\right) =\left\{
\begin{array}{c}
\left( \mathbf{2,2,Spin,2}\right) +(\mathbf{1},\mathbf{3},\mathbf{1},\mathbf{%
1})+\left( \mathbf{3},\mathbf{1},\mathbf{q+1},\mathbf{1}\right) \\
+ \\
\left( \mathbf{2,2,Spin,2}\right) +(\mathbf{3},\mathbf{1},\mathbf{1},\mathbf{%
1})+\left( \mathbf{1},\mathbf{3},\mathbf{q+1},\mathbf{1}\right) .%
\end{array}%
\right. \text{~}
\end{equation}
\end{itemize}

\subsection{\label{Hodge}Hodge Involution and Coset Cohomology}

A general property of the cosets $M_{N}^{D}$'s (\ref{Ggen}) resides in the
fact that the \textit{Hodge involution}\footnote{%
The involutive or anti-involutive property $\ast ^{2}\Lambda ^{d}=\pm
\Lambda ^{d}$ generally depends on the signature and the dimension of the
group manifold whose cohomology is under consideration. In this case, the
relevant group is $SO(D-2)=mcs\left( SL\left( D-2,\mathbb{R}\right) \right) $%
, and thus $\ast ^{2}\Lambda ^{d}=\Lambda ^{d}$ for $D-2=4n$, while $\ast
^{2}\Lambda ^{d}=-\Lambda ^{d}$ for $D-2=4n+2$ ($n\in \mathbb{N}$).}%
\begin{equation}
\ast :\Lambda ^{d}\longmapsto \ast \Lambda ^{d}=\Lambda ^{D-2-d}  \label{H}
\end{equation}%
acts as a symmetry of the coset cohomology, where the differential forms of
order $d$ are associated to $d$-fold antisymmetric irreps. $\Lambda ^{d}$ of
$SO(D-2)=mcs\left( SL(D-2,\mathbb{R})\right) $.

Note that, out of the possible forms belonging to the cohomology of $%
SO\left( D-2\right) =mcs\left( SL(D-2,\mathbb{R})\right) $, the coset $%
M_{N}^{D}$ (\ref{Ggen}) precisely singles out the physical massless $p>0$
forms of the corresponding supergravity theory with $\mathcal{N}=2N$
supersymmetries in $D$ (Lorentzian) space-time dimensions. Indeed, by
casting the Cartan decomposition of the cosets $M_{N}^{D}$'s (\ref{Ggen}) in
\ manifestly $SO(D-2)$-covariant way, the Lie algebra of $M_{N}^{D}$ itself
branches as%
\begin{equation}
\mathfrak{g}_{3}^{N}\ominus \left( \mathfrak{g}_{D}^{N}\oplus \mathfrak{sl}%
(D-2,\mathbb{R})\right) \sim \underset{\text{manifestly~}SO(D-2)\text{-cov.}}%
{\sum_{d}n_{d}\Lambda ^{d}+\sum_{d}n_{d}\ast \Lambda ^{d}},  \label{decomp}
\end{equation}%
where $\mathfrak{g}_{3}^{N}$ and $\mathfrak{g}_{D}^{N}$ respectively are the
Lie algebras of $G_{3}^{N}$ and $G_{D}^{N}$, and $n_{d}$ is the (real)
dimension of the relevant irreps. of the $U$-duality group $G_{D}^{N}$ in $D$
dimensions. Note that the r.h.s. of (\ref{decomp}) is manifestly invariant
under the Hodge involution $\ast $ (\ref{H}). Thus, the vanishing character (%
\ref{chichi}) of cosets $M_{N}^{D}$'s (\ref{Ggen}) trivially follows from
\begin{equation}
c\left( M_{N}^{D}\right) =\sum_{d}n_{d}\binom{D-2}{d}=\sum_{d}n_{d}\binom{D-2%
}{D-2-d}=nc\left( M_{N}^{D}\right) .  \label{decomp-2}
\end{equation}

By recalling formula (\ref{night-1}), $c\left( M_{N}^{D}\right) =nc\left(
M_{N}^{D}\right) $ can also be computed as the real dimension of the
\textquotedblleft $mcs$ counterparts" of cosets $M_{N}^{D}$'s (\ref{Ggen}),
namely of the cosets $\widehat{M}_{N}^{D}$ (\ref{Ggen-mcs}).\medskip

In maximal theories ($N=16$), by recalling the embedding (\ref{emb-mcs-N=16}%
) and Table 2, one can trace back the fact that only $d$-fold antisymmetric
irreps. $\Lambda ^{d}$'s occur in (\ref{decomp}) to the embedding%
\begin{equation}
\begin{array}{l}
SO(16)\supset \mathcal{R}_{D}^{16}\times SO(D-2)_{J} \\
\\
\mathbf{Adj}_{SO(16)}\equiv \mathbf{16}\times _{a}\mathbf{16}=\mathbf{Adj}%
_{SO(D-2)}+\sum_{d}n_{d}\Lambda ^{d}.%
\end{array}
\label{emb-mcs-N=16-2}
\end{equation}%
Namely, in $SO(D-2)_{J}$ the antisymmetric rank-$2$ tensor product of spinor
irreps. only contain antisymmetric $d$-fold irreps. (see \textit{e.g.} \cite%
{Spinor-Algebras}). We will consider here three examples, namely $D=11$ and $%
D=10$ (type IIA and IIB).

\begin{itemize}
\item[$(I)$] In maximal supergravity ($N=16$) in $D=11$, $d=3$ and $n_{d}=1$%
, thus (\ref{decomp}) and (\ref{decomp-2}) specifies as follows:%
\begin{eqnarray}
\mathfrak{e}_{8(8)}\ominus \mathfrak{sl}\left( 9,\mathbb{R}\right) &\sim &%
\underset{\text{manifestly~}SO(9)\text{-cov.}}{\Lambda ^{3}+\ast \Lambda
^{3}=\Lambda ^{3}+\Lambda ^{6}=\mathbf{84}+\mathbf{84}}; \\
&&  \notag \\
c\left( \frac{E_{8(8)}}{SL(9,\mathbb{R})}\right) &=&\binom{9}{3}=\binom{9}{6}%
=nc\left( \frac{E_{8(8)}}{SL(9,\mathbb{R})}\right) =84=\text{dim}_{\mathbb{R}%
}\left( \frac{SO(16)}{SO(9)}\right) .  \label{night-2}
\end{eqnarray}%
In terms of the Cartan decomposition of the maximal non-symmetric Riemannian
compact coset $\widehat{M}_{16}^{11}=SO(16)/SO(9)$, the result (\ref{night-2}%
) can be obtained as a consequence of the maximal non-symmetric embedding%
\footnote{%
The embedding (\ref{Dynkin-1}) actually follow from a Theorem due to Dynkin
\cite{D-1,D-2}, applied to the \textit{self-conjugate} spinor irrep. $%
\mathbf{16}$ of $SO(9)$:
\begin{equation*}
SO(9):\mathbf{16}\times _{s}\mathbf{16}=\Lambda ^{0}+\Lambda ^{1}+\Lambda
^{4}=\mathbf{1}+\mathbf{9}+\mathbf{126}.
\end{equation*}%
} (\textit{cfr.} (\ref{emb-mcs-N=16}) and Table 2)

\begin{equation}
\begin{array}{l}
\mathfrak{so}(16)\supset _{ns}\mathfrak{so}(9) \\
\\
\mathbf{16}=\mathbf{16} \\
\\
\underset{\mathbf{120}}{\mathbf{Adj}_{SO(16)}}\equiv \left( \mathbf{16\times
16}\right) _{a}=\underset{\mathbf{36}}{\mathbf{Adj}_{SO(9)}}+\underset{%
\mathbf{84}}{\Lambda ^{3}}.%
\end{array}
\label{Dynkin-1}
\end{equation}

\item[$(II)$] In maximal $D=10$ IIA supergravity, the relevant values are $%
d=1,2,3$ with $n_{1}=n_{2}=n_{3}=1$, and thus (\ref{decomp}) and (\ref%
{decomp-2}) specifies as follows:
\begin{eqnarray}
\mathfrak{e}_{8(8)}\ominus \left( \mathfrak{sl}\left( 8,\mathbb{R}\right)
\oplus \mathfrak{so}(1,1)\right) &\sim &\underset{\text{manifestly~}SO(8)%
\text{-cov.}}{\Lambda ^{1}+\Lambda ^{2}+\Lambda ^{3}+\ast \Lambda ^{1}+\ast
\Lambda ^{2}+\ast \Lambda ^{3}=\Lambda ^{1}+\Lambda ^{2}+\Lambda
^{3}+\Lambda ^{7}+\Lambda ^{6}+\Lambda ^{5}}  \notag \\
&=&\left( \mathbf{8}_{v}+\mathbf{28}+\mathbf{56}_{v}\right) +\left( \mathbf{8%
}_{v}+\mathbf{28}+\mathbf{56}_{v}\right) ; \\
&&  \notag \\
c\left( \frac{E_{8(8)}}{SO(1,1)\times SL(8,\mathbb{R})}\right) &=&8+\binom{8%
}{2}+\binom{8}{3}=\binom{8}{7}+\binom{8}{6}+\binom{8}{5}  \notag \\
&=&nc\left( \frac{E_{8(8)}}{SO(1,1)\times SL(8,\mathbb{R})}\right) =92=\text{%
dim}_{\mathbb{R}}\left( \frac{SO(16)}{SO(8)}\right) .  \label{night-33}
\end{eqnarray}%
In terms of the Cartan decomposition of the non-maximal non-symmetric
Riemannian compact coset $\widehat{M}_{16}^{10~IIA}=SO(16)/SO(8)$, the
result (\ref{night-33}) can be obtained as a consequence of the
next-to-maximal non-symmetric embedding ($\mathbf{Adj}=\Lambda ^{2}$; (%
\textit{cfr.} (\ref{emb-mcs-N=16}) and Table 2)
\begin{equation}
\begin{array}{l}
\mathfrak{so}(16)\supset _{ns}\mathfrak{so}(8) \\
\\
\mathbf{16}=\mathbf{8}_{s}+\mathbf{8}_{c} \\
\\
\underset{\mathbf{120}}{\mathbf{Adj}_{SO(16)}}\equiv \left( \mathbf{16\times
16}\right) _{a}=\left( \mathbf{8}_{s}+\mathbf{8}_{c}\right) \times
_{a}\left( \mathbf{8}_{s}+\mathbf{8}_{c}\right) \\
=\mathbf{8}_{s}\times _{a}\mathbf{8}_{s}+\mathbf{8}_{c}\times _{a}\mathbf{8}%
_{c}+\mathbf{8}_{s}\times \mathbf{8}_{c}=\underset{\mathbf{28}}{\mathbf{Adj}%
_{SO(8)}}+\underset{\mathbf{8}_{v}}{\Lambda _{v}^{1}}+\underset{\mathbf{28}}{%
\Lambda ^{2}}+\underset{\mathbf{56}_{v}}{\Lambda _{v}^{3}}.%
\end{array}
\label{Dynkin-2}
\end{equation}

\item[$(III)$] In maximal $D=10$ IIB supergravity, the relevant values are $%
d=2,4$ with $n_{2}=2n_{4}=2$, and thus (\ref{decomp}) and (\ref{decomp-2})
specifies as follows:
\begin{eqnarray}
\mathfrak{e}_{8(8)}\ominus \left( \mathfrak{sl}\left( 8,\mathbb{R}\right)
\oplus \mathfrak{sl}(2,\mathbb{R})\right) &\sim &\underset{\text{manifestly~}%
SO(8)\text{-cov.}}{2\Lambda ^{2}+\Lambda ^{4}+2\ast \Lambda ^{1}+\ast
\Lambda ^{4}=2\Lambda ^{2}+\Lambda ^{4}+2\Lambda ^{6}+\Lambda ^{4}}  \notag
\\
&=&\left( \left( \mathbf{28},\mathbf{2}\right) +\left( \mathbf{35}_{s},%
\mathbf{1}\right) \right) +\left( \left( \mathbf{28},\mathbf{2}\right)
+\left( \mathbf{35}_{c},\mathbf{1}\right) \right) ; \\
&&  \notag \\
c\left( \frac{E_{8(8)}}{SO(1,1)\times SL(8,\mathbb{R})}\right) &=&2\binom{8}{%
2}+\binom{8}{4}=2\binom{8}{6}+\binom{8}{4}  \notag \\
&=&nc\left( \frac{E_{8(8)}}{SO(1,1)\times SL(8,\mathbb{R})}\right) =91=\text{%
dim}_{\mathbb{R}}\left( \frac{SO(16)}{SO(8)\times SO(2)}\right) .  \notag \\
&&  \label{night-44}
\end{eqnarray}%
In terms of the Cartan decomposition of the non-maximal non-symmetric
Riemannian compact coset $\widehat{M}_{16}^{10~IIB}=SO(16)/\left(
SO(8)\times SO(2)\right) $, the result (\ref{night-44}) can be obtained as a
consequence of the next-to-maximal non-symmetric embedding (\textit{cfr.} (%
\ref{emb-mcs-N=16}) and Table 2):
\begin{equation}
\begin{array}{l}
\mathfrak{so}(16)\supset _{ns}\mathfrak{so}(8)\oplus \mathfrak{so}(2) \\
\\
\mathbf{16}=\left( \mathbf{8}_{s},\mathbf{2}\right) \\
\\
\underset{\mathbf{120}}{\mathbf{Adj}_{SO(16)}}\equiv \left( \mathbf{16\times
16}\right) _{a}=\left( \mathbf{8}_{s},\mathbf{2}\right) \times _{a}\left(
\mathbf{8}_{s},\mathbf{2}\right) =\left( \mathbf{8}_{s}\times _{a}\mathbf{8}%
_{s},\mathbf{2}\times _{s}\mathbf{2}\right) +\left( \mathbf{8}_{s}\times _{s}%
\mathbf{8}_{s},\mathbf{2}\times _{a}\mathbf{2}\right) \\
=\left( \mathbf{28},\mathbf{3}\right) +\left( \mathbf{1},\mathbf{1}\right)
+\left( \mathbf{35}_{s},\mathbf{1}\right) =\underset{\mathbf{28}_{0}}{%
\mathbf{Adj}_{SO(8),0}}+\underset{\mathbf{1}_{0}}{\mathbf{Adj}_{SO(2),0}}+%
\underset{\mathbf{28}_{2}}{\Lambda _{2}^{2}}+\underset{\mathbf{28}_{-2}}{%
\Lambda _{-2}^{2}}+\underset{\mathbf{35}_{s,0}}{\Lambda _{0}^{4}},%
\end{array}
\label{Dynkin-3}
\end{equation}%
where in the last step the subscripts denote the charges of the $D=10$ IIB $%
\mathcal{R}$-symmetry $\mathfrak{so}(2)$.
\end{itemize}

Finally, we present below the same analysis for other two \textit{%
\textquotedblleft pure"} supergravities:

\begin{itemize}
\item[$(IV)$] In $N=12$ supergravity (which shares the same bosonic sector
of the quaternionic \textit{magical} theory with $N=4$ \cite{GST}) in $D=5$
, $d=1$ and $n_{d}=15$, thus (\ref{decomp}) and (\ref{decomp-2}) specifies
as follows:%
\begin{eqnarray}
\mathfrak{e}_{7(-5)}\ominus \left( \mathfrak{su}^{\ast }\left( 6\right)
\oplus \mathfrak{sl}\left( 3,\mathbb{R}\right) \right) &\sim &\left(
14+1\right) \underset{\text{manifestly~}SO(3)\text{-cov.}}{\Lambda
^{1}+\left( 14+1\right) \ast \Lambda ^{1}=\left( 14+1\right) \Lambda
^{1}+\left( 14+1\right) \Lambda ^{2}}  \notag \\
&=&\left( 14+1\right) \mathbf{3+}\left( 14+1\right) \mathbf{3} \\
&&  \notag \\
c\left( \frac{E_{7(-5)}}{SU^{\ast }(6)\times SL(3,\mathbb{R})}\right)
&=&\left( 14+1\right) 3=\left( 14+1\right) \binom{3}{2}=nc\left( \frac{%
E_{7(-5)}}{SU^{\ast }(6)\times SL(3,\mathbb{R})}\right)  \notag \\
&=&45=\text{dim}_{\mathbb{R}}\left( \frac{SO(12)}{USp(6)}\right) .
\label{night-2!}
\end{eqnarray}%
In terms of the Cartan decomposition of the maximal non-symmetric Riemannian
compact coset $\widehat{M}_{12}^{5}=SO(12)/USp(6)$, the result (\ref%
{night-2!}) can be obtained as a consequence of the maximal non-symmetric
embedding%
\begin{equation}
\begin{array}{l}
\mathfrak{so}(12)\supset _{ns}\mathfrak{usp}(6)\oplus \mathfrak{su}(2) \\
\\
\mathbf{12}=\left( \mathbf{6,2}\right) \\
\\
\underset{\mathbf{66}}{\mathbf{Adj}_{SO(12)}}\equiv \left( \mathbf{12\times
12}\right) _{a}=\left( \mathbf{6,2}\right) \times _{a}\left( \mathbf{6},%
\mathbf{2}\right) =\left( \mathbf{6}\times _{a}\mathbf{6},\mathbf{2}\times
_{s}\mathbf{2}\right) +\left( \mathbf{6}\times _{s}\mathbf{6},\mathbf{2}%
\times _{a}\mathbf{2}\right) \\
=\left( \mathbf{14},\mathbf{3}\right) +\underset{\mathbf{Adj}_{SU(2)}}{%
\left( \mathbf{1},\mathbf{3}\right) }+\underset{\mathbf{Adj}_{USp(6)}}{%
\left( \mathbf{21},\mathbf{1}\right) },%
\end{array}%
\end{equation}%
where the $D=5$ massless spin algebra $\mathfrak{su}(2)$ is not modded out
in order to determine $\widehat{M}_{12}^{5}$, and it corresponds to the
\textquotedblleft extra" $USp(6)$ ($\mathcal{R}$-symmetry-)singlet, a
peculiar feature of this extended supergravity theory (which makes it
amenable to an $N=4$ interpretation).

\item[$(V)$] In $N=10$ supergravity in $D=4$, $d=1$ and $n_{d}=10$, thus (%
\ref{decomp}) and (\ref{decomp-2}) specifies as follows:%
\begin{eqnarray}
\mathfrak{e}_{6(-14)}\ominus \left( \mathfrak{su}\left( 5,1\right) \oplus
\mathfrak{sl}\left( 2,\mathbb{R}\right) \right) &\sim &10\underset{\text{%
manifestly~}SO(2)\text{-cov.}}{\Lambda ^{1}+10\ast \Lambda ^{1}=10\Lambda
^{1}+10\Lambda ^{1}}=\left( 10\right) \mathbf{2+}\left( 10\right) \mathbf{2}
\notag \\
&&  \notag \\
c\left( \frac{E_{6(-14)}}{SU(5,1)\times SL(2,\mathbb{R})}\right) &=&10\cdot
2=nc\left( \frac{E_{6(-14)}}{SU(5,1)\times SL(2,\mathbb{R})}\right)  \notag
\\
&=&40=\text{dim}_{\mathbb{R}}\left( \frac{SO(10)}{U(5)}\right) .
\label{night-3!}
\end{eqnarray}%
In terms of the Cartan decomposition of the maximal non-symmetric Riemannian
compact coset $\widehat{M}_{10}^{4}=SO(10)/U(5)$, the result (\ref{night-3!}%
) can be obtained as a consequence of the maximal symmetric embedding%
\begin{equation}
\begin{array}{l}
\mathfrak{so}(10)\supset _{s}\mathfrak{su}(5)\oplus \mathfrak{u}(1) \\
\\
\mathbf{10}=\mathbf{5}_{1}+\overline{\mathbf{5}}_{-1} \\
\\
\underset{\mathbf{45}}{\mathbf{Adj}_{SO(10)}}\equiv \left( \mathbf{10\times
10}\right) _{a}=\left( \mathbf{5}_{1}+\overline{\mathbf{5}}_{-1}\right)
\times _{a}\left( \mathbf{5}_{1}+\overline{\mathbf{5}}_{-1}\right) \\
=\mathbf{5}_{1}\times _{a}\mathbf{5}_{1}+\overline{\mathbf{5}}_{-1}\times
_{a}\overline{\mathbf{5}}_{-1}+\mathbf{5}_{1}\times \overline{\mathbf{5}}%
_{-1}=\mathbf{10}_{2}+\overline{\mathbf{10}}_{-2}+\mathbf{24}_{0}+\mathbf{1}%
_{0},%
\end{array}%
\end{equation}%
where the subscripts denote the charges with respect to the $D=4$ massless
spin algebra $\mathfrak{u}(1)$.
\end{itemize}

\section{\label{Sec-Conclusion}Conclusion}

In this paper we have analyzed some consequences of the super-Ehlers
structure of $N$-extended supergravity theories in $D\geqslant 4$ space-time
dimensions. As the Ehlers $SL(D-2,\mathbb{R})$ is an off-shell symmetry of
the Lagrangian \cite{Br-1,Br-2,Br-3}, so there should exist an Hamiltonian
formulation of light-cone supergravity in which $U$-duality $G_{N}^{D}$ is
an off-shell symmetry. Moreover, \textit{at least} for any amount of
supersymmetry $N\geqslant 4$, the Ehlers group can be regarded as the
commutant of $G_{N}^{D}$ itself inside the $U$-duality group $G_{N}^{3}$ in $%
D=3$.

The pseudo-Riemannian manifolds pertaining to the embedding of the
super-Ehlers group $G_{N}^{D}\times SL(D-2,\mathbb{R})$ into $G_{N}^{3}$,
namely the cosets $M_{N}^{D}$'s (\ref{Ggen}), have been found to exhibit an
interesting invariance under the Hodge involution (\ref{H}), acting on the
cohomology of $M_{N}^{D}$, which in turn singles out only the physical
massless $p>0$ forms of the corresponding supergravity theory, regarded as $%
p $-fold antisymmetric irreps. $\Lambda ^{p}$ of the massless spin group $%
SO(D-2)_{J}=mcs\left( SL(D-2,\mathbb{R})_{\text{Ehlers }}\right) $ in $D$
(Lorentzian) space-time dimensions.

The symmetry under\ the Hodge involution (\ref{H}) implies all the cosets $%
M_{N}^{D}$'s (\ref{Ggen}) to have a vanishing \textit{character}, namely to
have the same number of compact and non-compact generators : $c\left(
M_{N}^{D}\right) =nc\left( M_{N}^{D}\right) $. Such a number, along with its
manifestly $SO(D-2)_{J}$-covariant decomposition in terms of physical
massless $p>0$ forms, can be computed by considering the Cartan
decomposition of the cosets $\widehat{M}_{N}^{D}$'s (\ref{Ggen-mcs}), which
can be regarded as the \textquotedblleft $mcs$ counterpart" of $M_{N}^{D}$'s
(\ref{Ggen}). Indeed, the embedding of $SO(D-2)_{J}$ inside $H_{N}^{3}\equiv
mcs\left( G_{N}^{3}\right) $ is such that the generators of $\widehat{M}%
_{N}^{D}$ split only into antisymmetric tensor irreps. of $SO(D-2)_{J}$
itself, with multiplicities given by irreps. of $H_{N}^{D}\equiv mcs\left(
G_{N}^{D}\right) $.

The approach of this paper may be relevant for the analysis of the issue of
ultraviolet divergences in supergravity theories with maximal or non-maximal
supersymmetry, by exploiting the light-cone formulation, along the lines
\textit{e.g.} of \cite{GS,Br-1,Br-2,Br-3}.

\section*{Acknowledgements}

We would like to thank V. S. Varadarajan for an enlightening discussion on
the Hodge involution.

A.M. would like to thank the Department of Physics and Astronomy, University
of California at Los Angeles, where this project was completed, for kind
hospitality and stimulating environment.

This work was supported in part by DOE Grant DE-FG03-91ER40662.

The work of S.F. has been supported by the ERC Advanced Grant no. 226455,
Supersymmetry, Quantum Gravity and Gauge Fields (SUPERFIELDS).

\appendix

\section{Embeddings}

\label{App-Embeddings}

Let us start by recalling some useful definitions.

Given two semisimple Lie groups $G^{\prime }$ and $G$, generated by the Lie
algebras $\mathfrak{g}^{\prime },\mathfrak{g}$, respectively, if $G^{\prime
}\subset G$ (proper inclusion), we say that $G^{\prime }$ is \emph{maximal}
in $G$ iff there is no proper subalgebra $\mathfrak{g}^{\prime \prime }$ of $%
\mathfrak{g}$ containing $\mathfrak{g}^{\prime }$. If $G^{\prime }$ and $G$
are complex semisimple Lie groups such that $G^{\prime }\subset G$, the
embedding of $G^{\prime }$ into $G$ is \emph{regular} iff one can find a
basis of $\mathfrak{g}^{\prime }$ consisting of elements of a Cartan
subalgebra $\mathfrak{h}$ of $\mathfrak{g}$ and shift-generators $E_{\alpha
} $ corresponding to roots $\alpha $ of $\mathfrak{g}$ relative to $%
\mathfrak{h}$ \cite{D-1}. Regular subalgebras $\mathfrak{g}^{\prime }$ of a
semisimple Lie algebra $\mathfrak{g}$ can be constructed using the simple
procedure defined by Dynkin in \cite{D-1}: the Dynkin diagram of $\mathfrak{g%
}^{\prime }$ can be obtained as a truncation of the extended diagram of $%
\mathfrak{g}$. When considering \emph{real forms} $G^{\prime },\,G$ of
complex semisimple Lie groups $G_{\mathbb{C}}^{\prime },\,G_{\mathbb{C}}$,
we say that $G^{\prime }\subset G$ is regularly embedded in $G$ iff the
complexification $\mathfrak{g}_{\mathbb{C}}^{\prime }$ of $\mathfrak{g}%
^{\prime }$ is regularly embedded in the complexification $\mathfrak{g}_{%
\mathbb{C}}$ of $\mathfrak{g}$. The embedding of $G^{\prime }$ into $G$ is
\emph{symmetric} iff we can write $\mathfrak{g}=\mathfrak{g}^{\prime }\oplus
\mathfrak{p}$, such that $[\mathfrak{g}^{\prime },\,\mathfrak{p}]\subset
\mathfrak{p}$ and $[\mathfrak{p},\,\mathfrak{p}]\subset \mathfrak{g}^{\prime
}$. Finally the embedding is \emph{rank-preserving} iff $\mathrm{rank}(%
\mathfrak{g}^{\prime })=\mathrm{rank}(\mathfrak{g})$.

\subsection{The Embeddings $\mathrm{SL}(D-2,\mathbb{R})\times \mathrm{E}%
_{11-D(11-D)}\subset \mathrm{E}_{8(8)}$}

\paragraph{The $D=5$ case $\mathrm{SL}(3,\mathbb{R})\times \mathrm{E}%
_{6(6)}\subset \mathrm{E}_{8(8)}$}

\begin{figure}[h]
\begin{center}
\centerline{\includegraphics[width=0.5\textwidth]{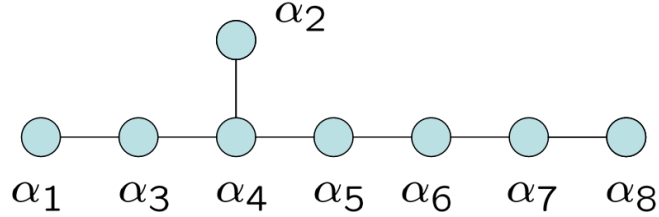}}
\end{center}
\caption{$\mathfrak{\ e}_{8(8)}$ Dynkin diagram.}
\label{e8}
\end{figure}

The embedding of $\mathfrak{sl}(3,\mathbb{R})\oplus \mathfrak{\ e}%
_{6(6)}\subset \mathfrak{\ e}_{8(8)}$ is regular and can be described using
Dynkin's construction \cite{D-1}. Let us number the simple roots of $%
\mathfrak{\ e}_{8(8)}$ so that the $D_{7}$ sub-Dynkin diagram consists of
the roots $\alpha _{2},\dots ,\alpha _{8}$, with $\alpha _{2}$ and $\alpha
_{3}$ on the two symmetric legs, and $\alpha _{1}$ is the $D_{7}$-spinor
weight attached to $\alpha _{3}$, see Fig. \ref{e8}.
The $\mathfrak{\ e}_{8(8)}$ Cartan matrix reads:
\begin{equation}
\langle \alpha _{i},\,\alpha _{j}\rangle =\left(
\begin{matrix}
2 & 0 & -1 & 0 & 0 & 0 & 0 & 0\cr0 & 2 & 0 & -1 & 0 & 0 & 0 & 0\cr-1 & 0 & 2
& -1 & 0 & 0 & 0 & 0\cr0 & -1 & -1 & 2 & -1 & 0 & 0 & 0\cr0 & 0 & 0 & -1 & 2
& -1 & 0 & 0\cr0 & 0 & 0 & 0 & -1 & 2 & -1 & 0\cr0 & 0 & 0 & 0 & 0 & -1 & 2
& -1\cr0 & 0 & 0 & 0 & 0 & 0 & -1 & 2\cr%
\end{matrix}%
\right) \,.
\end{equation}

In an orthonormal basis the simple roots $\alpha_i$ read:
\begin{align}
\alpha_1&=-\frac{1}{2}\,(\epsilon_1+\epsilon_2+\epsilon_3+\epsilon_4+%
\epsilon_5+\epsilon_6-\epsilon_7-\epsilon_8)\,,  \notag \\
\alpha_2 &=\epsilon_6+\epsilon_7\,\,;\,\,\,\alpha_3
=\epsilon_6-\epsilon_7\,\,;\,\,\,\alpha_4
=\epsilon_5-\epsilon_6\,\,;\,\,\,\alpha_5
=\epsilon_4-\epsilon_5\,\,;\,\,\,\alpha_6
=\epsilon_3-\epsilon_4\,\,;\,\,\,\alpha_7 =\epsilon_2-\epsilon_3\,;  \notag
\\
\alpha_8 & =\epsilon_1-\epsilon_2
\end{align}
Let us denote by $\Delta_+[\mathfrak{\ e}_{8(8)}]=\{\alpha=\sum_{i=1}^8
n^i\,\alpha_i\}$ the set of positive roots of $\mathfrak{\ e}_{8(8)}$. The $%
\mathfrak{\ e}_{6(6)}$ subalgebra is defined by the sub-Dynkin diagram
consisting of the simple roots $\alpha_a$, $a=1,\dots, 6$. The 36 positive $%
\mathfrak{\ e}_{6(6)}$-roots be denoted by $\gamma_A$, so that:
\begin{equation}
\Delta_+[\mathfrak{\ e}_{6(6)}]=\{\gamma_A=\sum_{a=1}^6 n_A^a\,\alpha_a\}\,.
\end{equation}
Furthermore let us consider the following positive roots $\beta_x$, $x=1,2,3$%
:
\begin{align}
\beta_1&=\alpha_8=\epsilon_1-\epsilon_2\,\,;\,\,\,\beta_2=2\alpha_1+3%
\alpha_2+4\alpha_3+6\alpha_4+5\alpha_5+4\alpha_6+3\alpha_7
+\alpha_8=\epsilon_2+\epsilon_8\,;  \notag \\
\beta_3&=2\alpha_1+3\alpha_2+4\alpha_3+6\alpha_4+5\alpha_5+4\alpha_6+3%
\alpha_7 +2\alpha_8=\epsilon_1+\epsilon_8\,.
\end{align}
One can easily verify that $\beta_x$ generate an $\mathfrak{sl}(3,\mathbb{R}%
) $-root space which is orthogonal to $\Delta_+[\mathfrak{\ e}_{6(6)}]$: $%
\beta_x\cdot \gamma_A=0$. We have then constructed an $\mathfrak{sl}(3,%
\mathbb{R})\oplus \mathfrak{\ e}_{6(6)}$ subalgebra of $\mathfrak{\ e}%
_{8(8)} $:
\begin{align}
\mathfrak{sl}(3,\mathbb{R})&=\mathrm{Span}(H_{\beta_1},\,H_{\beta_2},\,E_{%
\pm \beta_1},\,E_{\pm \beta_2},\,E_{\pm \beta_3})\,,  \notag \\
\mathfrak{\ e}_{6(6)}&=\mathrm{Span}(H_{\alpha_a},\,E_{\pm \gamma_A})_{%
{\tiny
\begin{matrix}
a=1,\dots, 6\cr A=1,\dots, 36%
\end{matrix}%
}}
\end{align}
Within $\mathfrak{sl}(3,\mathbb{R})\oplus \mathfrak{\ e}_{6(6)}$ we can
identify its maximal compact subalgebra $\mathfrak{so}(3)\oplus \mathfrak{\
usp}(8)$, which is a maximal subalgebra of $\mathfrak{so}(16)$:
\begin{align}
\mathfrak{so}(16)&=\mathrm{Span}(E_{\alpha}-E_{-\alpha})_{\alpha\in \Delta_+[%
\mathfrak{\ e}_{8(8)}]}\,,  \notag \\
\mathfrak{so}(3)&=\mathrm{Span}(E_{\beta_x}-E_{-\beta_x})_{x=1,2,3}\,,
\notag \\
\mathfrak{\ usp}(8)&=\mathrm{Span}(E_{ \gamma_A}-E_{-\gamma_A})_{A=1,\dots,
36}\,.
\end{align}
With respect to this $\mathrm{SO}(3)\times \mathrm{USp}(8)$ subgroup of $%
\mathrm{SO}(16)$ the coset space
\begin{equation}
\mathfrak{K}=\mathfrak{\ e}_{8(8)}\ominus \mathfrak{so}(16)=\mathrm{Span}%
(H_{\alpha_i},\,E_{\alpha}+E_{-\alpha})_{\alpha\in \Delta_+[\mathfrak{\ e}%
_{8(8)}]\,;\,i=1,\dots, 8}\,,
\end{equation}
should decompose as follows:
\begin{align}
\mathfrak{K}&=\mathfrak{K}[\mathfrak{sl}(3,\mathbb{R})]\oplus \mathfrak{K}[%
\mathfrak{\ e}_{6(6)}]\oplus \mathbf{(3,27)}\,,  \notag \\
\mathfrak{K}[\mathfrak{sl}(3,\mathbb{R})]&=\mathfrak{sl}(3,\mathbb{R}%
)\ominus \mathfrak{so}(3)=\mathrm{Span}(H_{\beta_1},\,H_{\beta_2},\,E_{
\beta_x}+E_{-\beta_x})_{x=1,2,3}=\mathbf{(5,1)}\,,  \notag \\
\mathfrak{K}[\mathfrak{\ e}_{6(6)}]&=\mathfrak{\ e}_{6(6)}\ominus \mathfrak{%
\ usp}(8)=\mathrm{Span}(H_{\alpha_a},\,E_{ \gamma_A}+E_{ -\gamma_A})_{{\tiny
\begin{matrix}
a=1,\dots, 6\cr A=1,\dots, 36%
\end{matrix}%
}}=\mathbf{(1,42)}\,,
\end{align}

\paragraph{Generalizing to $\mathrm{SL}(D-2,\mathbb{R})\times \mathrm{E}%
_{11-D(11-D)}\subset \mathrm{E}_{8(8)}$}

\begin{figure}[h]
\begin{center}
\centerline{\includegraphics[width=0.7\textwidth]{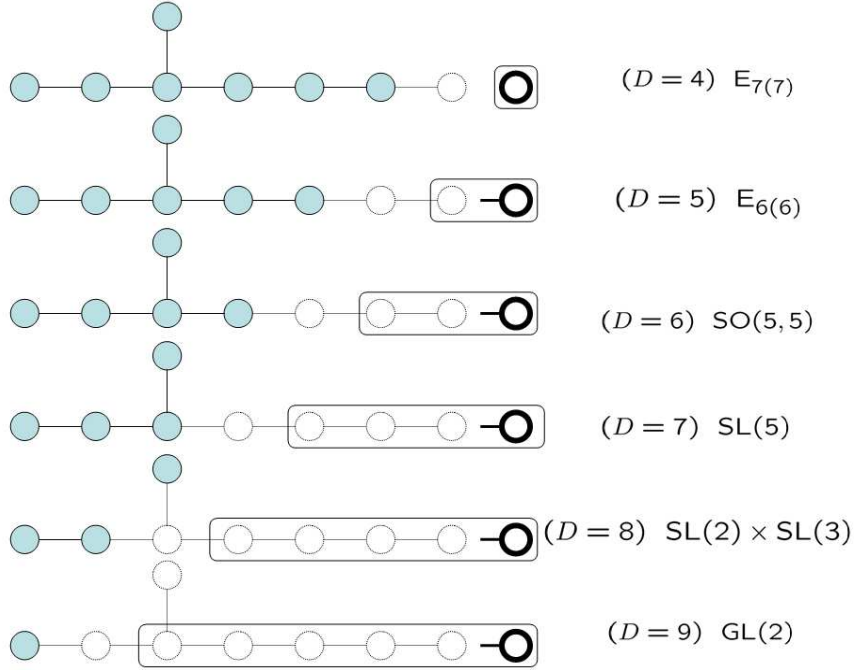}}
\end{center}
\caption{The filled circles define the $\mathfrak{\ e}_{11-D(11-D)}$
sub-Dynkin diagram, while the thick circle represents the exceptional root $-%
\protect\psi$, $\protect\psi=\protect\epsilon_1+\protect\epsilon_8$ being
the highest root of $\mathfrak{e}_8$, which, together with the other roots
in the rectangles, defines the Dynkin diagram of $\mathfrak{sl}(D-2,\mathbb{R%
})$.}
\label{es1}
\end{figure}

The above construction is extended to define the embedding of $\mathfrak{sl}%
(D-2,\mathbb{R})\oplus \mathfrak{\ e}_{11-D(11-D)}\subset \mathfrak{\ e}%
_{8(8)}$, $D\ge 4$, following the same recipe by Dynkin. The embedding of $%
\mathfrak{\ e}_{11-D(11-D)}$ is defined by deleting in the $\mathfrak{\ e}%
_{8(8)}$ Dynkin diagram the last $D-3$ simple roots to the right, namely $%
\alpha_{12-D},\,\dots, \alpha_8$, see Fig. \ref{es1}.
The set of positive roots of $\mathfrak{\ e}_{11-D(11-D)}$ reads:
\begin{align}
\Delta_+[\mathfrak{\ e}_{11-D(11-D)}]&=\{\gamma_A\}=\{\epsilon_a\pm
\epsilon_b,\,\, \left[\frac{\epsilon_8}{2} -\sum_{\alpha=1}^{D-3}\frac{%
\epsilon_\alpha}{2}+\left(\sum_{a=D-2}^{7}\pm \frac{\epsilon_a}{2}\right)_{%
{\tiny \mbox{odd }+}}\right]\}\,,
\end{align}
where those in square brackets are the weights of a chiral spinorial
representation of the $\mathfrak{so}(10-D,\,10-D)$ subalgebra of $\mathfrak{%
\ e}_{11-D(11-D)}$. The set of positive roots of $\mathfrak{sl}(D-2,\mathbb{R%
})$ reads:
\begin{align}
\Delta_+[\mathfrak{sl}(D-2,\mathbb{R})]&=\{\beta_x\}=\{\epsilon_\alpha-
\epsilon_\beta,\,\epsilon_\alpha+\epsilon_8\}\,,
\end{align}
where $\alpha,\,\beta=1,\dots, D-3$, $\beta>\alpha$ and $x=1,\dots,
(D-3)(D-2)/2$. One can easily verify that the two root systems are
orthogonal, namely: $\beta_x\cdot \gamma_A=0$.

This defines the $\mathfrak{sl}(D-2,\mathbb{R})\oplus \mathfrak{\ e}%
_{11-D(11-D)}$ subalgebra of $\mathfrak{\ e}_{8(8)}$:
\begin{align}
\mathfrak{sl}(D-2,\mathbb{R})&=\mathrm{Span}(H_{\alpha_{13-D}},\dots,H_{%
\alpha_{8}}, \,H_{\epsilon_{1}+\epsilon_8},\,E_{\pm \beta_x})_{\beta_x\in
\Delta_+[\mathfrak{sl}(D-2,\mathbb{R})] }\,,  \notag \\
\mathfrak{\ e}_{11-D(11-D)}&=\mathrm{Span}(H_{\alpha_a},\,E_{\pm \gamma_A})_{%
{\tiny
\begin{matrix}
a=1,\dots, 11-D\cr \gamma_A\in \Delta_+[\mathfrak{\ e}_{11-D(11-D)}]%
\end{matrix}%
}}\,\,,\,D=4,\dots, 8\,,  \notag \\
\mathfrak{\ e}_{2(2)}&=\mathrm{Span}(H_{\alpha_1}, H_{\lambda},\,E_{\pm
\alpha_1})\,\,,\,D=9\,,
\end{align}
where the generators $H_{\alpha_{13-D}},\dots,H_{\alpha_{8}}$, in the first
line, are not counted for $D=4$, for which the only Cartan generator of $%
\mathfrak{sl}(2,\mathbb{R})$ is $H_{\epsilon_{1}+\epsilon_8}$. In the $D=9$
case, the vector $\lambda$ in the last line is: $\lambda=\epsilon_7-%
\alpha_1/4$ and is orthogonal to the $\beta_x$ and to $\alpha_1$.

As far as the corresponding maximal compact subalgebra $\mathfrak{so}(D-2)
\oplus \mathfrak{H}_{D}$ is concerned, its can be constructed as follows:
\begin{align}
\mathfrak{so}(D-2)&=\mathrm{Span}(E_{\beta_x}-E_{-\beta_x})_{\beta_x\in
\Delta_+[\mathfrak{sl}(D-2,\mathbb{R})] }\,,  \notag \\
\mathfrak{H}_{D}&=\mathrm{Span}(E_{ \gamma_A}-E_{-\gamma_A})_{\gamma_A\in
\Delta_+[\mathfrak{\ e}_{11-D(11-D)}]}\,,
\end{align}
where $\mathfrak{H}_{D}$ is the maximal compact subalgebra of $\mathfrak{\ e}%
_{11-D(11-D)}$.

In the $D=10$ case we need to consider the type IIA and type IIB
descriptions in which the relevant subgroups of $E_{8(8)}$ are $\mathrm{SL}%
(8,\mathbb{R})\times \mathrm{SO}(1,1)$ and $\mathrm{SL}(8,\mathbb{R}%
)^{\prime }\times \mathrm{SL}(2,\,\mathbb{R})$, respectively.\footnote{%
We shall omit the prime in the following.} Their embeddings are illustrated
in Fig. \ref{sl8}. In the former case the $U$-duality group is $\mathrm{SO}%
(1,1)$ and is generated by the Cartan generator $\sum_{i=1}^8
H_{\epsilon_i}-2 H_{\epsilon_8}$.

\subsection{Other Embeddings}

\begin{figure}[h]
\begin{center}
\centerline{\includegraphics[width=0.7\textwidth]{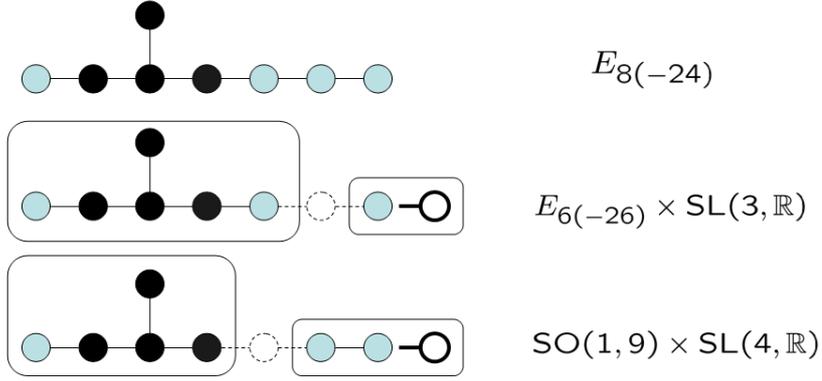}}
\end{center}
\caption{From top to bottom: Satake diagram of $\mathfrak{e}_{8(-24)}$ and
embeddings of the $\mathfrak{e}_{6(-26)}\oplus \mathfrak{sl}(3,\mathbb{R})$
and $\mathfrak{so}(1,9)\oplus \mathfrak{sl}(4,\mathbb{R})$ diagrams inside
the extension of the $\mathfrak{e}_{8(-24)}$ one.}
\label{satake}
\end{figure}

Embeddings considered here were also dealt with in \cite{Keu-2}. Here we
provide a detailed and explicit construction of a number of embeddings in
terms of the generators of the corresponding Lie algebras, using the
notation of \cite{Helgason}. Let us start discussing in detail the
embeddings of $E_{6(-26)}\times \mathrm{SL}(3,\mathbb{R})$ and $\mathrm{SO}%
(1,9)\times \mathrm{SL}(4,\mathbb{R})$ inside $E_{8(-24)}$. At the level of
the corresponding Lie algebras, these embeddings are illustrated in Figure %
\ref{satake}, where the \emph{Satake diagrams} of $\mathfrak{e}%
_{6(-26)}\oplus \mathfrak{sl}(3,\mathbb{R})$ and $\mathfrak{so}(1,9)\oplus
\mathfrak{sl}(4,\mathbb{R})$ are obtained from the $\mathfrak{e}_{8(-24)}$
one once again using Dynkin's procedure of extending the latter and
canceling a suitable simple root.
Let us briefly review the definition of Satake diagrams for non-split (i.e.
non-maximally-non-compact) Lie algebras and the construction of the $%
\mathfrak{e}_{6(-26)}\oplus \mathfrak{sl}(3,\mathbb{R})$ and $\mathfrak{so}%
(1,9)\oplus \mathfrak{sl}(4,\mathbb{R})$ generators in terms of a canonical
basis of the complex $\mathfrak{e}_{8}$. The latter consists of a basis $%
\{H_{\epsilon_i}\},\,i=1\,\dots , 8$, of Cartan generators, with respect to
which the $\mathfrak{e}_{8}$ roots are defined, and shift operators $%
E_\alpha,\,E_{-\alpha}$, $\alpha$ being the $120$ positive roots. The real
form $\mathfrak{e}_{8(-24)}$ is characterized by a Cartan subalgebra $%
\mathfrak{h}$ which splits into the direct sum of a subspace $\mathfrak{h}%
^{nc}$ of non-compact generators (i.e. generators which are odd with respect
to the Cartan involution $\tau$ \footnote{%
We can always find a suitable basis for the matrix representation of the
generators so that $\tau(M)=-M^\dagger$. This means that we shall regard
compactness and non-compactness of a generator to be synonyms, in any matrix
representation, of being anti-hermitian and hermitian, respectively.
Moreover, in our conventions, $E_{-\alpha}=-\tau(E_{\alpha})=E_\alpha^%
\dagger $.}) and a subspace $\mathfrak{h}^{c}$ of compact generators,
defined in terms of the $\{H_{\epsilon_i}\}$ as follows:
\begin{align}
\mathfrak{h}&=\mathfrak{h}^{nc}\oplus \mathfrak{h}^{c}\,\,;\,\,\,\mathfrak{h}%
^{c}=\mathrm{Span}(i\,H_{\alpha_2},\,i\,H_{\alpha_3},\,i\,H_{\alpha_4},\,i%
\,H_{\alpha_5})\,\,;\,\,\,\mathfrak{h}^{nc}=\mathrm{Span}(H_{\epsilon_1},%
\,H_{\epsilon_2},\,H_{\epsilon_3},\,H_{\epsilon_8})\,,
\end{align}
Note that $\mathfrak{h}^{c}$ is the Cartan subalgebra of an $\mathfrak{so}%
(8) $ subalgebra of $\mathfrak{e}_{8(-24)}$ whose Dynkin diagram is defined
by the black roots in Fig. \ref{satake}. The $\mathfrak{e}_8$ positive roots
split into a $12$-dimensional sub-space $\Delta_+^{(0)}[\mathfrak{e}_8]$ of
roots having null restriction to $\mathfrak{h}^{nc}$ and a $108$-dimensional
space $\bar{\Delta}_+[\mathfrak{e}_8]$ of roots with a non-trivial
restriction to $\mathfrak{h}^{nc}$:
\begin{equation}
\Delta_+[\mathfrak{e}_8]=\Delta_+^{(0)}[\mathfrak{e}_8]\oplus \bar{\Delta}_+[%
\mathfrak{e}_8]\,.
\end{equation}
The conjugation $\sigma$ with respect to $\mathfrak{e}_{8(-24)}$ is the
conjugation on the complex $\mathfrak{e}_{8}$ which leaves the elements of
the subalgebra $\mathfrak{e}_{8(-24)}$ invariant. It defines a
correspondence between $\mathfrak{e}_{8}$-roots $\alpha\leftrightarrow
\alpha^\sigma$ such that $\sigma(E_\alpha)\propto E_{\alpha^\sigma}$. The
couple of roots $\alpha\,,\alpha^\sigma$ satisfies the property:
\begin{equation}
\alpha_{|\mathfrak{h}^{nc}}=\alpha^\sigma{}_{|\mathfrak{h}%
^{nc}}\,\,;\,\,\,\,\alpha_{|\mathfrak{h}^{c}}=- \alpha^\sigma{}_{|\mathfrak{h%
}^{c}}\,.
\end{equation}
Clearly if $\alpha\in \Delta_+^{(0)}[\mathfrak{e}_8]$, $\alpha^\sigma=-%
\alpha $, while if $\alpha \in \bar{\Delta}_+[\mathfrak{e}_8]$ and $\alpha_{|%
\mathfrak{h}^{c}}=0$, we have $\alpha^\sigma=\alpha$. Thus if $\alpha\in
\bar{\Delta}_+[\mathfrak{e}_8]$, to each couple of nilpotent generators $%
E_\alpha$ and $\sigma(E_\alpha)$ in $\mathfrak{e}_{8}$, there corresponds a
couple of nilpotent generators in $\mathfrak{e}_{8(-24)}$ given by the $%
\sigma$-invariant combinations $i\,(E_\alpha-\sigma(E_\alpha)),\,E_\alpha+%
\sigma(E_\alpha)$, which can be both brought to an upper-triangular form,
for all $\alpha$. If, on the other hand, $\alpha\in \Delta_+^{(0)}[\mathfrak{%
e}_8]$, the same combinations define \emph{compact} $\mathfrak{so}(8)$
generators $i\,(E_\alpha+E_{-\alpha}),\,E_\alpha-E_{-\alpha}$.

To summarize, the $\mathfrak{e}_{8(-24)}$ generators can be expressed in
terms of the $\mathfrak{e}_8$ canonical basis as follows:
\begin{align}
\mathfrak{e}_{8(-24)}&=\mathfrak{h}\oplus \mathfrak{l}_+\oplus\mathfrak{l}%
_-\oplus \mathfrak{m}_0\,,  \notag \\
\mathfrak{l}_+&=\mathrm{Span}\left[i\,(E_\alpha-\sigma(E_\alpha)),\,E_%
\alpha+\sigma(E_\alpha)\right]_{(\alpha,\,\alpha^\sigma)\in \bar{\Delta}_+[%
\mathfrak{e}_8]}\,,  \notag \\
\mathfrak{l}_-&= \mathrm{Span}\left[i\,(E_{-\alpha}-\sigma(E_{-\alpha})),%
\,E_{-\alpha}+\sigma(E_{-\alpha})\right]_{(\alpha,\,\alpha^\sigma)\in \bar{%
\Delta}_+[\mathfrak{e}_8]}\,,  \notag \\
\mathfrak{m}_0&= \mathrm{Span}\left[i\,(E_\alpha+E_{-\alpha}),\,E_%
\alpha-E_{-\alpha}\right]_{\alpha\in \Delta_+^{(0)}[\mathfrak{e}_8]}\,,
\end{align}
The $112$-dimensional solvable Lie algebra $\mathfrak{s}_0=\mathfrak{h}%
^{nc}\oplus \mathfrak{l}_+$ is the one defined by the Iwasawa decomposition
of $\mathfrak{e}_{8(-24)}$ with respect to $\mathfrak{e}_{7(-133)}\oplus%
\mathfrak{su}(2)$, and its generators, in a suitable basis, can all be
represented by upper-triangular matrices. The centralizer of $\mathfrak{h}%
^{nc}$ is the $\mathfrak{so}(8)$ subalgebra given by $\mathfrak{h}^{c}\oplus
\mathfrak{m}_0$ and is also contained inside the subalgebras $\mathfrak{e}%
_{6(-26)}$ and $\mathfrak{so}(1,9)$, as it is apparent from Fig. \ref{satake}%
.

The $\mathfrak{e}_{6(-26)}$ generators in terms of the above $\mathfrak{e}%
_{8(-24)}$ ones are easily written:
\begin{align}
\mathfrak{e}_{6(-26)}&=\mathfrak{h}^{\prime }\oplus \mathfrak{l}^{\prime
}_+\oplus\mathfrak{l}^{\prime }_-\oplus \mathfrak{m}_0\,,  \notag \\
\mathfrak{l}^{\prime }_+&=\mathrm{Span}\left[i\,(E_\alpha-\sigma(E_\alpha)),%
\,E_\alpha+\sigma(E_\alpha)\right]_{(\alpha,\,\alpha^\sigma)\in \bar{\Delta}%
_+[\mathfrak{e}_8]\cap\Delta_+[\mathfrak{e}_6]}\,,  \notag \\
\mathfrak{l}^{\prime }_-&= \mathrm{Span}\left[i\,(E_{-\alpha}-\sigma(E_{-%
\alpha})),\,E_{-\alpha}+\sigma(E_{-\alpha})\right]_{(\alpha,\,\alpha^\sigma)%
\in \bar{\Delta}_+[\mathfrak{e}_8]\cap\Delta_+[\mathfrak{e}_6]}\,,  \notag \\
\mathfrak{m}_0&= \mathrm{Span}\left[i\,(E_\alpha+E_{-\alpha}),\,E_%
\alpha-E_{-\alpha}\right]_{\alpha\in \Delta_+^{(0)}[\mathfrak{e}_8]}\,,
\end{align}
where $\Delta_+[\mathfrak{e}_6]$ are the $\mathfrak{e}_6$-positive roots in
the $\mathfrak{e}_8$-root system, while
\begin{equation}
\mathfrak{h}^{\prime }=\mathfrak{h}^{\prime nc}\oplus \mathfrak{h}%
^{c}\,\,\,;\,\,\,\,\mathfrak{h}^{\prime nc}=\mathrm{Span}(H_{\epsilon_1+%
\epsilon_2-\epsilon_8},\,H_{\epsilon_3})\,.
\end{equation}
The $\mathfrak{sl}(3,\mathbb{R})$ subalgebra commuting with $\mathfrak{e}%
_{6(-26)}$ has the following form:
\begin{equation}
\mathfrak{sl}(3,\mathbb{R})=\mathrm{Span}\left[H_{\epsilon_1-\epsilon_2},%
\,H_{-\epsilon_1-\epsilon_8},\,E_{\pm \beta_x}\right]\,,  \notag \\
\{\beta_x\}=\{\epsilon_1-\epsilon_2,\,\epsilon_8+\epsilon_1,\,\epsilon_8+%
\epsilon_2\}\,,
\end{equation}
note that $\beta_x^\sigma=\beta_x$.

\begin{figure}[h]
\begin{center}
\centerline{\includegraphics[width=0.7\textwidth]{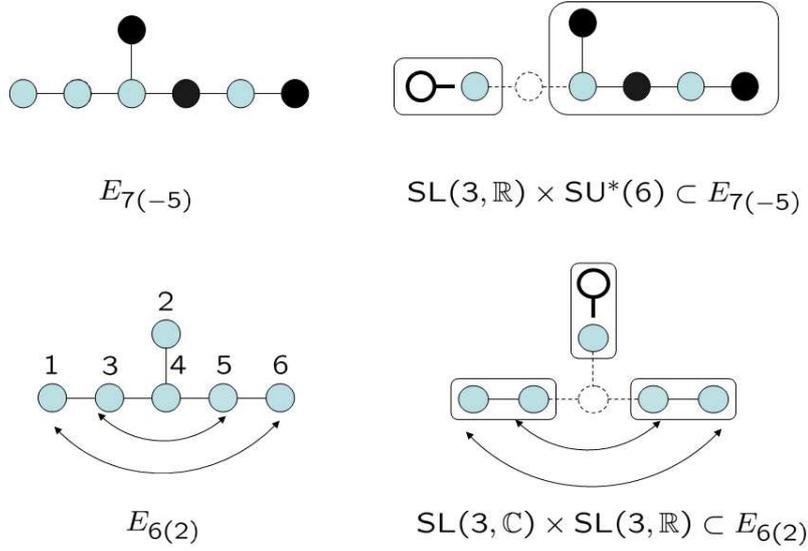}}
\end{center}
\caption{Embeddings $\mathrm{SL}(3,\mathbb{R})\times \mathrm{SU}^*(6)\subset
E_{7(-5)}$ and $\mathrm{SL}(3,\mathbb{C})\times \mathrm{SL}(3,\mathbb{R})
\subset E_{6(2)}$. The thick circle is, as usual, the opposite of the
highest root of the corresponding algebra.}
\label{satake2}
\end{figure}

Finally the $\mathfrak{so}(1,9)\subset \mathfrak{e}_{6(-26)} $ generators
read:
\begin{align}
\mathfrak{so}(1,9)&=\mathfrak{h}^{\prime \prime }\oplus \mathfrak{l}^{\prime
\prime }_+\oplus\mathfrak{l}^{\prime \prime }_-\oplus \mathfrak{m}_0\,,
\notag \\
\mathfrak{l}^{\prime \prime }_+&=\mathrm{Span}\left[i\,(E_\alpha-\sigma(E_%
\alpha)),\,E_\alpha+\sigma(E_\alpha)\right]_{(\alpha,\,\alpha^\sigma)\in
\bar{\Delta}_+[\mathfrak{e}_8]\cap\Delta_+[\mathfrak{so}(10)]}\,,  \notag \\
\mathfrak{l}^{\prime \prime }_-&= \mathrm{Span}\left[i\,(E_{-\alpha}-%
\sigma(E_{-\alpha})),\,E_{-\alpha}+\sigma(E_{-\alpha})\right]%
_{(\alpha,\,\alpha^\sigma)\in \bar{\Delta}_+[\mathfrak{e}_8]\cap\Delta_+[%
\mathfrak{so}(10)]}\,,  \notag \\
\mathfrak{m}_0&= \mathrm{Span}\left[i\,(E_\alpha+E_{-\alpha}),\,E_%
\alpha-E_{-\alpha}\right]_{\alpha\in \Delta_+^{(0)}[\mathfrak{e}_8]}\,,
\end{align}
where $\Delta_+[\mathfrak{so}(10)]$ are the roots of the complex $\mathfrak{%
so}(10)$ algebra within $\mathfrak{e}_8$-root system, and
\begin{equation}
\mathfrak{h}^{\prime \prime }=\mathfrak{h}^{\prime\prime \, nc}\oplus
\mathfrak{h}^{c}\,\,\,;\,\,\,\,\mathfrak{h}^{\prime\prime \, nc}=\mathrm{Span%
}(H_{\epsilon_1+\epsilon_2+\epsilon_3-\epsilon_8})\,.
\end{equation}
The $\mathfrak{sl}(4,\mathbb{R})$ subalgebra commuting with $\mathfrak{so}%
(1,9)$ is described by the following generators:
\begin{equation}
\mathfrak{sl}(4,\mathbb{R})=\mathrm{Span}\left[H_{\epsilon_1-\epsilon_2},%
\,H_{\epsilon_2-\epsilon_3},\,H_{-\epsilon_1-\epsilon_8},\,E_{\pm \beta_x}%
\right]\,,  \notag \\
\{\beta_x\}=\{\epsilon_\alpha-\epsilon_\beta,\,\epsilon_8+\epsilon_\alpha%
\}_{\alpha<\beta, \,\alpha,\beta=1,2,3}\,,
\end{equation}
By the same token we can prove other embeddings, like $\mathrm{SL}(3,\mathbb{%
R})\times \mathrm{SU}^*(6)\subset E_{7(-5)}$ and $\mathrm{SL}(3,\mathbb{C}%
)\times \mathrm{SL}(3,\mathbb{R}) \subset E_{6(2)}$, see Fig. \ref{satake2}.
In the latter case there is a subtlety which is not apparent from the
truncation of the extended Satake diagram: The bottom-right diagram in Fig. %
\ref{satake2} would naively suggest that the roots $\alpha_1,\alpha_3,%
\alpha_5,\alpha_6$ define two commuting $\mathfrak{sl}(3,\mathbb{R})$
subalgebras. This is however not the case since, as represented by the lower
arrows, the conjugation $\sigma$ corresponding to the real form $\mathfrak{e}%
_{6(2)}$ inside the complex $\mathfrak{e}_{6}$, maps $\alpha_1$ and $%
\alpha_3 $ into $\alpha_1^\sigma=\alpha_6$ and $\alpha_3^\sigma=\alpha_5$,
respectively. As a consequence of this the $\mathfrak{e}_{6}$ shift
generators corresponding to the two orthogonal $\mathfrak{sl}(3,\mathbb{R})$
root spaces are mixed together in $\sigma$-invariant combinations inside $%
\mathfrak{e}_{6(2)}$, which make the shift generators of a $\mathfrak{sl}(3,%
\mathbb{C})$ subalgebra. This subalgebra also contains the two non-compact
combinations $H_{\alpha_1}+H_{\alpha_6},\,H_{\alpha_2}+H_{\alpha_5}$ and the
two compact combinations $i(H_{\alpha_1}-H_{\alpha_6}),\,i\,(H_{%
\alpha_2}-H_{\alpha_5})$ of the $\mathfrak{e}_{6}$ Cartan generators.

\begin{figure}[h]
\begin{center}
\centerline{\includegraphics[width=0.7\textwidth]{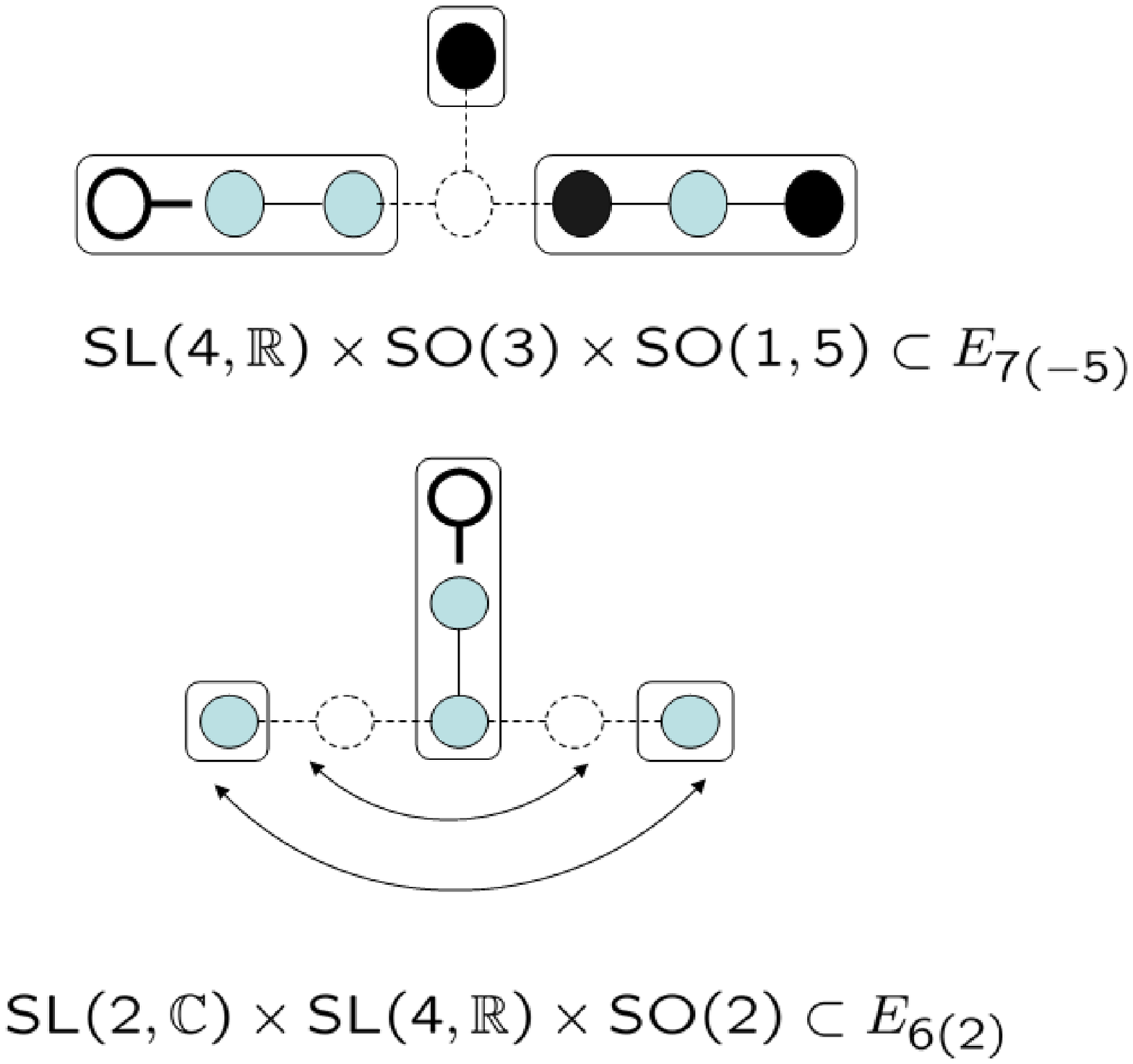}}
\end{center}
\caption{Embeddings $\mathrm{SL}(4,\mathbb{R})\times \mathrm{SO}(3)\times%
\mathrm{SO}(1,5) \subset E_{7(-5)}$ and $\mathrm{SL}(2,\mathbb{C})\times%
\mathrm{SL}(4,\mathbb{R})\times \mathrm{SO}(2) \subset E_{6(2)}$.}
\label{satake3}
\end{figure}
In Fig. \ref{satake3} the embeddings $\mathrm{SL}(4,\mathbb{R})\times
\mathrm{SO}(3)\times\mathrm{SO}(1,5) \subset E_{7(-5)}$ and $\mathrm{SL}(2,%
\mathbb{C})\times\mathrm{SL}(4,\mathbb{R})\times \mathrm{SO}(2) \subset
E_{6(2)}$ are illustrated.

\subsection{General Features}

\label{GF} One can generalize the above discussion and show that, as a
general feature of the embeddings considered in this work, the $\mathfrak{g}%
_N^3$ algebra, and its super-Ehlers subalgebra $\mathfrak{g}_N^D\oplus
\mathfrak{sl}(D-2)$ can be written in the forms:
\begin{align}
\mathfrak{g}_N^3&=\mathfrak{h}\oplus \mathfrak{l}_+\oplus\mathfrak{l}%
_-\oplus \mathfrak{m}_0\,\,;\,\,\,\, \mathfrak{g}_N^D\oplus \mathfrak{sl}%
(D-2)=\mathfrak{h}\oplus \hat{\mathfrak{l}}_+\oplus\hat{\mathfrak{l}}%
_-\oplus \mathfrak{m}_0\,.  \notag \\
\end{align}
Note that, as a consequence of the regularity of the embedding and
properties (\ref{rank1}), (\ref{rank2}), their Cartan subalgebras
\begin{align}
\mathfrak{h}&=\mathfrak{h}^{nc}\oplus \mathfrak{h}^{c}\,,
\end{align}
can be chosen to coincide, where $\mathrm{dim}(\mathfrak{h}^{nc})$ is the
non-compact rank of the two groups. This is implicit in Dynkin's
construction of the $\mathfrak{g}_N^D\oplus \mathfrak{sl}(D-2)$ algebra by
truncating the extended diagram of $\mathfrak{g}_N^3$. Moreover the
centralizer of $\mathfrak{h}^{nc}$, which is the compact algebra $\mathfrak{h%
}^c\oplus \mathfrak{m}_0$, is common to the two algebras:
\begin{equation}
\mathfrak{h}^c\oplus \mathfrak{m}_0\subset\mathfrak{g}_N^3\bigcap\left[%
\mathfrak{g}_N^D\oplus \mathfrak{sl}(D-2)\right]\,.
\end{equation}
For a split (maximally non-compact) $\mathfrak{g}_N^3$, $\mathfrak{h}^c=%
\mathfrak{m}_0=\varnothing$ and $\alpha^\sigma=\alpha$.

The nilpotent spaces $\mathfrak{l}_\pm,\,\hat{\mathfrak{l}}_\pm$ have the
form:
\begin{align}
\mathfrak{l}_\pm&=\mathrm{Span}\left[i\,(E_{\pm\alpha}-\sigma(E_{\pm%
\alpha})),\,E_{\pm\alpha}+\sigma(E_{\pm\alpha})\right]_{(\alpha,\,\alpha^%
\sigma)\in \bar{\Delta}_+[\mathfrak{g}_N^3]}\,,  \notag \\
\hat{\mathfrak{l}}_\pm&=\mathrm{Span}\left[i\,(E_{\pm\alpha}-\sigma(E_{\pm%
\alpha})),\,E_{\pm\alpha}+\sigma(E_{\pm\alpha})\right]_{(\alpha,\,\alpha^%
\sigma)\in \bar{\Delta}_+[\mathfrak{g}_N^3]\bigcap\Delta_+[\mathfrak{g}%
_N^D\oplus \mathfrak{sl}(D-2)]}\,,
\end{align}
where, as usual, $\bar{\Delta}_+[\mathfrak{g}_N^3]$ denotes the set of
positive roots of the (complexification of) $\mathfrak{g}_N^3$ with
non-trivial restriction to $\mathfrak{h}^{nc}$, and $\Delta_+[\mathfrak{g}%
_N^D\oplus \mathfrak{sl}(D-2)]$ the set of positive roots of the
(complexification of) $\mathfrak{g}_N^D\oplus \mathfrak{sl}(D-2)$, which is
a subset of ${\Delta}_+[\mathfrak{g}_N^3]$. Thus in general we have:
\begin{equation}
\hat{\mathfrak{l}}_\pm \subset \mathfrak{l}_\pm\,.
\end{equation}
We can then write the coset space as follows:
\begin{equation}
\mathfrak{g}_N^3\ominus [\mathfrak{g}_N^D\oplus \mathfrak{sl}(D-2)]=%
\mathfrak{N}^+\oplus \mathfrak{N}^-\,,  \label{Npm}
\end{equation}
where $\mathfrak{N}^\pm=\mathfrak{l}_\pm\ominus \hat{\mathfrak{l}}_\pm$.
Semisimplicity of $\mathfrak{g}_N^3$ and $\mathfrak{g}_N^D$ implies that $%
\mathrm{dim}(\mathfrak{l}_+)=\mathrm{dim}(\mathfrak{l}_-)$ and $\mathrm{dim}(%
\hat{\mathfrak{l}}_+)=\mathrm{dim}(\hat{\mathfrak{l}}_-)$, so that $\mathrm{%
dim}(\mathfrak{N}^+)=\mathrm{dim}(\mathfrak{N}^-)$. More precisely, in a
suitable basis, for each strictly-upper-triangular matrix $M_+$ representing
an element in $\mathfrak{N}^+$, its (strictly-lower-triangular)
hermitian-conjugate $M_-=M_+^\dagger=-\tau(M_+)$ represents an element in $%
\mathfrak{N}^-$: The former is given by a generator either of the form $%
i\,(E_{\alpha}-\sigma(E_{\alpha}))$ or $E_{\alpha}+\sigma(E_{\alpha})$, for
some $\alpha\in \bar{\Delta}_+[\mathfrak{g}_N^3]\ominus\bar{\Delta}_+[%
\mathfrak{g}_N^D\oplus \mathfrak{sl}(D-2)]$, the latter will either be $%
-i\,(E_{-\alpha}-\sigma(E_{-\alpha}))$ or $E_{-\alpha}+\sigma(E_{-\alpha})$,
corresponding to the same $\alpha$. Thus if $\{L^+_\ell\}$, $\ell=1,\dots,
\mathrm{dim}(\mathfrak{N}^\pm)$, is a basis of $\mathfrak{N}^+$, $%
\{L^-_\ell\}=\{-\tau(L^+_\ell)\}$ is a basis of $\mathfrak{N}^-$ and we can
also write the coset space in the form:
\begin{equation}
\mathfrak{g}_N^3\ominus [\mathfrak{g}_N^D\oplus \mathfrak{sl}(D-2)]=%
\mathfrak{N}^c\oplus \mathfrak{N}^{nc}\,,  \label{Npm2}
\end{equation}
where
\begin{equation}
\mathfrak{N}^{nc}=\mathrm{Span}(L^+_\ell+L^-_\ell)\,\,;\,\,\,\mathfrak{N}%
^{c}=\mathrm{Span}(L^+_\ell-L^-_\ell)\,,
\end{equation}
which are the eigenspaces of $\tau$ on $\mathfrak{N}^+\oplus \mathfrak{N}^-$
corresponding to the eigenvalues $-1$ and $+1$, respectively. These
subspaces define representations with respect to the compact group $\mathrm{%
SO}(D-2)\times\, mcs (G_N^D)$. With respect to the $G^3_N$-invariant scalar
product on $\mathfrak{g}_N^3$, $\mathfrak{N}^{c}$ and $\mathfrak{N}^{nc}$
have negative and positive signatures, respectively. Since
\begin{equation}
\mathrm{dim}(\mathfrak{N}^{c})=\mathrm{dim}(\mathfrak{N}^{nc})\,,
\end{equation}
the manifold $M_N^D$ in (\ref{Ggen}) has vanishing character, being
\begin{equation*}
c(M_N^D)=\mathrm{dim}(\mathfrak{N}^{c})=\mathrm{dim}(\mathfrak{N}%
^{nc})=nc(M_N^D)\,,
\end{equation*}
as also proven in Sect \ref{Hodge}. We shall come back on this issue in
Appendix \ref{pdld}.

\section{$\mathfrak{so}(8,8)$ Outer Automorphisms and Dual Subalgebras of $%
\mathfrak{e}_{8(8)}$}

\label{App-Spinor-Polarizations} 
\begin{figure}[h]
\begin{center}
\centerline{\includegraphics[width=0.7\textwidth]{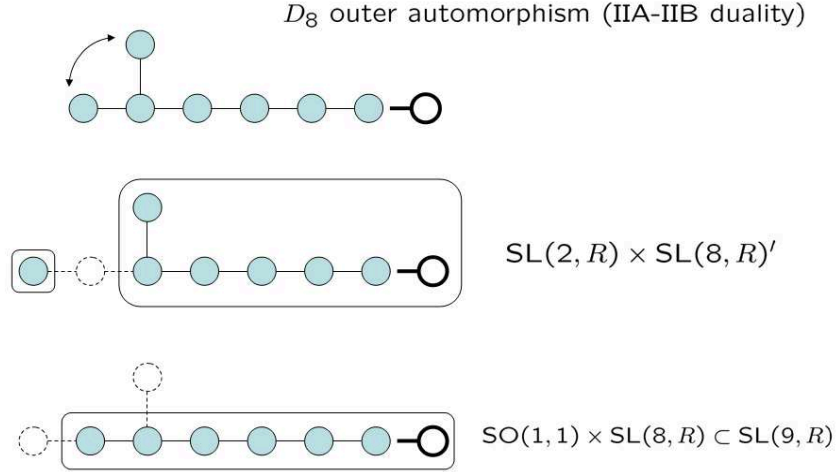}}
\end{center}
\caption{Outer automorphism of the $D_8$ subalgebra of $\mathfrak{e}_8$ and
two inequivalent $\mathfrak{sl}(8,\mathbb{R})$ subalgebras of $\mathfrak{e}%
_{8(8)}$.}
\label{sl8}
\end{figure}

Consider in the maximal $D=3$ theory the
effect of an $\mathrm{O}(8,8)$ ``reflection'' of the form:
\begin{equation}
\mathcal{O}_k=\left(%
\begin{matrix}
\mathbf{1}_8-\mathbf{D}_{k} & \mathbf{D}_{k}\cr {\bf D}_{k} & \mathbf{1}_8-%
\mathbf{D}_{k}%
\end{matrix}%
\right)\,,  \label{outaut}
\end{equation}
where each block is an $8\times 8$ matrix and $\mathbf{D}_{k}$ is the
zero-matrix except for only an \emph{odd number} $k$ of $1$s along the
diagonal. Such transformation, which belongs to the $\mathrm{O}(8)$ subgroup
of $\mathrm{O}(8,8)$, is an outer automorphism of the $D_8$ algebra whose
effect, modulo Weyl transformations of the same algebra, is to interchange $%
\alpha_2$ with $\alpha_3$ in Fig. \ref{e8}. While it is a symmetry of the $%
D_8$ Dynkin diagram, it is not a symmetry of the $\mathfrak{e}_{8(8)}$ one,
as it changes the $\mathrm{SO}(8,8)$-chirality of the $\alpha_1$ root, which
is a $D_8$-spinorial weight \cite{BT-1}. In particular this outer
automorphism may map inequivalent subalgebras $\mathfrak{g},\,\mathfrak{g}%
^{\prime }$ of $\mathfrak{so}(8,8)$ into one another. This is the case of
subalgebras $\mathfrak{g}$ (and thus $\mathfrak{g}^{\prime }$) which are the
direct sum of commuting $A_k$-algebras with odd rank $k$. In mathematical
language such \emph{dual} subalgebras are said to be \emph{linearly
equivalent}, i.e. in any matrix representation they are equivalent through
conjugation by means of a matrix, which is however not necessarily a
representation of an $\mathrm{SO}(8,8)$ element, as it is the case for the
outer automorphisms. Equivalence therefore implies linear equivalence though
the reverse implication is not true. With respect to $\mathfrak{g}$ and $%
\mathfrak{g}^{\prime }$, a same spinorial representation of $\mathfrak{so}%
(8,8)$, and thus the adjoint representation of the whole $\mathfrak{e}%
_{8(8)} $, will branch differently. They are clearly inequivalent $\mathfrak{%
e}_{8(8)}$-subalgebras. Examples are given in \cite{minchenko}: $\mathfrak{g}%
=\mathfrak{sl}(8),\,\mathfrak{sl}(6)\oplus\mathfrak{sl}(2),\,\mathfrak{sl}%
(4)\oplus\mathfrak{sl}(4),$ etc., see Fig \ref{sl8}.

What has been said for $\mathfrak{so}(8,8)$ also holds for the $\mathfrak{so}%
^*(16)$ and $\mathfrak{so}(16)$ subalgebras of $\mathfrak{e}_{8(8)}$. For
instance there are two inequivalent $\mathfrak{u}(8),\,\mathfrak{u}^{\prime
}(8)$ in either $\mathfrak{so}^*(16)$ or $\mathfrak{so}(16)$. One contains
the $R$-symmetry algebras $\mathfrak{su}(8),\,\mathfrak{usp}(8), $ etc. of
the higher dimensional parent maximal supergravities, the other dual
subalgebras $\mathfrak{su}^{\prime }(8),\,\mathfrak{usp}^{\prime }(8), $
etc. which are not contained in the chain of exceptional duality algebras $%
\mathfrak{e}_{7(7)},\,\mathfrak{e}_{6(6)}$ etc.

Let us briefly recall the relation between outer automorphisms of $\mathfrak{%
so}(8,8)$ and dualities. Consider the toroidal reduction of the $D=11$
theory down to $D=3$ (in the Einstein frame). The Kaluza-Klein ansatz for
the metric reads:
\begin{equation}
G^{(11)}_{\hat{\mu}\hat{\nu}}=\left(%
\begin{matrix}
e^{2 \xi}\,g^{(3)}_{\mu\nu}+G_{pq}\,G^p_\mu\,G^q_\nu & G_{np}\,G^p_\mu\cr %
G_{mp}\,G^p_\nu & G_{mn}%
\end{matrix}%
\right)\,\,;\,\,\,\xi=-\frac{1}{2}\,\log\left(\mathrm{det}(G_{mn})\right)\,,
\end{equation}
where $\hat{\mu},\,\hat{\nu}=0,\dots,10$, $\mu,\nu=0,1,2$, $m,n=3,\dots, 10$
and the internal metric of $T^8$ is conveniently written as follows:
\begin{equation}
\mathbf{G}=(G_{mp})=\mathbf{E}\mathbf{E}^T=\hat{\mathbf{E}}\mathbf{D}^2\hat{%
\mathbf{E}}^T\,,
\end{equation}
where $\mathbf{E}=(E_m{}^a)$ is the vielbein of the coset $\mathrm{GL}(8,%
\mathbb{R})/\mathrm{SO}(8)$, $a=3,\dots,10 $, written as the product of a
matrix $\hat{\mathbf{E}}$ which only depends on the axionic moduli
associated with the off-diagonal components of the metric times the diagonal
matrix $\mathbf{D}=(D_m{}^a)=(e^{\sigma_a}\,\delta_m^a)$. The exponentials $%
e^{\sigma_a}$ can be viewed as the internal radii $R_a$. The bosonic section
of the $D=3$ Lagrangian reads\footnote{%
We adopt the mostly plus signature for the metric.}:
\begin{align}
e^{-1}\,\mathcal{L}_3&=\frac{R}{2}- \frac{1}{2}\,\partial_\mu \vec{h}%
\cdot\partial^\mu\vec{h}-\frac{1}{2}\sum_{a<b}e^{2(\sigma_b-\sigma_a)}
P_{\mu a}{}^b P^\mu{}_{a}{}^b-\frac{1}{2}\sum_a e^{-2(\sigma_a-\xi)}\,F_{\mu
a}F^\mu{}_{ a}-  \notag \\
&-\frac{1}{4}\sum_{a,b} e^{2(\sigma_a+\sigma_b+\xi)}\,F_{\mu}{}^{ab}F^{\mu
ab}-\frac{1}{12}\sum_{a,b,c} e^{-2(\sigma_a+\sigma_b+\sigma_c)}\,F_{\mu
abc}F^\mu{}_{ abc}\,,
\end{align}
where $P_{\mu a}{}^b\equiv (\hat{\mathbf{E}}^{-1}\partial_\mu \hat{\mathbf{E}%
})_a{}^b$ and the dialtonic vector $\vec{h}$ has the following form in the $%
(\epsilon_i)$ orthonormal basis:
\begin{equation}
\vec{h}=\sum_{a=3}^9 \left(\sigma_a+\frac{\sigma_{10}}{2}\right)\,%
\epsilon_{a-2}+\left(\frac{\sigma_{10}}{2}+\sum_{a=3}^9
\sigma_a\right)\epsilon_{8}\,.
\end{equation}
The field strengths $F_{\mu a}$ and $F_{\mu}{}^{ab}$ are associated with the
scalars $\chi_n$ and $\chi^{mn}$ dual in $D=3$ to the vectors $G^m_\mu$ and $%
A_{\mu mn}$ respectively, while $F_{\mu abc}$ is the one pertaining to the
scalars $A_{mnp}$. In these conventions, the lower (or upper) internal $%
\mathrm{SO}(8)$-indices $a,b,c$ of these field strengths are related to the $%
\mathrm{SL}(8,\mathbb{R})$ indices $m,n,p$ by means of $\hat{\mathbf{E}}$
(or $\hat{\mathbf{E}}^{-1}$). For instance:
\begin{equation}
F_{\mu abc}=\hat{E}_a{}^{m}\hat{E}_b{}^{n}\hat{E}_c{}^{p}\,F_{\mu
mnp}\,\,;\,\,\,F_{\mu mnp}=\partial_\mu A_{mnp} \,.
\end{equation}
The above Lagrangian can also be written in the more compact form:
\begin{align}
e^{-1}\,\mathcal{L}_3&=\frac{R}{2}-\frac{1}{2}\,\partial_\mu \vec{h}%
\cdot\partial^\mu\vec{h}-\frac{1}{2}\sum_{\alpha\in \Delta_+[\mathfrak{e}%
_{8(8)}]} e^{-2 \alpha\cdot \vec{h}}\,\Phi_\mu^{(\alpha)}\Phi^{\mu\,(%
\alpha)}\,,  \label{lag3fin}
\end{align}
where the one-forms $\Phi_\mu^{(\alpha)}$ are associated with each of the $%
\mathfrak{e}_{8(8)}$-positive roots $\alpha$ \cite%
{Andrianopoli:1996bq,Andrianopoli:1996zg,Cremmer:1997ct}.\footnote{%
The representation (\ref{lag3fin}) of the $D=3$ Lagrangian applies to all $%
D=3$ supergravities. In the general (non necessarily maximal) case, $\vec{h}$
is a suitable dilaton-dependent vector in the $\mathfrak{h}^{nc}$ subspace
of the Cartan subalgebra of $\mathfrak{g}^3_N$, while $\alpha$ are the
restrictions to $\mathfrak{h}^{nc}$ of the $\mathfrak{g}^3_N$ positive roots
(\emph{restricted roots}, see \cite{Helgason}).} It is useful to express the
various radial moduli $\sigma_a$ in terms of the corresponding fields $\hat{%
\sigma}_a$ in the $D=10$ \emph{string frame}:
\begin{equation}
\sigma_a=\hat{\sigma}_a-\frac{\phi}{3}\,,\,\,a=3,\dots,
9\,\,\,;\,\,\,\,\sigma_{10}=\frac{2}{3}\,\phi\,,
\end{equation}
we find:
\begin{equation}
\vec{h}=\sum_{i=1}^8\,h_i\,\epsilon_i=\sum_{a=3}^9\hat{\sigma}%
_a\,\epsilon_{a-2}+\left(-2\phi+\sum_{a=3}^9 \hat{\sigma}_a\right)%
\epsilon_{8}\,.
\end{equation}
The outer automorphism $\mathcal{O}_k$ in (\ref{outaut}) has the effect of
changing the sign to an odd number of $\epsilon_a$, or, equivalently, to
their coefficients in $\vec{h}$:
\begin{equation}
\epsilon_{i_\ell}\rightarrow -\epsilon_{i_\ell}\,\,\,;\,\,\,\ell=1,\dots,
k\,.
\end{equation}
To see this let us consider the effect of $\mathcal{O}_k$ on the dilatonic
part of the coset representative of $\mathrm{O}(8,8)/[\mathrm{O}(8)\times
\mathrm{O}(8)]$, which has the following form:
\begin{equation}
\mathbf{D}(\vec{h})=\left(%
\begin{matrix}
(e^{h_i}\delta_i{}^j) & \mathbf{0}\cr {\bf 0} & (e^{-h_i}\delta_i{}^j)%
\end{matrix}%
\right)\,.
\end{equation}
We see that:
\begin{equation}
\mathcal{O}_k^{-1}\mathbf{D}(\vec{h})\mathcal{O}_k=\mathbf{D}(\vec{h}%
^{\prime })\,,
\end{equation}
where $h_{i_\ell}^{\prime }=-h_{i_\ell},\,h_{i\neq i_\ell}^{\prime
}=h_{i\neq i_\ell}$, $\ell=1,\dots, k$. If $i_\ell$ run between $1$ and $7$,
this transformation amounts to a $T$-duality along the internal directions $%
y^{i_\ell+2}$ \cite{Lu:1996ge,BT-1}:
\begin{equation}
R^{\prime }_{i_\ell+2}=e^{\hat{\sigma}^{\prime }_{i_\ell+2}}=e^{-\hat{\sigma}%
_{i_\ell+2}}=\frac{1}{R_{i_\ell+2}}\,\,;\,\,\,\phi^{\prime }=\phi-
\sum_{\ell=1}^k\hat{\sigma}_{i_\ell+2}\,.
\end{equation}
These transformations map type IIA into type IIB theory. If $k=1$ and $%
i_\ell=8$ then there is an $S$-duality involved: $\hat{\sigma}^{\prime }_i=%
\hat{\sigma}_i$ and $\phi^{\prime }=-\phi+\sum_{a=3}^9\hat{\sigma}_{a}$.

Instead of considering inequivalent T-dual subalgebras $\mathfrak{g},\,%
\mathfrak{g}^{\prime }\subset \mathfrak{so}(8,8)$ within a same $\mathfrak{e}%
_{8(8)}$ algebra, we may adopt an equivalent point of view and consider a
same subalgebra $\mathfrak{g}\subset \mathfrak{so}(8,8)$ within two $%
\mathfrak{e}_{8(8)}$ algebras, called in \cite{BT-1} $\mathfrak{e}^+_{8(8)}$
and $\mathfrak{e}^-_{8(8)}$,\footnote{%
Actually in \cite{BT-1} only the $D=4$ theory was considered, the $T$%
-duality group being $\mathrm{O}(6,6)$ in this case, and the algebras $%
\mathfrak{e}^\pm_{7(7)}$ defined.} defined respectively by completing the $%
\mathfrak{so}(8,8)$ Dynkin diagram with spinorial weights of different
chiralities, namely attaching the weight $\alpha_1$ to $\alpha_3$, as in
Fig. \ref{e8}, or a weight $\alpha_1^{\prime }$ to $\alpha_2$, defined as
follows:
\begin{equation}
\alpha_1=-\frac{1}{2}\,\left(\sum_{i=1}^8\epsilon_i\right)+\epsilon_7+%
\epsilon_8\,\,\overset{\mbox{T-duality along $y^9$}}{\longrightarrow}
\alpha_1^{\prime }=-\frac{1}{2}\,\left(\sum_{i=1}^8\epsilon_i\right)+%
\epsilon_8\,.
\end{equation}
This is useful if, for instance, we fix the $\mathfrak{g}=\mathfrak{gl}(8,%
\mathbb{R})\subset \mathfrak{so}(8,8)$ group to be the same in the type IIA
and type IIB settings. Then the different $\mathfrak{gl}(8,\mathbb{R})$%
-weights defining the dimensionally reduced type IIA and type IIB forms are
obtained by branching the adjoint representations of $\mathfrak{e}^+_{8(8)}$
and $\mathfrak{e}^-_{8(8)}$, respectively, with respect to the common $%
\mathfrak{gl}(8,\mathbb{R})$, \cite{BT-1}.

The doubling of the equivalence classes inside a $D_n$ algebra into dual
pairs, discussed above, does not occur if the subalgebra is the sum of
commuting algebras in the case in which either all of them are of type $A_k$
with \emph{even} rank $k$, or at least one of them is of type $D$ \cite{D-1}%
. This is consistent with the fact observed in Subsect. \ref{maxim}, that
the $\mathrm{SL}(7,\mathbb{R})$ $D=9$ Ehlers subgroups of $\mathrm{SO}(8,8)$
which pertain to the type IIA and IIB descriptions are equivalent. The same
rule guarantees that, in $D=6$, the $\mathrm{SO}(5,5)\times \mathrm{SL}(4,%
\mathbb{R})$ subgroups of $\mathrm{SO}(8,8)$ in the type IIA and IIB
settings, are equivalent.

\section{Poincar\'{e} Duality and Level Decomposition}

\label{pdld} Consider now the branching of the adjoint representation of $%
\mathfrak{g}_N^3$ with respect to $\mathrm{SL}(D-2)\times G^D_N$:
\begin{eqnarray}
\mathbf{Adj}_{G_N^3}&\rightarrow & (\mathbf{Adj}_{\mathrm{SL}(D-2)},\mathbf{1%
})\oplus (\mathbf{1},\mathbf{Adj}_{G_N^D})\bigoplus_d\,\mathfrak{N}_d\,,
\notag \\
\mathfrak{N}_d&=&\left[(\Lambda^d,\mathcal{R}_d)\oplus (*\Lambda^d,\mathcal{R%
}^{\prime }_d)\right]\,,  \label{Npm3}
\end{eqnarray}
where it is understood that if $(\Lambda^d,\mathcal{R}_d)=(*\Lambda^d,%
\mathcal{R}^{\prime }_d)$, they are counted just once in $\mathfrak{N}_d$.
In light of our discussion in Appendix \ref{GF}, we can write the coset
space as the carrier of a representation $\bigoplus_d\,\mathfrak{N}_d$,
namely rewrite eq. (\ref{Npm}) as follows:
\begin{equation}
\mathfrak{g}_N^3\ominus (\mathfrak{g}^D_N\oplus \mathfrak{sl}(D-2))=%
\mathfrak{N}^+\oplus \mathfrak{N}^-=\bigoplus_d\,\mathfrak{N}_d\,.
\end{equation}
In fact each subspace $\mathfrak{N}_d$ splits into conjugate nilpotent
subalgebras as follows:
\begin{equation}
\mathfrak{N}_d=\mathfrak{N}_d^+\oplus \mathfrak{N}_d^-\,\,,\,\,\,\mathfrak{N}%
_d^+=\mathfrak{N}_d\bigcap \mathfrak{N}^+\,\,\,,\,\,\,\,\mathfrak{N}_d^-=%
\mathfrak{N}_d\bigcap \mathfrak{N}^-=\tau(\mathfrak{N}_d^+) \,,
\end{equation}
this being a consequence of the property: $\tau(\mathfrak{N}_d)=\mathfrak{N}%
_d$. Each nilpotent subalgebra $\mathfrak{N}_d^+$ or $\mathfrak{N}_d^-$
separately defines a representation with respect to (the adjoint action of) $%
G^D_N$ and the subgroup $\mathrm{GL}(D-3)\subset \mathrm{SL}(D-2)$, though
not with respect to $\mathrm{SL}(D-2)$ itself. We can decompose each space $%
\mathfrak{N}_d$ into eigenspaces of the Cartan involution $\tau$, consisting
of compact and non-compact generators:
\begin{align}
\mathfrak{N}_d &=\mathfrak{N}_d^{c}\oplus \mathfrak{N}_d^{nc}\,\,\,,\,\,\,%
\mathfrak{N}_d^{c}=\mathfrak{N}^{c}\bigcap\mathfrak{N}_d\,\,\,,\,\,\,
\mathfrak{N}_d^{nc}=\mathfrak{N}^{nc}\bigcap\mathfrak{N}_d\,.
\end{align}
These subspaces define representations with respect to the compact group $%
\mathrm{SO}(D-2)\times\, mcs (G_N^D)$ and, moreover
\begin{equation}
\mathrm{dim}(\mathfrak{N}_d^{c})=\mathrm{dim}(\mathfrak{N}_d^{nc})\,.
\end{equation}
For the sake of simplicity, let us consider a split (maximally non-compact) $%
\mathfrak{g}_N^3$. Then each $\mathfrak{N}_d$ will be generated by shift
operators corresponding to a certain set of positive roots $\alpha^{(d)}$
and their negatives:
\begin{equation}
\mathfrak{N}_d=\mathrm{Span}(E_{\alpha^{(d)}},\,E_{-\alpha^{(d)}})_{%
\alpha^{(d)}\in \Delta_+[\mathfrak{g}_N^3]}\,,
\end{equation}
and the conjugate nilpotent subalgebras are $\mathfrak{N}_d^+=\mathrm{Span}%
(E_{\alpha^{(d)}})$ and $\mathfrak{N}_d^-=\mathrm{Span}(E_{-\alpha^{(d)}})$.
The eigenspaces $\mathfrak{N}_d^{nc},\,\mathfrak{N}_d^{c}$ of the Cartan
involution, consisting of compact and non-compact generators read:
\begin{align}
\mathfrak{N}_d^{c}&=\mathrm{Span}(E_{\alpha^{(d)}}-E_{-\alpha^{(d)}})_{%
\alpha^{(d)}\in \Delta_+[\mathfrak{g}_N^3]}\,\,;\,\,\,\, \mathfrak{N}_d^{nc}=%
\mathrm{Span}(E_{\alpha^{(d)}}+E_{-\alpha^{(d)}})_{\alpha^{(d)}\in \Delta_+[%
\mathfrak{g}_N^3]}\,.
\end{align}
Each positive root $\alpha^{(d)}$ corresponds to a $D=3$ scalar field in the
Lagrangian (\ref{lag3fin}). For a given $d$ the roots $\alpha^{(d)}$ are
defined by the \emph{level decomposition} of the $\mathfrak{g}_N^3$-roots
with respect to the root which is truncated out of its extended diagram in
order to define the $\mathfrak{g}^D_N\oplus \mathfrak{sl}(D-2)$-subdiagram.%
\footnote{%
In the non-split case, one should consider the level decomposition of the
restricted roots. Level decompositions are a common procedure in the $E_{10}$
and $E_{11}$ approaches to maximal supergravity \cite%
{Kleinschmidt:2004dy,Bergshoeff:2006qw}.}

Let us illustrate this procedure in the maximal theory. As shown in Appendix %
\ref{App-Embeddings}, the $\mathfrak{e}_{11-D(11-D)}\times \mathfrak{sl}%
(D-2) $ diagram is obtained by deleting from the $\mathfrak{e}_{8(8)}$
-extended Dynkin diagram the root $\alpha_{12-D}$. The $\mathfrak{sl}(D-2)$
subalgebra is defined by the simple roots $\alpha_{13-D},\dots,
\alpha_8,\,-\psi$, $\psi=\epsilon_1+\epsilon_8$ being the $\mathfrak{e}%
_{8(8)}$ highest root, while its $\mathfrak{gl}(D-3)$ subalgebra only by the
roots $\alpha_{13-D},\dots, \alpha_8$. Writing a generic $\mathfrak{e}%
_{8(8)} $ positive root in the simple root basis:
\begin{equation}
\alpha=\sum_{i=1}^8 n_i\,\alpha_i\,,
\end{equation}
the positive integer $n_i$ defines the \emph{level} of $\alpha$ with respect
to $\alpha_i$. Let us consider the level-decomposition with respect to the
root $\alpha_{12-D}$ for dimensions $D<9$, namely the values of $n_{12-D}$
defining the roots $\alpha^{(d)}$.\footnote{%
More precisely, the level $n^i$ is the grading of the generator $E_\alpha$
with respect to the $\mathrm{SO}(1,1)$ generator $H_{\lambda^i}$ (i.e. $%
[H_{\lambda^i},E_\alpha]=n^i\,E_\alpha$), $\lambda^i$ being the $\mathfrak{g}%
_N^3$ simple weights. The level decomposition is defined by the Cartan
generator which is orthogonal to the Cartan subalgebra of $\mathfrak{g}%
^D_N\oplus \mathfrak{gl}(D-3)$ (and therefore commutes with $\mathfrak{g}%
^D_N\oplus \mathfrak{gl}(D-3)$). In the maximal theory, for $D<9$, the
relevant Cartan generator is $H_{\lambda^{12-D}}$ and thus the level to
consider is $n_{12-D}$. For $D=9$ the generator is $H_{\lambda^{2}}+H_{%
\lambda^{3}}$ and so we shall consider the decomposition with respect to the
integer $n=n_2+n_3$. In the type IIA $D=10$ description, the generator is $%
H_{\lambda^{1}}+2\,H_{\lambda^{2}}$ and the decomposition will be effected
with respect to $n=n_1+2\, n_2$.}

\paragraph{$D=4$.}

In the case of $D=4$ we have $63$ roots with $n_8=0$, corresponding to the $%
\mathfrak{e}_{7(7)}$-positive roots. The level $n_8=1$ roots are $56$ and
are the $\alpha^{(1)}$-roots whose shift generators $E_{\pm\alpha^{(1)}}$
define the carrier space of the $\mathfrak{N}_{d=1}=(\mathbf{1},\mathbf{56})$
representation. The level $n_8=2$ root defines, with its negative, the shift
generators in the quotient $\mathfrak{sl}(D-2)\ominus \mathfrak{gl}(D-3)=%
\mathfrak{sl}(2)\ominus \mathfrak{gl}(1)$, which are the two shift
generators of the Ehlers group.

\paragraph{$D=5$.}

Consider now the $D=5$ case. There are $37$ level-$n_7=0$ roots
corresponding to the positive roots of $\mathfrak{e}_{6(6)}\oplus \mathfrak{%
gl}(2)$. The level-$n_7=1$ roots are $54$ and define in $\mathfrak{N}%
_{d=1}^+ $ a subspace in the $(\mathbf{2},\mathbf{{27})}$-representation of $%
\mathrm{SL}(D-3)\times E_{6(6)}=\mathrm{SL}(2)\times E_{6(6)}$, while the $%
27 $ level-$n_7=2$ roots define a subspace in the $(\mathbf{1},\mathbf{{27}%
^{\prime })}$ with respect to the same group. The space $\mathfrak{N}%
_{d=1}^- $ will be the carrier of the conjugate representations. Together,
the level $n_7=1,2$ roots and their negatives define the space $\mathfrak{N}%
_{d=1}=\mathfrak{N}_{d=1}^+\oplus \mathfrak{N}_{d=1}^-=(\mathbf{3},\mathbf{{%
27})\oplus (3^{\prime },{27}^{\prime })}$, and are collectively denoted by $%
\alpha^{(1)} $. Finally the $2$ level-$n_7=3$ roots, with their negative,
define the generators of the coset $\mathfrak{sl}(D-2)\ominus \mathfrak{gl}%
(D-3)=\mathfrak{sl}(3)\ominus \mathfrak{gl}(2)$.

\paragraph{$D=6$.}

As far as the $D=6$ case is concerned, the $23$ level-$n_6=0$ roots are
positive roots of $\mathfrak{gl}(3)\oplus \mathfrak{so}(5,5) $, while the $%
48 $ level-$n_6=1$ and the $16$ level-$n_6=3$ roots define generators in $%
\mathfrak{N}_{d=1}^+$ transforming in the $(\mathbf{3},\mathbf{{16})}$ and $(%
\mathbf{1},\mathbf{{16}^{\prime })}$ of $\mathrm{SL}(D-3)\times \mathrm{SO}%
(5,5)=\mathrm{SL}(3)\times \mathrm{SO}(5,5)$, respectively. These are the $%
\alpha^{(1)}$ roots, which, together with their negatives, define the $%
\mathfrak{N}_{d=1}=(\mathbf{4},\mathbf{{16})\oplus (4^{\prime },{16}^{\prime
})}$ space. The roots $\alpha^{(2)}$ ($d=2$) are $30$ and have $n_6=2$. The
corresponding space $\mathfrak{N}_{d=2}^+$ is the carrier of a $(\mathbf{3},%
\mathbf{{10})}$ representation of $\mathrm{SL}(3)\times \mathrm{SO}(5,5)$,
while $E_{\pm \alpha^{(2)}}$ generate the $\mathfrak{N}_{d=2}=(\mathbf{6},%
\mathbf{{10})}$. Finally the $3$ level-$n_6=4$ roots, with their negative,
define the generators of the coset $\mathfrak{sl}(D-2)\ominus \mathfrak{gl}%
(D-3)=\mathfrak{sl}(4)\ominus \mathfrak{gl}(3)$.

A similar pattern occurs in the higher-$D$ cases.

\paragraph{$D=7$.}

For $D=7$, we have $16$ roots with $n_5=0$ which are the roots of $\mathfrak{%
gl}(4)\oplus \mathfrak{sl}(5) $. The $40$ $n_5=1$ and the $10$ $n_5=4$ roots
define subspaces of $\mathfrak{N}_{d=1}^+$ in the $(\mathbf{4},\mathbf{10}%
^{\prime }) $ and $(\mathbf{1},\mathbf{10})$ of $\mathrm{SL}(D-3)\times
\mathrm{SL}(5)=\mathrm{SL}(4)\times \mathrm{SL}(5)$, respectively. These are
the $\alpha^{(1)}$-roots and the space $\mathfrak{N}_{d=1}=\mathfrak{N}%
_{d=1}^+\oplus \mathfrak{N}_{d=1}^-$ is the carrier of the representation $(%
\mathbf{5},\mathbf{10}^{\prime })\oplus (\mathbf{5}^{\prime },\mathbf{10}) $
of $\mathrm{SL}(5)\times \mathrm{SL}(5)$. The $\alpha^{(2)}$-roots consist
in the $30$ level-$n_5=2$ and the $20$ level-$n_5=3$ roots defining the
representations $(\mathbf{6},\mathbf{5}) $ and $(\mathbf{4}^{\prime },%
\mathbf{5}^{\prime })$ of $\mathrm{SL}(4)\times \mathrm{SL}(5)$ in $%
\mathfrak{N}_{d=2}^+$, respectively. The space $\mathfrak{N}_{d=2}=\mathfrak{%
N}_{d=2}^+\oplus \mathfrak{N}_{d=2}^-$ then defines the representation $(%
\mathbf{10},\mathbf{5})\oplus (\mathbf{10}^{\prime },\mathbf{5}^{\prime }) $
of $\mathrm{SL}(5)\times \mathrm{SL}(5)$. Finally the $4$ level-$n_5=5$
roots, with their negative, define the generators of the coset $\mathfrak{sl}%
(D-2)\ominus \mathfrak{gl}(D-3)=\mathfrak{sl}(5)\ominus \mathfrak{gl}(4)$.

\paragraph{$D=8$.}

In the $D=8$ case the $14$ $n_4=0$ roots are the positive roots of $%
\mathfrak{gl}(5)\oplus \mathfrak{sl}(2)\oplus \mathfrak{sl}(3) $. The $%
\alpha^{(1)}$s consist of the $30$ $n_4=1$ and the $6$ $n_4=5$ roots
defining the $\mathrm{SL}(5)\times \mathrm{SL}(2)\times \mathrm{SL}(3)$%
-representations $(\mathbf{5},\mathbf{2},\mathbf{3}^{\prime })\oplus (%
\mathbf{1},\mathbf{2},\mathbf{3})$ in $\mathfrak{N}_{d=1}^+$ which, together
with the conjugate representations in $\mathfrak{N}_{d=1}^-$, complete the $%
\mathfrak{N}_{d=1}=(\mathbf{6},\mathbf{2},\mathbf{3}^{\prime })\oplus (%
\mathbf{6}^{\prime },\mathbf{2},\mathbf{3})$ of $\mathrm{SL}(6)\times
\mathrm{SL}(2)\times \mathrm{SL}(3)$. The $\alpha^{(2)}$s are defined by the
$30$ $n_4=2$ and $15$ $n_4=4$ roots, corresponding to the representation $(%
\mathbf{10},\mathbf{1},\mathbf{3})\oplus (\mathbf{5}^{\prime },\mathbf{1},%
\mathbf{3}^{\prime })$ in $\mathfrak{N}_{d=2}^+$, so that $\mathfrak{N}%
_{d=2}=\mathfrak{N}_{d=2}^+\oplus \mathfrak{N}_{d=2}^-=(\mathbf{15},\mathbf{1%
},\mathbf{3})\oplus (\mathbf{15}^{\prime },\mathbf{1},\mathbf{3}^{\prime })$
of $\mathrm{SL}(6)\times \mathrm{SL}(2)\times \mathrm{SL}(3)$. The $%
\alpha^{(3)}$ roots are the 20 ones with $n_4=3$. They define the $\mathfrak{%
N}_{d=3}^+$ space in the $(\mathbf{10},\mathbf{2},\mathbf{1})$ of $\mathrm{SL%
}(5)\times \mathrm{SL}(2)\times \mathrm{SL}(3)$ which, together with $%
\mathfrak{N}_{d=3}^-$, complete the $\mathfrak{N}_{d=3}=(\mathbf{20},\mathbf{%
2},\mathbf{1})$ of $\mathrm{SL}(6)\times \mathrm{SL}(2)\times \mathrm{SL}(3)$%
. The remaining 5 roots with $n_4=6$, with their negative, define the
generators of the coset $\mathfrak{sl}(D-2)\ominus \mathfrak{gl}(D-3)=%
\mathfrak{sl}(6)\ominus \mathfrak{gl}(5)$.

\paragraph{$D=9$.}

The same analysis applies to $D=9$, although in this case we shall consider
the sum $n=n_2+n_3$. There are 16 roots with $n=0$, which are the positive
roots of the algebra $\mathfrak{gl}(6)\oplus \mathfrak{gl}(2) $. The $%
\alpha^{(1)}$ roots consist of the 18 with $n=1$ and the $3$ with $n=6$,
defining the $\mathrm{SL}(6)\times \mathrm{GL}(2)$-representations $(\mathbf{%
6},\mathbf{2}_{+3}+\mathbf{1}_{-4})\oplus (\mathbf{1},\mathbf{2}_{-3}+%
\mathbf{1}_{+4})$ in $\mathfrak{N}_{d=1}^+$ which, together with the
conjugate representations in $\mathfrak{N}_{d=1}^-$, complete the $\mathfrak{%
N}_{d=1}=(\mathbf{7},\mathbf{2}_{+3}+\mathbf{1}_{-4})\oplus (\mathbf{7}%
^{\prime },\mathbf{2}_{-3}+\mathbf{1}_{+4})$ of $\mathrm{SL}(7)\times
\mathrm{GL}(2)$. The $\alpha^{(2)}$s are the 30 roots with $n=2$ and the 12
with $n=5$ defining the $\mathrm{SL}(6)\times \mathrm{GL}(2)$%
-representations $(\mathbf{15},\mathbf{2}_{-1})\oplus (\mathbf{6}^{\prime },%
\mathbf{2}_{+1})$ in $\mathfrak{N}_{d=2}^+$, so that $\mathfrak{N}_{d=2}=%
\mathfrak{N}_{d=2}^+\oplus \mathfrak{N}_{d=2}^-=(\mathbf{21},\mathbf{2}%
_{-1})\oplus (\mathbf{21}^{\prime },\mathbf{2}_{+1})$. Next we have to
consider the 20 $n=3$ and the $15$ $n=4$ roots which make the $\alpha^{(3)}$
and define the $\mathrm{SL}(6)\times \mathrm{GL}(2)$-representations $(%
\mathbf{20},\mathbf{1}_{+2})\oplus (\mathbf{15}^{\prime },\mathbf{1}_{-2})$
in $\mathfrak{N}_{d=3}^+$, so that $\mathfrak{N}_{d=3}=\mathfrak{N}%
_{d=3}^+\oplus \mathfrak{N}_{d=3}^-=(\mathbf{35},\mathbf{1}_{+2})\oplus (%
\mathbf{35}^{\prime },\mathbf{1}_{-2})$. Finally the 6 $n=7$ roots, with
their negative, define the generators of the coset $\mathfrak{sl}%
(D-2)\ominus \mathfrak{gl}(D-3)=\mathfrak{sl}(7)\ominus \mathfrak{gl}(6)$.

\paragraph{$D=10$, IIB.}

In the $D=10$ case we have to distinguish between the type IIA and IIB
theories. In the type IIB setting we need consider the level $n_3$ with
respect to $\alpha_3$. The $22$ roots with $n_3=0$ are the positive roots of
$\mathfrak{gl}(7)\oplus \mathfrak{sl}(2)$, $\mathfrak{sl}(2)$ being the $U$%
-duality group. In this case we only have $d=2,4$. The $\alpha^{(2)}$s
consist of the 42 $n_3=1$ and the 14 $n_3=3$ defining the $\mathrm{SL}%
(7)\times \mathrm{SL}(2)$-representations $(\mathbf{21},\mathbf{2})\oplus (%
\mathbf{7}^{\prime },\mathbf{2})$ in $\mathfrak{N}_{d=2}^+$, so that $%
\mathfrak{N}_{d=2}=\mathfrak{N}_{d=2}^+\oplus \mathfrak{N}_{d=2}^-=(\mathbf{%
28},\mathbf{2})\oplus (\mathbf{28}^{\prime },\mathbf{2})$ of $\mathrm{SL}%
(8)\times \mathrm{SL}(2)$. Next we have the 35 roots with $n_3=2$, which are
the $\alpha^{(4)}$s and define the $(\mathbf{35},\mathbf{1})$ of $\mathrm{SL}%
(7)\times \mathrm{SL}(2)$ in $\mathfrak{N}_{d=4}^+$, so that $\mathfrak{N}%
_{d=4}=\mathfrak{N}_{d=4}^+\oplus \mathfrak{N}_{d=4}^-=(\mathbf{70},\mathbf{1%
})$. There are 7 $n_3=4$ roots which, with their negative, define the
generators of the coset $\mathfrak{sl}(D-2)\ominus \mathfrak{gl}(D-3)=%
\mathfrak{sl}(8)\ominus \mathfrak{gl}(7)$.

\paragraph{$D=10$, IIA.}

As far as the type IIA description is concerned, the level to consider for
the decomposition is the sum $n=n_1+2\,n_2$. In this case we only have $%
d=1,2,3$. There are 21 $n=0$ roots which are the positive roots of $%
\mathfrak{gl}(7)\oplus \mathfrak{so}(1,1)$, $\mathfrak{so}(1,1)$ being the $%
U $-duality algebra. The $\alpha^{(1)}$ roots consist of the 7 roots with $%
n=1$ and the single $n=7$ root defining the $\mathrm{SL}(7)\times \mathrm{SO}%
(1,1) $-representation $\mathbf{7}_{+3}\oplus \mathbf{1}_{-3}$ in $\mathfrak{%
N}_{d=1}^+$ which, together with the conjugate representations in $\mathfrak{%
N}_{d=1}^-$, complete the $\mathfrak{N}_{d=1}=\mathbf{8}_{+3}\oplus \mathbf{8%
}_{-3}$ of $\mathrm{SL}(8)\times \mathrm{SO}(1,1)$. Next we consider the $21$
roots with $n=2$ and the $7$ with $n=6$, whose shift operators generating $%
\mathfrak{N}_{d=2}^+$ transform in the $\mathbf{21}_{-2}\oplus \mathbf{7}%
^{\prime }_{+2}$ with respect to $\mathrm{SL}(7)\times \mathrm{SO}(1,1)$.
These roots define then the $\alpha^{(2)}$ and $\mathfrak{N}_{d=2}=\mathfrak{%
N}_{d=2}^+\oplus \mathfrak{N}_{d=2}^-=\mathbf{28}_{-2}\oplus \mathbf{28}%
^{\prime }_{+2}$ of $\mathrm{SL}(8)\times \mathrm{SO}(1,1)$. The $%
\alpha^{(3)}$s consist of the 35 $n=3$ and the $21$ $n=6$ roots defining the
$\mathrm{SL}(7)\times \mathrm{SO}(1,1)$-representation $\mathbf{35}%
_{+1}\oplus \mathbf{21}^{\prime }_{-1}$ in $\mathfrak{N}_{d=3}^+$ which,
together with the conjugate representations in $\mathfrak{N}_{d=3}^-$,
complete the $\mathfrak{N}_{d=3}=\mathbf{56}_{+1}\oplus \mathbf{56}^{\prime
}_{-1}$. There are no roots with $n=5$ while those with $n=8$ are 7 and,
with their negative, define the generators of the coset $\mathfrak{sl}%
(D-2)\ominus \mathfrak{gl}(D-3)=\mathfrak{sl}(8)\ominus \mathfrak{gl}(7)$.

\paragraph{$D=11$.}

We end this analysis with the $D=11$ case discussed in the previous Section.
The relevant level decomposition is with respect to the root $\alpha_2$.
With $n_2=0$ we have the positive roots of $\mathfrak{gl}(8)$. In this case
we only have $d=3$ and the $\alpha^{(3)}$-roots consist of the 56 $n_2=1$
and the 28 $n_2=2$ ones defining the $\mathrm{SL}(8)\times \mathrm{SO}(1,1)$%
-representation $\mathbf{56}_{+1}\oplus \mathbf{28}^{\prime }_{+2}$ in $%
\mathfrak{N}_{d=3}^+$ which, together with the conjugate representations in $%
\mathfrak{N}_{d=3}^-$, completes the $\mathfrak{N}_{d=3}=\mathbf{84}\oplus
\mathbf{84}^{\prime }$. Finally the 8 $n_2=3$ roots, with their negative,
define the generators of the coset $\mathfrak{sl}(D-2)\ominus \mathfrak{gl}%
(D-3)=\mathfrak{sl}(9)\ominus \mathfrak{gl}(8)$.

\end{document}